\documentclass[draftclsnofoot, onecolumn]{IEEEtran}
\usepackage{cite}
\usepackage{amsfonts}
\usepackage{amssymb}
\usepackage{amsmath}
\usepackage{mathrsfs}
\usepackage{bm}
\usepackage{graphicx}
\usepackage{subfigure}

\IEEEaftertitletext{\vspace{-1.5\baselineskip}}

\newtheorem{theorem}{Theorem}
\newtheorem{assumption}{Assumption}

\newtheorem{lemma}{Lemma}
\newtheorem{remark}{Remark}

\begin{document}

\title{Performance of Network-Assisted Full-Duplex for Cell-Free Massive MIMO}
\author{Dongming~Wang, Menghan Wang, Pengcheng~Zhu, Jiamin Li, Jiangzhou~Wang, and~Xiaohu~You
\thanks{D. Wang, M. Wang, P. Zhu, J. Li and X. You are with the National Mobile Communications Research Laboratory, Southeast University, Nanjing 210096, China.}
\thanks{J. Wang is with the School of Engineering and Digital Arts, University of Kent, Canterbury CT2 7NT, U.K..}}

\maketitle

\begin{abstract}
Network assisted full-duplex (NAFD) is a spatial-division duplex technique for future wireless networks with cell-free massive multiple-input multiple-output (CF massive MIMO) network,
where a large number of remote antenna units (RAUs), either using half-duplex or full-duplex, jointly support truly flexible duplex
including time-division duplex, frequency-division duplex and full duplex on demand of uplink and downlink traffic
by using network MIMO methods. With NAFD, bi-directional data rates of the wireless network could be increased and end-to-end delay could be reduced.
In this paper, the spectral efficiency of NAFD communications in CF massive MIMO network with imperfect channel state information (CSI)
is investigated under spatial correlated channels. Based on large dimensional random matrix theory, the deterministic equivalents
for the uplink sum-rate with minimum-mean-square-error (MMSE) receiver as well as the downlink sum-rate with zero-forcing (ZF) and regularized
zero-forcing (RZF) beamforming are derived. Numerical results show that under various environmental settings, the deterministic equivalents are accurate in both a large-scale system and system with a finite number of antennas.
It is also shown that with the downlink-to-uplink interference cancellation, the uplink spectral efficiency of CF
massive MIMO with NAFD could be improved. The spectral efficiencies of NAFD with different duplex configurations such as in-band full-duplex, and half-duplex are compared. With the same total numbers of transmit and receive antennas, NAFD with half-duplex RAUs offers a higher spectral efficiency. To alleviate the uplink-to-downlink interference, a novel genetic algorithm based user scheduling strategy (GAS) is proposed.
Simulation results show that the achievable downlink sum-rate by using the GAS is greatly improved compared to that by using the random user scheduling.
\end{abstract}


\section{Introduction}
In recent years, with the rapid development of mobile internet, the global wireless data services have shown explosive growth. The diverse and personalized mobile services, such as high-speed uplink (UL) and downlink (DL) wireless data services, and delay sensitive wireless services, result in higher requirements for duplex techniques in next generation mobile communication systems.

In fourth-generation (4G) mobile communication systems, fixed duplexes such as time-division duplex (TDD) and frequency-division duplex (FDD) have been adopted. Fifth-generation (5G) new radio (NR) has been specified to support flexible duplex which can dynamically allocate uplink and downlink resources and then improve the utilization of resources \cite{Erik_Dahlman}. Recently, in-band full-duplex has attracted a great deal of attention\cite{asabharwal_inband_2014}. By using advanced self-interference cancellation, a full-duplex wireless transceiver may transmit and receive simultaneously in the same frequency band, called co-frequency co-time full-duplex (CCFD), and double the spectral efficiency for wireless networks\cite{Choi_JI,asabharwal_inband_2014}. While this is true for point-to-point links, it is not necessarily true for large-scale networks due to increased cross link interference (CLI). Stochastic geometry based analysis \cite{AlAmmouri,BiZhou} has shown the susceptibility of UL-to-DL interference and the negative impact that in-band full duplex communication can impose on the uplink transmission.

In \cite{osimeone_fullduplex_2014}, the performance of cloud radio access network (C-RAN) with CCFD remote antenna units (RAUs) was studied by using an information theoretic approach and the analytical results confirmed the significant potential advantages of the C-RAN architecture for in-band full-duplex systems. Coordinated multi-point transmissions (CoMP) for in-band wireless full-duplex (CoMPflex) was proposed in \cite{hthomsen_compflex_2016}, by considering emulation of a full-duplex cellular base-station (BS) and using two spatially separated and coordinated half-duplex BSs. The simulation results in \cite{hthomsen_full_2016} have shown that due to the increased density of BSs, the outage performances of both uplink and downlink in CoMPflex are better than that in cellular systems with CCFD BSs. Large-scale antenna systems, including massive MIMO systems and large-scale distributed antenna systems (DAS)\cite{xyou_cooperative_2010,ldai_comparative_2011,hzhu_performance_2011,jwang_asymptotic_2015}, could enable duplexing approaches over the spatial domain \cite{fboccardi_why_2016,Xiaochen_2017,yxin_bidirectional_2017}. Considering massive MIMO and distributed antennas, a novel transmission strategy was proposed in \cite{yxin_bidirectional_2017,yxin_antenna_2017} to improve both uplink and downlink traffic by flexibly adjusting the number of receiving and transmitting RAUs. However, due to the DL-to-UL interference, the spatial degree of freedom was not fully exploited in \cite{Xiaochen_2017,yxin_bidirectional_2017}.

CLI is also a challenging problem for supporting flexible duplex in 5G NR \cite{CMCC}, as well as the hybrid deployment of half-duplex and CCFD BSs.
For the future mobile systems such as beyond 5G or sixth generation (6G) systems, in order to support truly flexible duplex including dynamic TDD, flexible FDD and CCFD on demand of uplink and downlink traffic, we must solve the CLI problem from both network architecture and wireless transmission.

Recently, a novel concept called cell-free (CF) massive multiple-input multiple-output (MIMO) was proposed to overcome the inter-cell interference by innovating the cellular architecture\cite{Ngo_Spawc_2015,Ngo_CF_SC_2017}.
Compared to small-cell network, CF massive MIMO potentially has a large spectral efficiency\cite{Ngo_CF_SC_2017}. From the point of view of baseband transmission, the performance gain of CF massive MIMO comes from the joint processing of a large number of geographically distributed RAUs\cite{Wang_VTC_2015}. Our recent prototyping system in large-scale distributed MIMO has demonstrated that the data rate of 10Gbps could be achieved by a 128x128 large-scale distributed MIMO (or CF massive MIMO) with 100MHz bandwidth\cite{Fengyuan_2018}. Different from the existing works, in this paper, we propose a network-assisted full duplex (NAFD) to unify the flexible duplex, hybrid-duplex, full-duplex, and other duplex methods \cite{hthomsen_compflex_2016,Xiaochen_2017,yxin_bidirectional_2017,KangSong} under the CF massive MIMO network, and solve the CLI problem to achieve truly flexible duplex, which is essential in the 5G NR\cite{CMCC} and 6G systems. To demonstrate the superiority of NAFD with CF massive MIMO, the performance comparisons between NAFD and the existing duplexing methods are worth studying.

Channel state information (CSI) plays a very important role in the uplink and downlink transmission of CF massive MIMO, as well as the CLI cancellation of NAFD. In \cite{koh2018feasibility}, pilot design for CCFD large-scale MIMO under C-RAN architecture was studied, and analysis of spectral efficiency with imperfect CSI was also given. Considering the non-ideal CSI is crucial to demonstrate the performance of NAFD with imperfect interference cancellation for both uplink and downlink.

In this paper, a unified system model is given for the NAFD by considering imperfect CSI and the impact of spatial correlations. For downlink transmission, both zero forcing (ZF) and regularized ZF (RZF) precoders are considered. After the DL-to-UL interference cancellation, it is assumed that an minimum-mean-square-error (MMSE) joint detection is adopted. To mitigate the UL-to-DL interference, joint scheduling is studied. The main contributions of this paper are three-fold:
\begin{itemize}
  \item First, under CF massive MIMO network, the proposed NAFD is quite general and the derived results can be applied to its special cases such as CCFD massive MIMO, CoMPflex, hybrid full-duplex and half-duplex, and full-duplex C-RAN. The performance comparisons are given for NAFD with half-duplex RAUs, CCFD-CRAN, and CCFD massive MIMO. The results demonstrate that due to the increasing density of RAUs, NAFD with half-duplex RAUs has best performance among these three schemes.
  \item Second, a large system analysis is given for the spectral efficiency of NAFD with the consideration of imperfect CSI and the impact of spatial correlations. With MMSE receiver for uplink transmission as well as ZF or RZF precoders for downlink transmission, the deterministic equivalents of sum-rate for the CF massive MIMO with NAFD are provided based on large dimensional random matrix theory. Given statistical channel knowledge, the spectral efficiency of the system can be approximated without knowing the actual channel realization. Numerical simulations show that the deterministic equivalents are accurate in both large-scale system and system with a finite number of antennas.
  \item Third, a novel genetic algorithm based user scheduling strategy (GAS) is proposed, which provides a low complexity scheme to alleviate the UL-to-DL interference.
\end{itemize}

The remainder of this paper is organized as follows. Section II presents the transmission model and channel model of CF massive MIMO with NAFD. In Section III, we give the deterministic equivalents of the sum-rates of downlink transmission with ZF and RZF precoders. In Section IV, after mitigating the DL-to-UL interference, we derive the deterministic equivalents of the sum-rate for uplink with MMSE receiver. Section V introduces the GAS aiming at minimizing the UL-to-DL interference. Simulation results and theoretical comparisons are given in Section VI. Finally, Section VII concludes the paper. Most technical proofs are presented in the Appendix. In these proofs, we apply several important lemmas collected in Appendices.

The notation adopted in this paper conforms to the following convention. Uppercase and lowercase boldface letters are used to denote matrices and vectors, respectively. An $M \times M$ identity matrix is denoted by $\bm{I}_M$. We use $\left[ {\bm{A}} \right]_{k,l}$ to denote the $\left( {k,l} \right)$th entry of the matrix $\bm A$, and $a_k$ denotes the $k$th entry of the column vector ${\bm{a}}$. $\left|  \cdot  \right|$ denotes the absolute value of a scalar. ${\left[  \cdot  \right]^{\text{T}}}$ and ${\left[  \cdot  \right]^{\text{H}}}$ represent the transpose and the Hermitian transpose of a vector or a matrix, respectively. ${\mathbb{C}^{m \times n}}$ denotes the set of $m \times n$ complex valued matrices. $\text{E}\left[  \cdot  \right]$ and $\text{cov}\left[  \cdot  \right]$ represent mathematical expectation and covariance, respectively. $\text{tr}\left[  \cdot  \right]$ calculates the trace of matrix. ${\lambda _{\min }}\left( {\bm{A}} \right)$ is the minimum eigenvalue of the Hermitian matrix ${\bm{A}}$. $\text{diag}\left( {\bm{x}} \right)$ denotes a diagonal matrix with ${\bm{x}}$ along its main diagonal. The distribution of a circularly symmetric complex Gaussian (CSCG) random variable with zero mean and variance ${\sigma ^2}$ is denoted as ${\cal C}{\cal N}\left( {0,{\sigma ^2}} \right)$. $\left| \mathcal{K} \right|$ represents the cardinality of set $\mathcal{K}$. $\xrightarrow{{a.s.}}$ denotes the almost sure (a.s.) convergence.

\begin{figure}
\centering
\includegraphics[scale=.25]{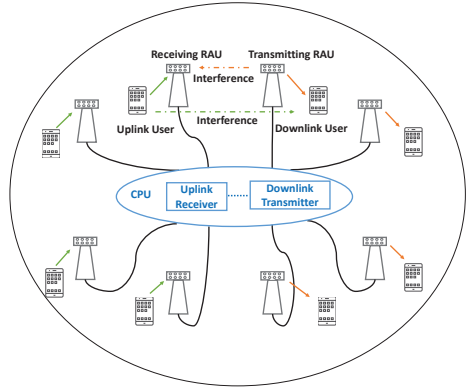}
\caption{Sytem configuration of CF massive MIMO with NAFD.}
\label{Figure1}
\end{figure}

\section{System Model}
In CF massive MIMO, an implementation of a truly flexible duplex is illustrated in Fig.\ref{Figure1}, where there are no cells but both uplink and downlink are jointly served in the same time-frequency resources by distributed RAUs. Each RAU is connected to the central processing unit (CPU) through fronthaul link, such as fiber optics cable or mmwave radio link, and then the baseband processing is jointly carried out at the CPU. Each RAU has a transceiver device performing either transmitting, receiving or both transmitting and receiving simultaneously which are decided by CPU according to the traffic load of the whole network. For CCFD RAUs, the self-interference of intra-RAU could be cancelled in analogy domain. \emph{To give a unified description, for CCFD RAUs, we could virtually look it as two RAUs, one for uplink reception and one for downlink transmission}.
In CPU, the residual DL-to-UL interference of inter-RAU (including CCFD RAUs and half-duplex RAUs) could be cancelled in digital domain. From this aspect, in-band full-duplex could be achieved under CF massive MIMO with existing half-duplex hardware devices. \emph{This is the reason it is named as NAFD.}

Although user equipment can be with CCFD transceiver, to reduce the implementation complexity, only half-duplex transceiver is considered at user equipment. Each user could operate in flexible duplex mode, such that the evolution of legacy systems can be achieved efficiently. To mitigate UL-to-DL interference, joint scheduling at CPU or interference cancellation at user equipments could be employed.

\begin{figure}[htbp]
\centering
\subfigure[NAFD for dynamic TDD.]{
\begin{minipage}[t]{0.4\linewidth}
\centering
\includegraphics[scale=.28]{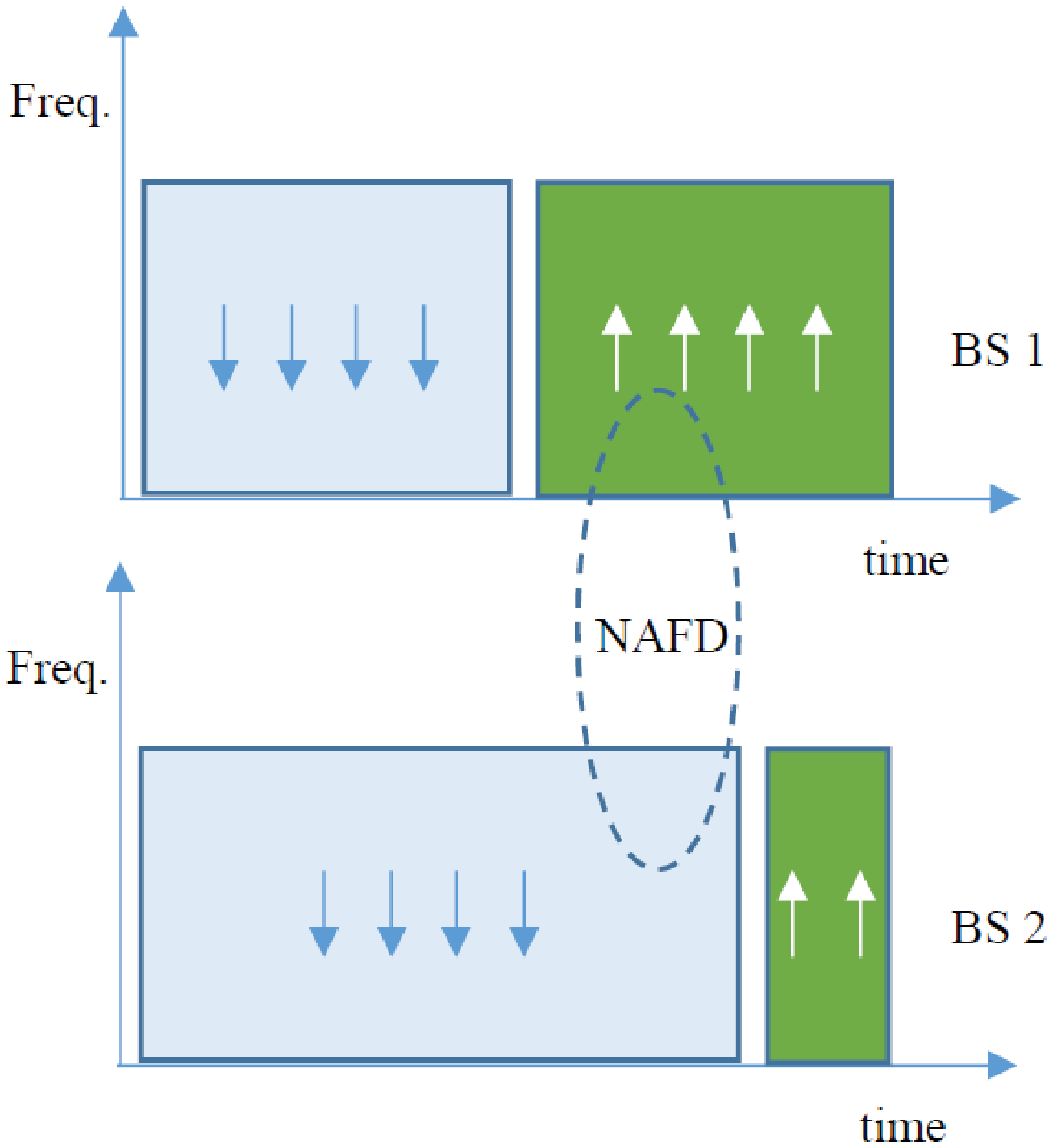}
\label{fig1a}
\end{minipage}
}
\subfigure[NAFD for flexible FDD.]{
\begin{minipage}[t]{0.4\linewidth}
\centering
\includegraphics[scale=.28]{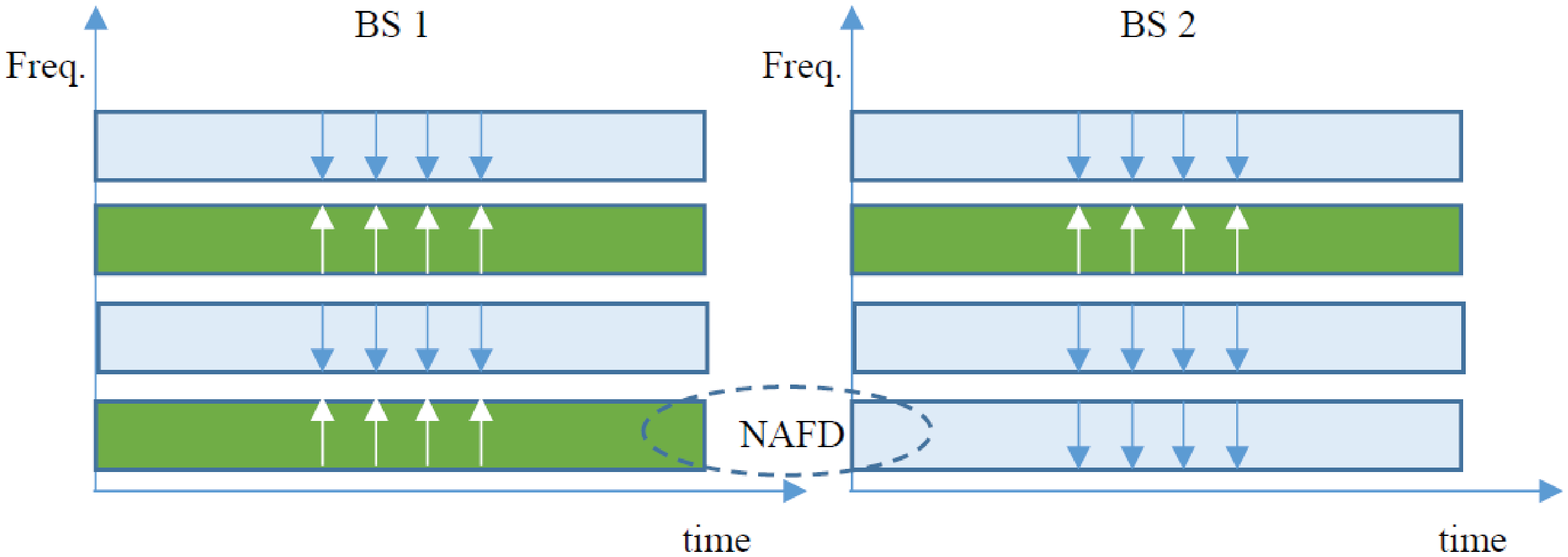}
\label{fig1b}
\end{minipage}
}
\centering
\caption{NAFD for flexible duplex.}
\end{figure}

CF massive MIMO with NAFD can support flexible duplex for 5G NR. Taking dynamic TDD as an example in Fig.\ref{fig1a}, the DL-to-UL can be cancelled with NAFD. Most importantly, under CF massive MIMO architecture, its implementation can be scaled to a large number of RAUs. Fig.\ref{fig1b} shows an application of NAFD for flexible FDD.

In conclusion, with DL-to-UL interference cancellation and UL-to-DL interference mitigation, CF massive MIMO with NAFD can realize truly flexible duplex including dynamic TDD, flexible FDD and full-duplex, and then utilize the UL/DL resource efficiently and flexibly. In the following, we will give the signal models for both uplink and downlink transmissions, as well as the equivalent channel model with imperfect CSI.

%

\subsection{Transmission Model}\label{sec:transmission_model}
As shown in Fig.\ref{Figure1}, it is assumed that there are ${N_{\text{U}}}$ RAUs performing uplink transmission and ${N_{\text{D}}}$ RAUs performing downlink transmission, and each RAU is equipped with $M$ antennas. Each user is assumed with single antenna. ${K_{\text{U}}}$ and ${K_{\text{D}}}$ denote the numbers of user equipments (UEs) active in the uplink and downlink, respectively.

In the uplink, the signal received by the CPU can be written as
\begin{equation}\label{eq:trans_up_model}
{{\bm{y}}_{{\text{ul}}}} = \sqrt {{p_{{\text{ul}},i}}} {{\bm{g}}_{{{\text{ul}},i}}}{x_i} + \sum\limits_{j = 1,j \ne i}^{{K_{\text{U}}}} {\sqrt {{p_{{\text{ul}},j}}} {{\bm{g}}_{{{\text{ul}},j}}}{x_j}}  + \sum\limits_{k = 1}^{{K_{\text{D}}}} {{{\bm{G}}_{\text{I}}}{{\bm{w}}_{k}}{s_k}}  + {{\bm{z}}_{{\text{ul}}}},
\end{equation}
where ${p_{{\text{ul}},i}}$ is the transmission power of the $i$th UE active in the uplink, ${{\bm{g}}_{{{\text{ul}},i}}} = {\left[ {{\bm{g}}_{{{\text{ul}},i,1}}^{\text{T}},\cdots,{\bm{g}}_{{{\text{ul}},i,}{{N}_{\text{U}}}}^{\text{T}}} \right]^{\text{T}}} \in {\mathbb{C}^{M{N_{\text{U}}} \times 1}},$ ${\bm{g}}_{{{\text{ul}},i,n}}^{} \in {\mathbb{C}^{M \times 1}}$ is the channel vector between the $i$th UE and the $n$th receiving RAU, ${{\bm{G}}_{{\text{ul}}}} \triangleq \left[ {{{\bm{g}}_{{{\text{ul}},1}}},\cdots,{{{\bm{g}}_{{\text{ul}},{{K}_{\text{U}}}}}}} \right] \in {\mathbb{C}^{M{N_{\text{U}}} \times {K_{\text{U}}}}}$ defines the uplink channel matrix, ${x_i}$ is the data symbol transmitted by the $i$th UE with $\text{E}\left( {{x_i}x_i^{\text{H}}} \right) = 1$.

The co-existence of downlink and uplink transmissions causes DL-to-UL interference. Let
$${{\bm{G}}_{\text{I}}} = [{{\bm{g}}_{{{\text{I}},1}}},\cdots,{{\bm{g}}_{{{\text{I}},M}{{N}_{\text{D}}}}}] \in {\mathbb{C}^{M{N_{\text{U}}} \times M{N_{\text{D}}}}}$$
be the qusi-static channels between transmitting RAUs and receiving RAUs, wherein
$${{\bm{g}}_{{{\text{I}},i}}} = {\left[ {{\bm{g}}_{{{\text{I}},i,1}}^{\text{T}},\cdots,{\bm{g}}_{{{\text{I}},i,}{{N}_{\text{U}}}}^{\text{T}}} \right]^{\text{T}}} \in {\mathbb{C}^{M{N_{\text{U}}} \times 1}}$$
and ${{\bm{g}}_{{{\text{I}},i,j}}} \in {\mathbb{C}^{M \times 1}}$ is the channel vector between the $i$th antenna of transmitting RAUs and the $j$th receiving RAU.  ${{\bm{w}}_{k}} \in {\mathbb{C}^{M{N_{\text{D}}} \times 1}}$ is the precoding vector between transmitting RAUs and the $k$th UE. ${\bm{W}} \triangleq [{{\bm{w}}_{1}}\cdots,{{\bm{w}}_{{{K}_{\text{D}}}}}] \in {\mathbb{C}^{M{N_{\text{D}}} \times {K_{\text{D}}}}}$ represents the precoding matrix between transmitting RAUs and all active UEs in the downlink. ${s_k}$ is the data symbol sending to the $k$th active UE in the downlink with $\text{E}\left( {{s_k}s_k^{\text{H}}} \right) = 1$. ${{\bm{z}}_{{\text{ul}}}} \sim \mathcal{C}\mathcal{N}\left( {{\bm{0}},\sigma _{\text{ul}}^2{{\bm{I}}_{{M}{{N}_{\text{U}}}}}} \right)$ is the complex additive white Gaussian noise (AWGN) vector.

In the downlink, the signal received by the $k$th UE can be given by
\begin{equation}\label{eq:trans_down_model}
{y_{{\text{dl}},k}} = {\bm{g}}_{{{\text{dl}},k}}^{\text{H}}{{\bm{w}}_{k}}{s_k} + \sum\limits_{j = 1,j \ne k}^{{K_{\text{D}}}} {{\bm{g}}_{{{\text{dl}},k}}^{\text{H}}{{\bm{w}}_{j}}{s_j}}  + \sum\limits_{i = 1}^{{K_{\text{U}}}} {\sqrt {{p_{{\text{ul}},i}}} {u_{k,i}}{x_i}}  + {z_{{\text{dl}},k}},
\end{equation}
where ${\bm{g}}_{{{\text{dl}},k}}^{\text{H}} = \left[ {{\bm{g}}_{{{\text{dl}},k,1}}^{\text{H}},\cdots,{\bm{g}}_{{{\text{dl}},k,}{{N}_{\text{D}}}}^{\text{H}}} \right] \in {\mathbb{C}^{1 \times M{N_{\text{D}}}}}$, ${\bm{g}}_{{{\text{dl}},k,n}}^{\text{H}} \in {\mathbb{C}^{1 \times M}}$ is the channel vector between the $n$th transmitting RAU and the $k$th UE. ${{\bm{G}}_{{\text{dl}}}} \triangleq [{{\bm{g}}_{{{\text{dl}},1}}},\cdots,{{\bm{g}}_{{{\text{dl}},}{{K}_{\text{D}}}}}] \in {\mathbb{C}^{M{N_{\text{D}}} \times {K_{\text{D}}}}}$ defines the channel between transmitting RAUs and all active UEs in the downlink. ${u_{k,i}}$ denotes the UL-to-DL interference channel between the $i$th UE active in the uplink and the $k$th UE active in the downlink. The additive noise is modelled as ${{z}_{{\text{dl}},k}} \sim \mathcal{C}\mathcal{N}\left( {0,\sigma _{\text{dl}}^2} \right)$.

Let ${{\bm{W}}_i} = [{{\bm{w}}_{{1,i}}},\cdots,{{\bm{w}}_{{{K}_{\text{D}}}{,i}}}] \in {\mathbb{C}^{M \times {K_{\text{D}}}}}$ denote the overall precoding matrix of the $i$th transmitting RAU. Assume that the $i$th transmitting RAU satisfies the following transmit power constraint:
\begin{equation}\label{eq:power_constraint}
\text{tr}\left( {{{\bm{W}}_{i}}{\bm{W}}_{i}^{\text{H}}} \right) = \text{tr}\left( {{{\bm{E}}_{i}}{\bm{WW}}_{}^{\text{H}}{{\bm{E}}_{i}}} \right) \leqslant MP,
\end{equation}
for $i = 1,\cdots,{N_{\text{D}}}$, where $P > 0$ is the parameter that decides on the per-antenna power budget and
$${{\bm{E}}_{i}} \triangleq \text{diag}\left( {0,\cdots,0,\underbrace {1,\cdots,1}_M,0,\cdots,0} \right).$$
Although the conventional sum-power constraint can achieve better performance, per-RAU power constraint adopted is more practical. Related discussions of the two power constraints can be seen in \cite{osomekh_cooperative_2009}.

\begin{remark}
The transmission models of CF massive MIMO with NAFD shown in (\ref{eq:trans_up_model}) and (\ref{eq:trans_down_model}) are quite general, with full-duplex massive MIMO, CoMPflex, and CCFD C-RAN as its special cases. For example, in the CCFD massive MIMO discussed in \cite{yli_spectral_2017}, there was one RAU (i.e. ${N_{\text{U}}}={N_{\text{D}}}=1$) and the receiving RAU and the transmitting RAU are co-located. For C-RAN based full-duplex discussed in \cite{osimeone_fullduplex_2014}, the numbers of receiving and transmitting RAUs were equal (i.e. ${N_{\text{U}}}={N_{\text{D}}}$) and one receiving RAU and one transmitting RAU are paired and co-located.
\end{remark}

\subsection{Channel Model}\label{sec:channel_model}
In practical MIMO, due to limited angular spread and insufficient antenna spacing, the impact of spatial correlations has to be considered. For this purpose, the channel vector between the $i$th UE and receiving RAUs is modeled as
\begin{equation}\label{eq:up_channel}
{{\bm{g}}_{{{\text{ul}},i}}} = {\bm{T}}_{{{\text{ul}},i}}^{\frac{{1}}{{2}}}{{\bm{h}}_{{{\text{ul}},i}}},
\end{equation}
where ${{\bm{h}}_{{{\text{ul}},i}}} = {\left[ {{\bm{h}}_{{{\text{ul}},i,1}}^{\text{T}},\cdots,{\bm{h}}_{{{\text{ul}},i,}{{N}_{\text{U}}}}^{\text{T}}} \right]^{\text{T}}} \in {\mathbb{C}^{M{N_{\text{U}}} \times 1}}$ and ${{\bm{h}}_{{{\text{ul}},i,n}}} \in {\mathbb{C}^{M \times 1}} \sim \mathcal{C}\mathcal{N}\left( {0,\frac{1}{M}{{\bm{I}}_{M}}} \right)$. ${{\bm{T}}_{{{\text{ul}},i}}} = \text{diag}({{\bm{T}}_{{{\text{ul}},i,1}}},\cdots,$ ${{\bm{T}}_{{{\text{ul}},i,}{{N}_{\text{U}}}}}) \in {\mathbb{C}^{M{N_{\text{U}}} \times M{N_{\text{U}}}}}$, ${{\bm{T}}_{{{\text{ul}},i,n}}} \in {\mathbb{C}^{M \times M}}$ is a deterministic nonnegative-definite matrix, which characterizes the pathloss and the spatial correlations of the transmitted signals across the antenna elements of the $n$th receiving RAU.

Similarly, the channel vector between the transmitting RAUs and the $i$th UE active in the downlink, the interference channel between the $i$th antenna of transmitting RAUs and all receiving RAUs, the interference channel between the $i$th UE active in the uplink and the $k$th UE active in the downlink are, respectively, modelled as
\begin{eqnarray}
{{\bm{g}}_{{{\text{dl}},i}}} = {\bm{T}}_{{{\text{dl}},i}}^{\frac{{1}}{{2}}}{{\bm{h}}_{{{\text{dl}},i}}}, \;\; {{\bm{g}}_{{{\text{I}},i}}}  = {\bm{T}}_{{{\text{I}},i}}^{\frac{{1}}{{2}}}{{\bm{h}}_{{{\text{I}},i}}}, \;\; {u_{k,i}} = T_{k,i}^{\frac{1}{2}}{h_{k,i}}.
\end{eqnarray}

Assuming that only an imperfect CSI of the true channel is available, we model the uplink channel by \cite{cwang_adaptive_2006, tyoo_MIMO_2004}
\begin{equation}\label{eq:es_up_channel}
{{\hat{\bm g}}_{{{\text{ul}},i}}} = {\bm{T}}_{{{\text{ul}},i}}^{\frac{{1}}{{2}}}\left( {{{\bm{\Lambda }}_{{{\text{ul}},i}}}{{\bm{h}}_{{{\text{ul}},i}}} + {{\bm{\Omega }}_{{{\text{ul}},i}}}{{\bm{z}}_{{{\text{ulp}},i}}}} \right) = {\bm{T}}_{{{\text{ul}},i}}^{\frac{{1}}{{2}}}{{\hat{\bm h}}_{{{\text{ul}},i}}},
\end{equation}
where ${{\bm{z}}_{{{\text{ulp}},i}}} \in {\mathbb{C}^{M{N_{\text{U}}} \times 1}}$ has the same statistical properties as ${{\bm{h}}_{{{\text{ul}},i}}}$ but is independent from ${{\bm{h}}_{{{\text{ul}},i}}}$ and ${{\bm{z}}_{{\text{ul}}}}$. Define
\begin{eqnarray}
\label{eq:phi_ul}
{\varphi _{{\text{ul}},i,n}}\!\!\!&\triangleq&\!\!\! \sqrt {1 - \tau _{{\text{ul}},i,n}^2}, \\
\label{eq:A_ul}
{{\bm{\Lambda }}_{{{\text{ul}},i}}}\!\!\!&\triangleq&\!\!\! \text{diag}\left( {{\varphi _{{\text{ul}},i,1}}{{\bm{I}}_{M}},\cdots,{\varphi _{{\text{ul}},i,{N_{\text{U}}}}}{{\bm{I}}_{M}}} \right), \\
\label{eq:O_ul}
{{\bm{\Omega }}_{{{\text{ul}},i}}}\!\!\!&\triangleq&\!\!\! \text{diag}\left( {{\tau _{{\text{ul}},i,1}}{{\bm{I}}_{M}},\cdots,{\tau _{{\text{ul}},i,{N_{\text{U}}}}}{{\bm{I}}_{M}}} \right),
\end{eqnarray}
where the parameter $\tau _{{\text{ul}},i,n}^{} \in \left[ {0,1} \right]$ denotes the accuracy or quality of the channel estimation. Specifically, $\tau _{{\text{ul}},i,n}^{} = 0$ corresponds to perfect CSI, whereas for $\tau _{{\text{ul}},i,n}^{} = 1$ the CSI is completely uncorrelated to the true channel.

In an analogous fashion, the imperfect estimation of downlink channel and the interference channel between transmitting RAUs and receiving RAUs respectively are
\begin{eqnarray}
\label{eq:es_down_channel}
{{\hat{\bm g}}_{{{\text{dl}},i}}}\!\!\! &=&\!\!\! {\bm{T}}_{{{\text{dl}},i}}^{\frac{{1}}{{2}}}\left( {{{\bm{\Lambda }}_{{{\text{dl}},i}}}{{\bm{h}}_{{{\text{dl}},i}}} + {{\bm{\Omega }}_{{{\text{dl}},i}}}{{\bm{z}}_{{{\text{dlp}},i}}}} \right) = {\bm{T}}_{{{\text{dl}},i}}^{\frac{{1}}{{2}}}{{\hat{\bm h}}_{{{\text{dl}},i}}}, \\
\label{eq:es_RAUI_channel}
{{\hat{\bm g}}_{{{\text{I}},i}}}\!\!\! &=&\!\!\! {\bm{T}}_{{{\text{I}},i}}^{\frac{{1}}{{2}}}\left( {{{\bm{\Lambda }}_{{{\text{I}},i}}}{{\bm{h}}_{{{\text{I}},i}}} + {{\bm{\Omega }}_{{{\text{I}},i}}}{{\bm{z}}_{{{\text{I}},i}}}} \right) = {\bm{T}}_{{{\text{I}},i}}^{\frac{{1}}{{2}}}{{\hat{\bm h}}_{{{\text{I}},i}}}.
\end{eqnarray}
The channel model (\ref{eq:es_RAUI_channel}) between transmitting RAUs and receiving RAUs includes intra-RAU and inter-RAU channels. For the inter-RAU DL-to-UL channels, they can be estimated by pilot signals with very low overhead due to the quasi-static property of the links. For a full-duplex RAU, it would be appropriate to deal with the self-interference in analogy domain. Actually, (\ref{eq:es_RAUI_channel}) also models the residual self-interference for intra-RAU.

\section{Deterministic Equivalent Sum-Rate for Downlink}\label{sec:DE_DL}
Assuming that only imperfect CSI is available and channels have independent spatial correlations, this section introduces deterministic approximations of the downlink ergodic sum-rates under RZF and ZF precoding.

\subsection{Regularized Zero-Forcing Precoding}
Consider the RZF precoding matrix\cite{mjoham_transmit_2002,cbpeel_vector_2005}
\begin{equation}\label{eq:RZF_matrix}
{{\bm{W}}_{{\text{rzf}}}} = \xi {\left( {{{{\hat{\bm G}}}_{{\text{dl}}}}{\hat{\bm G}}_{{\text{dl}}}^{\text{H}} + \alpha {{\bm{I}}_{{M}{{N}_{\text{D}}}}}} \right)^{ - 1}}{{\hat{\bm G}}_{{\text{dl}}}} = \xi {\bm{C}}_{{\text{dl}}}^{{\text{-1}}}{{\hat{\bm G}}_{{\text{dl}}}},
\end{equation}
the $k$th column of which is
\begin{equation}\label{eq:RZF_vector}
{{\bm{w}}_{{{\text{rzf}},k}}} = \xi {\left( {{{{\hat{\bm G}}}_{{\text{dl}}}}{\hat{\bm G}}_{{\text{dl}}}^{\text{H}} + \alpha {{\bm{I}}_{{M}{{N}_{\text{D}}}}}} \right)^{ - 1}}{{\hat{\bm g}}_{{{\text{dl}},k}}} = \xi {\bm{C}}_{{\text{dl}}}^{{\text{-1}}}{{\hat{\bm g}}_{{{\text{dl}},k}}},
\end{equation}
where ${{\bm{C}}_{{\text{dl}}}} \triangleq {{\hat{\bm G}}_{{\text{dl}}}}{\hat{\bm G}}_{{\text{dl}}}^{\text{H}} + \alpha {{\bm{I}}_{{M}{{N}_{\text{D}}}}}$, ${{\hat{\bm G}}_{{\text{dl}}}} \triangleq [{{\hat{\bm g}}_{{{\text{dl}},1}}},\cdots,{{\hat{\bm g}}_{{{\text{dl}},}{{K}_{\text{D}}}}}] \in {\mathbb{C}^{M{N_{\text{D}}} \times {K_{\text{D}}}}}$ is the channel estimation available at the transmitter. $\alpha  > 0$ represents the regularization parameter, $\xi $ is a normalization scalar to
fulfill the per-RAU transmit power constraint (\ref{eq:power_constraint}). Such precoding is a practical linear precoding scheme to control inter-user interference and increase the downlink sum-rate by choosing the proper $\alpha$ \cite{cbpeel_vector_2005}. It includes two well-known precoding schemes with $\alpha  = 0$ being the ZF precoding, and $\alpha  \to \infty $ being the matched-filter precoding.

From the per-RAU power constraint (\ref{eq:power_constraint}), we obtain $\xi _i^2$ as
\begin{equation}\label{eq:xi_i_2}
\xi _i^2 = \frac{{MP}}{{\text{tr}\left( {{{\bm{W}}_{{{\text{rzf}},i}}}{\bm{W}}_{{{\text{rzf}},i}}^{\text{H}}} \right)}} = \frac{{\frac{P}{{{N_{\text{D}}}}}}}{{\frac{1}{{M{N_{\text{D}}}}}\text{tr}\left( {{{\bm{E}}_{i}}{\bm{C}}_{{\text{dl}}}^{{\text{-1}}}{{{\hat{\bm G}}}_{{\text{dl}}}}{\hat{\bm G}}_{{\text{dl}}}^{\text{H}}{\bm{C}}_{{\text{dl}}}^{{\text{-1}}}{{\bm{E}}_{i}}} \right)}},
\end{equation}
for $i = 1,\cdots,{N_{\text{D}}}$. To satisfy (\ref{eq:power_constraint}), we set ${\xi ^2} = \mathop {\min }\limits_i \left\{ {\xi _i^2} \right\}$ \cite{jzhang_large_2013}. Then, the SINR at the $k$th UE under RZF precoding takes the form
\begin{equation}\label{eq:SINR_dl}
{\gamma _{{\text{dl}},{\text{rzf}},k}} = \frac{{{{\left| {{\bm{g}}_{{{\text{dl}},k}}^{\text{H}}{\bm{C}}_{{\text{dl}}}^{{\text{-1}}}{{{\hat{\bm g}}}_{{{\text{dl}},k}}}} \right|}^2}}}{{\sum\limits_{j = 1,j \ne k}^{{K_{\text{D}}}} {{{\left| {{\bm{g}}_{{{\text{dl}},k}}^{\text{H}}{\bm{C}}_{{\text{dl}}}^{{\text{-1}}}{{{\hat{\bm g}}}_{{{\text{dl}},j}}}} \right|}^2}}  + {v_{\text{dl}}}}},
\end{equation}
where
\begin{equation}\label{eq:v_dl}
{v_{\text{dl}}} \triangleq \frac{{\sum\limits_{j = 1}^{{K_{\text{U}}}} {{p_{{\text{ul}},j}}{{\left| {{u_{k,j}}} \right|}^2}}  + \sigma _{\text{dl}}^2}}{{{\xi ^2}}} = \left( {\sum\limits_{j = 1}^{{K_{\text{U}}}} {{p_{{\text{ul}},j}}{{\left| {{u_{k,j}}} \right|}^2}}  + \sigma _{\text{dl}}^2} \right)\frac{{{N_{\text{D}}}}}{P}\mathop {\max }\limits_i \frac{1}{{M{N_{\text{D}}}}}\text{tr}\left( {{{\bm{E}}_{i}}{\bm{C}}_{{\text{dl}}}^{{\text{-1}}}{{{\hat{\bm G}}}_{{\text{dl}}}}{\hat{\bm G}}_{{\text{dl}}}^{\text{H}}{\bm{C}}_{{\text{dl}}}^{{\text{-1}}}{{\bm{E}}_{i}}} \right).
\end{equation}

Then, the downlink ergodic sum-rate with RZF beamforming can be defined as
\begin{equation}\label{eq:R_dl_rzf}
{R_{{\text{dl}},{\text{rzf}},{\text{sum}}}} \triangleq \sum\limits_{k = 1}^{{K_{\text{D}}}} {{\text{E}_{{{{\hat{\bm G}}}_{{\text{dl}}}}}}\left\{ {{{\log }_2}\left( {1 + {\gamma _{{\text{dl}},{\text{rzf}},k}}} \right)} \right\}}.
\end{equation}

To derive a deterministic equivalent of the sum-rate for RZF precoding, we consider the large-system regime where $M$, ${K_{\text{D}}}$ and ${K_{\text{U}}}$ approach to infinity at fixed ratios ${\beta _{{\text{dl}}}} = \frac{M}{{{K_{\text{D}}}}}$ and ${\beta _{{\text{ul}}}} = \frac{M}{{{K_{\text{U}}}}}$, such that $0 < \lim \inf {\beta _{{\text{dl}}}} \leqslant \lim \sup {\beta _{{\text{dl}}}} < \infty $, $0 < \lim \inf {\beta _{{\text{ul}}}} \leqslant \lim \sup {\beta _{{\text{ul}}}} < \infty $. For ease of notation, we use $\mathcal{N} \to \infty $ to denote the quantity in such limit. Besides, the following assumptions are required in our derivations.
\begin{assumption}\label{asm:g_dl}
For the channel ${{\hat{\bm g}}_{{{\text{dl}},i}}}$ in (\ref{eq:es_down_channel}), where $i = 1,\cdots,{K_{\text{D}}}$, we have the following hypotheses\cite{swagner_large_2012}:

1) ${{\bm{h}}_{{{\text{dl}},i}}}$ and ${{\bm{z}}_{{{\text{dlp}},i}}}$ have finite eighth-order moment.

2) ${\bm{T}}_{{{\text{dl}},i}}^{}$ has uniformly bounded spectral norm.
\end{assumption}

The deterministic equivalent ${\overline R_{{\text{dl}},{\text{rzf}},{\text{sum}}}}$ of ergodic sum-rate under RZF precoding ${R_{{\text{dl}},{\text{rzf}},{\text{sum}}}}$ is given in the following theorem.

\begin{theorem}\label{thm:R_dl_rzf}
Letting Assumption \ref{asm:g_dl} hold true and as $\mathcal{N} \to \infty $, we have
\begin{equation}\label{eq:R_dl_rzf_as}
\frac{1}{{{K_{\text{D}}}}}\left( {{R_{{\text{dl}},{\text{rzf}},{\text{sum}}}} - {{\overline R}_{{\text{dl}},{\text{rzf}},{\text{sum}}}}} \right)\xrightarrow{{a.s.}}0,
\end{equation}
where
\begin{eqnarray}
\label{eq:R_dl_rzf_DE}
{\overline R_{{\text{dl}},{\text{rzf}},{\text{sum}}}}\!\!\! &=&\!\!\! \sum\limits_{k = 1}^{{K_{\text{D}}}} {{{\log }_2}\left( {1 + {{\overline \gamma }_{{\text{dl}},{\text{rzf}},k}}} \right)}, \\
\label{eq:r_dl_rzf_DE}
{\overline \gamma _{{\text{dl}},{\text{rzf}},k}}\!\!\! &=&\!\!\! \frac{{{{\left( {\overline u_{{\text{dl}},k}^{\text{(2)}}} \right)}^2}}}{{{{\left( {1 + \overline u_{{\text{dl}},k}^{\text{(1)}}} \right)}^2}\left( {{{\overline u}_{{\text{dl}},k}} + {{\overline v}_{\text{dl}}}} \right)}},
\end{eqnarray}
with
\begin{subequations} \label{eq:rzf_six}
\begin{align}
\label{eq:u_dl_k_1}
& \overline u_{{\text{dl}},k}^{\text{(1)}} = \sum\limits_{n = 1}^{{N_{\text{D}}}} {\frac{1}{M}{\text{tr}}\left( {{{\bm{T}}_{{{\text{dl}},k,n}}}{{\bm{\Psi }}_{{{\text{dl}},n}}}} \right)}, \\
\label{eq:u_dl_k_2}
& \overline u_{{\text{dl}},k}^{\text{(2)}} = \sum\limits_{n = 1}^{{N_{\text{D}}}} {\frac{{{\varphi _{{\text{dl}},k,n}}}}{M}\text{tr}\left( {{{\bm{T}}_{{{\text{dl}},k,n}}}{{\bm{\Psi }}_{{{\text{dl}},n}}}} \right)}, \\
\label{eq:v_dl_DE}
& {\overline v_{\text{dl}}} = \frac{{\left( {\sum\limits_{j = 1}^{{K_{\text{U}}}} {{p_{{\text{ul}},j}}T_{k,j}^{}}  + \sigma _{\text{dl}}^2} \right)}}{{MP}}\mathop {\max }\limits_i \left[ {\text{tr}{{\bm{\Psi }}_{{{\text{dl}},i}}} - \alpha \text{tr}{\bm{\Psi }}_{{{\text{dl}},i}}^{\text{2}} + \alpha \sum\limits_{j = 1}^{{K_{\text{D}}}} {{{\dot c}_{{\text{dl}},j,i}}} \text{tr}\left( {{{\bm{\Psi }}_{{{\text{dl}},i}}}{{\bm{T}}_{{{\text{dl}},j,i}}}{{\bm{\Psi }}_{{{\text{dl}},i}}}} \right)} \right], \\
\label{eq:u_dl_k}
& \overline u_{{\text{dl}},k}^{} = \overline u_{{\text{dl}},k}^{\text{(1)}} - \alpha \overline{\dot{u}}_{{\text{dl}},k}^{\text{(1)}} - \frac{{2\overline u_{{\text{dl}},k}^{\text{(2)}}\left( {\overline u_{{\text{dl}},k}^{\text{(2)}} - \alpha \overline{\dot{u}}_{{\text{dl}},k}^{\text{(2)}}} \right)}}{{1 + \overline u_{{\text{dl}},k}^{\text{(1)}}}} + \frac{{{{\left( {\overline u_{{\text{dl}},k}^{\text{(2)}}} \right)}^2}\left( {\overline u_{{\text{dl}},k}^{\text{(1)}} - \alpha \overline{\dot{u}}_{{\text{dl}},k}^{\text{(1)}}} \right)}}{{{{\left( {1 + \overline u_{{\text{dl}},k}^{\text{(1)}}} \right)}^2}}}, \\
\label{eq:u_dl_k_1_dot}
& \overline{\dot{u}}_{{\text{dl}},k}^{\text{(1)}} = \sum\limits_{n = 1}^{{N_{\text{D}}}} {\frac{1}{M}\left[ {{\text{tr}}\left( {{{\bm{T}}_{{{\text{dl}},k,n}}}{\bm{\Psi }}_{{{\text{dl}},n}}^2} \right) - \sum\limits_{j = 1}^{{K_{\text{D}}}} {{{\dot c}_{{\text{dl}},j,n}}} \text{tr}\left( {{{\bm{T}}_{{{\text{dl}},k,n}}}{{\bm{\Psi }}_{{{\text{dl}},n}}}{{\bm{T}}_{{{\text{dl}},j,n}}}{{\bm{\Psi }}_{{{\text{dl}},n}}}} \right)} \right]}, \\
\label{eq:u_dl_k_2_dot}
& \overline{\dot{u}}_{{\text{dl}},k}^{\text{(2)}} = \sum\limits_{n = 1}^{{N_{\text{D}}}} {\frac{{{\varphi _{{\text{dl}},k,n}}}}{M}\left[ {{\text{tr}}\left( {{{\bm{T}}_{{{\text{dl}},k,n}}}{\bm{\Psi }}_{{{\text{dl}},n}}^2} \right) - \sum\limits_{j = 1}^{{K_{\text{D}}}} {{{\dot c}_{{\text{dl}},j,n}}} \text{tr}\left( {{{\bm{T}}_{{{\text{dl}},k,n}}}{{\bm{\Psi }}_{{{\text{dl}},n}}}{{\bm{T}}_{{{\text{dl}},j,n}}}{{\bm{\Psi }}_{{{\text{dl}},n}}}} \right)} \right]}.
\end{align}
\end{subequations}

In (\ref{eq:u_dl_k_1})-(\ref{eq:u_dl_k_2_dot}), ${e_{{\text{dl}},k,i}}$ ($k = 1,\cdots,{K_{\text{D}}},i = 1,\cdots,{N_{\text{D}}}$) form the unique solutions of the following ${K_{\text{D}}} \times {N_{\text{D}}}$ equations
\begin{eqnarray}
\label{eq:pha_dl}
{{\bm{\Psi }}_{{{\text{dl}},i}}}\!\!\! &=&\!\!\! {\left( {\frac{1}{M}\sum\limits_{j = 1}^{{K_{\text{D}}}} {\frac{{{{\bm{T}}_{{{\text{dl}},j,i}}}}}{{1 + \sum\limits_{m = 1}^{{N_{\text{D}}}} {{e_{{\text{dl}},j,m}}} }} + \alpha {{\bm{I}}_{M}}} } \right)^{ - 1}}, \\
\label{eq:e_dl}
{e_{{\text{dl}},k,i}}\!\!\! &=&\!\!\! \frac{1}{M}\text{tr}\left( {{{\bm{T}}_{{{\text{dl}},k,i}}}{{\bm{\Psi }}_{{{\text{dl}},i}}}} \right).
\end{eqnarray}

Besides, ${{\dot{\bm C}}_{{\text{dl}}}} = \left[ {{{\dot c}_{{\text{dl}},k,i}}} \right] \in {\mathbb{C}^{{K_{\text{D}}} \times {N_{\text{D}}}}}$ is a solution to the linear equation given by
\begin{equation}\label{eq:li_eq_Crzf}
{{\bm{\Theta }}_{{\text{dl}}}}\text{vec}\left( {{{{\dot{\bm C}}}_{{\text{dl}}}}} \right) = \text{vec}\left( {{{\bm{\Gamma }}_{{\text{dl}}}}} \right),
\end{equation}
wherein ${{\bm{\Theta }}_{{\text{dl}}}} = \left[ {{{\bm{\Theta }}_{{{\text{dl}},ik}}}} \right] \in {\mathbb{C}^{{N_{\text{D}}}{K_{\text{D}}} \times {N_{\text{D}}}{K_{\text{D}}}}}$ and ${{\bm{\Gamma }}_{{\text{dl}}}} \in {\mathbb{C}^{{K_{\text{D}}} \times {N_{\text{D}}}}}$ take the form
\begin{subequations}
\begin{align}
\label{eq:theta_dl}
& {\left[ {{{\bm{\Theta }}_{{{\text{dl}},ik}}}} \right]_{jl}} = \left\{ {\begin{array}{*{20}{l}}
  { - \frac{1}{{{M^2}}}\frac{1}{{{{\left( {1 + \sum\limits_{m = 1}^{{N_{\text{D}}}} {{e_{{\text{dl}},k,m}}} } \right)}^2}}}\text{tr}\left( {{{\bm{T}}_{{{\text{dl}},k,j}}}{{\bm{\Psi }}_{{{\text{dl}},j}}}{{\bm{T}}_{{{\text{dl}},l,j}}}{{\bm{\Psi }}_{{{\text{dl}},j}}}} \right),}&{\left( {i,k} \right) \ne \left( {j,l} \right)} \\
  {1 - \frac{1}{{{M^2}}}\frac{1}{{{{\left( {1 + \sum\limits_{m = 1}^{{N_{\text{D}}}} {{e_{{\text{dl}},k,m}}} } \right)}^2}}}\text{tr}{{\left( {{{\bm{T}}_{{{\text{dl}},k,i}}}{{\bm{\Psi }}_{{{\text{dl}},i}}}} \right)}^2},}&{\left( {i,k} \right) = \left( {j,l} \right)}
\end{array}} \right., \\
& {\left[ {{{\bm{\Gamma }}_{{\text{dl}}}}} \right]_{k,i}} =  - \frac{1}{M}\sum\limits_{j = 1}^{{N_{\text{D}}}} {\frac{1}{M}\frac{1}{{{{\left( {1 + \sum\limits_{m = 1}^{{N_{\text{D}}}} {{e_{{\text{dl}},k,m}}} } \right)}^2}}}} {\text{tr}}\left( {{{\bm{T}}_{{{\text{dl}},k,j}}}{\bm{\Psi }}_{{{\text{dl}},j}}^{\text{2}}} \right).
\end{align}
\end{subequations}
\end{theorem}

\begin{IEEEproof}
To begin with, the downlink SINR ${\gamma _{\text{dl},\text{rzf},k}}$ in (\ref{eq:SINR_dl}) consists of three terms: 1) ${v_{\text{dl}}}$; 2) ${\left| {{\bm{g}}_{{\text{dl},k}}^{\text{H}}{\bm{C}}_{{\text{dl}}}^{{\text{-1}}}{{{\hat{\bm g}}}_{{\text{dl},k}}}} \right|^2}$; 3) $\sum\limits_{j = 1,j \ne k}^{{K_{\text{D}}}} {{{\left| {{\bm{g}}_{{\text{dl},k}}^{\text{H}}{\bm{C}}_{{\text{dl}}}^{{\text{-1}}}{{{\hat{\bm g}}}_{{\text{dl},j}}}} \right|}^2}} $. We should derive the deterministic equivalents for each of these three terms, which together constitute the final expression for ${\overline \gamma _{\text{dl},\text{rzf},k}}$.

According to (\ref{eq:v_dl}), ${v_{\text{dl}}}$ is comprised of two components: 1) $\sum\limits_{j = 1}^{{K_{\text{U}}}} {{p_{\text{ul},j}}{{\left| {{u_{k,j}}} \right|}^2}} $; 2) $\frac{1}{{M{N_{\text{D}}}}}{\text{tr}}\left( {{{\bm{E}}_{i}}{\bm{C}}_{{\text{dl}}}^{{\text{-1}}}{{{\hat{\bm G}}}_{{\text{dl}}}}{\hat{\bm G}}_{{\text{dl}}}^{\text{H}}} {{\bm{C}}_{{\text{dl}}}^{{\text{-1}}}{{\bm{E}}_{i}}} \right)$. Firstly, using the law of large numbers, we have
\begin{equation} \label{a1_a2}
\frac{1}{{{K_{\text{U}}}}}\sum\limits_{j = 1}^{{K_{\text{U}}}} {{p_{\text{ul},j}}{{\left| {{u_{k,j}}} \right|}^2}}  - \frac{1}{{{K_{\text{U}}}}}\sum\limits_{j = 1}^{{K_{\text{U}}}} {{p_{\text{ul},j}}T_{k,j}^{}} \xrightarrow{{a.s.}}0.
\end{equation}
Secondly, utilizing the decompisition
\begin{equation} \label{a1_a3}
{{\hat{\bm G}}_{{\text{dl}}}}{\hat{\bm G}}_{{\text{dl}}}^{\text{H}}{\left( {{{{\hat{\bm G}}}_{{\text{dl}}}}{\hat{\bm G}}_{{\text{dl}}}^{\text{H}} + \alpha {{\bm{I}}_{{M}{{N}_{\text{D}}}}}} \right)^{ - 2}} = {\left( {{{{\hat{\bm G}}}_{{\text{dl}}}}{\hat{\bm G}}_{{\text{dl}}}^{\text{H}} + \alpha {{\bm{I}}_{{M}{{N}_{\text{D}}}}}} \right)^{ - 1}} - \alpha {\left( {{{{\hat{\bm G}}}_{{\text{dl}}}}{\hat{\bm G}}_{{\text{dl}}}^{\text{H}} + \alpha {{\bm{I}}_{{M}{{N}_{\text{D}}}}}} \right)^{ - 2}},
\end{equation}
we have
\begin{equation} \label{a1_a4}
\frac{1}{{M{N_{\text{D}}}}}{\text{tr}}\left( {{{\bm{E}}_i}{\bm{C}}_{{\text{dl}}}^{{\text{-1}}}{{{\hat{\bm G}}}_{{\text{dl}}}}{\hat{\bm G}}_{{\text{dl}}}^{\text{H}}{\bm{C}}_{{\text{dl}}}^{{\text{-1}}}{{\bm{E}}_i}} \right) = \frac{1}{{M{N_{\text{D}}}}}{\text{tr}}\left( {{{\bm{E}}_{i}}{\bm{C}}_{{\text{dl}}}^{{\text{-1}}}} \right) - \alpha \frac{1}{{M{N_{\text{D}}}}}{\text{tr}}\left( {{{\bm{E}}_{i}}{\bm{C}}_{{\text{dl}}}^{{\text{-2}}}} \right),
\end{equation}
where $\frac{1}{{M{N_{\text{D}}}}}{\text{tr}}\left( {{{\bm{E}}_i}{\bm{C}}_{{\text{dl}}}^{{\text{-2}}}} \right)$ is the derivative of $\frac{1}{{M{N_{\text{D}}}}}{\text{tr}}\left( {{{\bm{E}}_i}{\bm{C}}_{{\text{dl}}}^{{\text{-1}}}} \right)$ with respect to $\alpha $. Similar to \cite[Lemma 1]{jzhang_large_2013}, letting Assumption \ref{asm:g_dl} hold true and as $\mathcal{N} \to \infty $, we obtain
\begin{equation}
\label{a1_a1}
{v_{\text{dl}}} - {\overline v_{\text{dl}}}\xrightarrow{{a.s.}}0,
\end{equation}
where ${\overline v_{\text{dl}}}$ is explicitly expressed in (\ref{eq:v_dl_DE}).

Furthermore, similar to \cite[Lemma 2]{jzhang_large_2013} and \cite[Lemma 3]{jzhang_large_2013}, letting Assumption \ref{asm:g_dl} hold true and as $\mathcal{N} \to \infty $, we can derive that
\begin{align}
\label{a1_a9}
{\bm{g}}_{{\text{dl}},k}^{\text{H}}{\bm{C}}_{{\text{dl}}}^{{\text{-1}}}{{\hat{\bm g}}_{{\text{dl}},k}} - \frac{{\overline u_{{\text{dl}},k}^{\text{(2)}}}}{{1 + \overline u_{{\text{dl}},k}^{\text{(1)}}}}&\xrightarrow{{a.s.}}0, \\
\label{a1_a12}
\sum\limits_{j = 1,j \ne k}^{{K_{\text{D}}}} {{{\left| {{\bm{g}}_{{\text{dl}},k}^{\text{H}}{\bm{C}}_{{\text{dl}}}^{{\text{-1}}}{{{\hat{\bm g}}}_{{\text{dl}},j}}} \right|}^2}}  - {\overline u_{{\text{dl}},k}}&\xrightarrow{{a.s.}}0,
\end{align}
where $\overline u_{{\text{dl}},k}^{\text{(1)}}$, $\overline u_{{\text{dl}},k}^{\text{(2)}}$ and ${\overline u_{{\text{dl}},k}}$ are given by (\ref{eq:u_dl_k_1}), (\ref{eq:u_dl_k_2}) and (\ref{eq:u_dl_k}), respectively.

From (\ref{a1_a1}), (\ref{a1_a9}) and (\ref{a1_a12}), we derive the deterministic equivalent ${\overline \gamma _{\text{dl},\text{rzf},k}}$, as shown in (\ref{eq:r_dl_rzf_DE}), of downlink SINR under RZF precoding ${\gamma _{\text{dl},\text{rzf},k}}$. With the definition of the deterministic equivalent \cite[Definition 6.1]{rcouillet_random_2011} as well as the continuous mapping theorem \cite{pbillingsley_probability_1985}, we know that when $\mathcal{N} \to \infty $, $\log \left( {1 + {\gamma _{\text{dl},\text{rzf},k}}} \right) - \log \left( {1 + {{\overline \gamma }_{\text{dl},\text{rzf},k}}} \right)\xrightarrow{{a.s.}}0$. By replacing the instantaneous ${\gamma _{\text{dl},\text{rzf},k}}$ with its large system approximation ${\overline \gamma _{\text{dl},\text{rzf},k}}$, the deterministic equivalent ${\overline R_{\text{dl},\text{rzf},\text{sum}}}$ of downlink ergodic sum-rate under RZF precoding ${R_{\text{dl},\text{rzf},\text{sum}}}$ is derived as in (\ref{eq:R_dl_rzf_DE}). When $\mathcal{N} \to \infty $, it satisfies that \cite{swagner_large_2012}
\begin{equation}
\frac{1}{{{K_{\text{D}}}}}\left( {{R_{\text{dl},\text{rzf},\text{sum}}} - {{\overline R}_{\text{dl},\text{rzf},\text{sum}}}} \right)\xrightarrow{{a.s.}}0.
\end{equation}
This completes the proof.
\end{IEEEproof}

Though the procedure refers to \cite[Appendix A]{jzhang_large_2013}, the considered CF massive MIMO with NAFD is more complex and the impact of UL-to-DL interference is required to be taken into account. In Theorem \ref{thm:R_dl_rzf}, ${\overline v_{\text{dl}}}$ represents UL-to-DL interference plus noise. It can be seen from (\ref{eq:r_dl_rzf_DE}) that noise and the interference are the two main factors in reducing the downlink sum-rate. From information-theoretic viewpoint, \cite{osimeone_fullduplex_2014} proposed that successive interference cancellation at UEs can be performed. However, the success of the interference cancellation depends on the inter-user channel capacity and the uplink data rate, which also requires a high implementation complexity. Therefore, the mitigation of UL-to-DL interference is studied later in this paper by designing a proper user scheduling scheme. Furthermore, when statistical channel knowledge such as ${{\bm{T}}_{{{\text{dl}},k,n}}}$, ${\tau _{{\text{dl}},k,n}}$ and $\sigma _{\text{dl}}^2$ are given, without knowing the actual channel realization, the downlink ergodic sum-rate with RZF beamforming can be approximated by Theorem \ref{thm:R_dl_rzf}.

\subsection{Zero-Forcing Precoding}
Imposing $\alpha  = 0$, the RZF precoding matrix in (\ref{eq:RZF_matrix}) reduces to the ZF precoding matrix which follows
\begin{equation}\label{eq:ZF_matrix}
{{\bm{W}}_{{\text{zf}}}} = \xi {{\hat{\bm G}}_{{\text{dl}}}}{\left( {{\hat{\bm G}}_{{\text{dl}}}^{\text{H}}{{{\hat{\bm G}}}_{{\text{dl}}}}} \right)^{ - 1}},
\end{equation}
the $k$th column of which is
\begin{equation}\label{eq:ZF_vector}
{{\bm{w}}_{{{\text{zf}},k}}} = \xi {{\hat{\bm G}}_{{\text{dl}}}}{\left( {{\hat{\bm G}}_{{\text{dl}}}^{\text{H}}{{{\hat{\bm G}}}_{{\text{dl}}}}} \right)^{ - 2}}{\hat{\bm G}}_{{\text{dl}}}^{\text{H}}{{\hat{\bm g}}_{{{\text{dl}},k}}} = \xi {\underline{\bm {C} }}_{{\text{dl}}}^{{\text{-1}}}{{\hat{\bm g}}_{{{\text{dl}},k}}},
\end{equation}
where ${\underline{\bm {C} }}_{{\text{dl}}}^{{\text{-1}}} \triangleq {{\hat{\bm G}}_{{\text{dl}}}}{\left( {{\hat{\bm G}}_{{\text{dl}}}^{\text{H}}{{{\hat{\bm G}}}_{{\text{dl}}}}} \right)^{ - 2}}{\hat{\bm G}}_{{\text{dl}}}^{\text{H}}$. $\xi $ is a normalization scalar to fulfill the per-RAU transmit power constraint (\ref{eq:power_constraint}), from which we can obtain
\begin{equation}\label{eq:xi_i_2_ZF}
\xi _i^2 \triangleq \frac{{MP}}{{\text{tr}\left( {{{\bm{W}}_{{{\text{zf}},i}}}{\bm{W}}_{{{\text{zf}},i}}^{\text{H}}} \right)}} = \frac{{\frac{P}{{{N_{\text{D}}}}}}}{{\frac{1}{{M{N_{\text{D}}}}}\text{tr}\left( {{{\bm{E}}_{i}}{\underline{\bm{C} }}_{{\text{dl}}}^{{\text{-1}}}{{{\hat{\bm G}}}_{{\text{dl}}}}{\hat{\bm G}}_{{\text{dl}}}^{\text{H}}{\underline{\bm{C} }}_{{\text{dl}}}^{{\text{-1}}}{{\bm{E}}_{i}}} \right)}},
\end{equation}
for $i = 1,\cdots,{N_{\text{D}}}$. To satisfy (\ref{eq:power_constraint}), we set ${\xi ^2} = \mathop {\min }\limits_i \left\{ {\xi _i^2} \right\}$. Then, the SINR at the $k$th UE under ZF precoding takes the form
\begin{equation}\label{eq:SINR_dl_ZF}
{\gamma _{{\text{dl}},{\text{zf}},k}} = \frac{{{{\left| {{\bm{g}}_{{{\text{dl}},k}}^{\text{H}}{\underline{\bm{C} }}_{{\text{dl}}}^{{\text{-1}}}{{{\hat{\bm g}}}_{{{\text{dl}},k}}}} \right|}^2}}}{{\sum\limits_{j = 1,j \ne k}^{{K_{\text{D}}}} {{{\left| {{\bm{g}}_{{{\text{dl}},k}}^{\text{H}}{\underline{\bm{C} }}_{{\text{dl}}}^{{\text{-1}}}{{{\hat{\bm g}}}_{{{\text{dl}},j}}}} \right|}^2}}  + {v_{\text{dl}}}}},
\end{equation}
where
\begin{equation}\label{eq:v_dl_ZF}
{v_{\text{dl}}} \triangleq \frac{{\sum\limits_{j = 1}^{{K_{\text{U}}}} {{p_{{\text{ul}},j}}{{\left| {{u_{k,j}}} \right|}^2}}  + \sigma _{\text{dl}}^2}}{{{\xi ^2}}} = \left( {\sum\limits_{j = 1}^{{K_{\text{U}}}} {{p_{{\text{ul}},j}}{{\left| {{u_{k,j}}} \right|}^2}}  + \sigma _{\text{dl}}^2} \right)\frac{{{N_{\text{D}}}}}{P}\mathop {\max }\limits_i \frac{1}{{M{N_{\text{D}}}}}\text{tr}\left( {{{\bm{E}}_{i}}{\underline{\bm{C} }}_{{\text{dl}}}^{{\text{-1}}}{{{\hat{\bm G}}}_{{\text{dl}}}}{\hat{\bm G}}_{{\text{dl}}}^{\text{H}}{\underline{\bm{C} }}_{{\text{dl}}}^{{\text{-1}}}{{\bm{E}}_{i}}} \right).
\end{equation}

Then, the downlink ergodic sum-rate with ZF beamforming is given by
\begin{equation}\label{eq:R_dl_zf}
{R_{{\text{dl}},{\text{zf}},{\text{sum}}}} \triangleq \sum\limits_{k = 1}^{{K_{\text{D}}}} {{\text{E}_{{{{\hat{\bm G}}}_{{\text{dl}}}}}}\left\{ {{{\log }_2}\left( {1 + {\gamma _{{\text{dl}},{\text{zf}},k}}} \right)} \right\}}.
\end{equation}

To derive a deterministic equivalent of the sum-rate for ZF precoding, we require the following assumptions.
\begin{assumption}\label{asm:eigen}
There exists $\varepsilon  > 0$ such that ${\lambda _{\min }}\left( {{\hat{\bm G}}_{{\text{dl}}}^{\text{H}}{{{\hat{\bm G}}}_{{\text{dl}}}}} \right) > \varepsilon $ with probability 1.
\end{assumption}

\begin{assumption}\label{asm:e}
For $k = 1,\cdots,{K_{\text{D}}}$, $n = 1,\cdots,{N_{\text{D}}}$, ${\underline{e} _{{\text{dl}},k,n}} = {\lim _{\alpha  \to 0}}\alpha {e_{{\text{dl}},k,n}}$ exists and ${\underline{e} _{{\text{dl}},k,n}} > \varepsilon _{}^{}$ $\forall k,n$ for some $\varepsilon  > 0$.
\end{assumption}

The deterministic equivalent ${\overline R_{{\text{dl}},{\text{zf}},{\text{sum}}}}$ of ergodic sum-rate under ZF precoding ${R_{{\text{dl}},{\text{zf}},{\text{sum}}}}$ is provided in the following theorem.

\begin{theorem}\label{thm:R_dl_zf}
Letting Assumptions {\ref{asm:g_dl}}-{\ref{asm:e}} hold true and as $\mathcal{N} \to \infty $, we have
\begin{equation}\label{eq:R_dl_zf_as}
\frac{1}{{{K_{\text{D}}}}}\left( {{R_{{\text{dl}},{\text{zf}},{\text{sum}}}} - {{\overline R}_{{\text{dl}},{\text{zf}},{\text{sum}}}}} \right)\xrightarrow{{a.s.}}0,
\end{equation}
where
\begin{eqnarray}
\label{eq:R_dl_zf_DE}
{\overline R_{{\text{dl}},{\text{zf}},{\text{sum}}}} = \sum\limits_{k = 1}^{{K_{\text{D}}}} {{{\log }_2}\left( {1 + {{\overline \gamma }_{{\text{dl}},{\text{zf}},k}}} \right)}, \\
\label{eq:r_dl_zf_DE}
{\overline \gamma _{{\text{dl}},{\text{zf}},k}} = \frac{{{{\left( {\underline{\overline u} _{{\text{dl}},k}^{\text{(2)}}} \right)}^2}}}{{{{\left( {\underline{\overline u} _{{\text{dl}},k}^{\text{(1)}}} \right)}^2}\left( {{{\underline{\overline u} }_{{\text{dl}},k}} + {{\underline{\overline v} }_{\text{dl}}}} \right)}},
\end{eqnarray}
with
\begin{subequations} \label{eq:zf_six}
\begin{align}
\label{eq:zf_six_a}
& \underline{\overline u} _{{\text{dl}},k}^{\text{(1)}} = \sum\limits_{n = 1}^{{N_{\text{D}}}} {\frac{1}{M}\text{tr}\left( {{{\bm{T}}_{{{\text{dl}},k,n}}}{{{\underline{\bm{\Psi } }}}_{{{\text{dl}},n}}}} \right)}, \\
\label{eq:zf_six_b}
& \underline{\overline u} _{{\text{dl}},k}^{\text{(2)}} = \sum\limits_{n = 1}^{{N_{\text{D}}}} {\frac{{{\varphi _{{\text{dl}},k,n}}}}{M}\text{tr}\left( {{{\bm{T}}_{{{\text{dl}},k,n}}}{{{\underline{\bm{\Psi } }}}_{{{\text{dl}},n}}}} \right)}, \\
\label{eq:zf_six_c}
& {\underline{\overline v} _{\text{dl}}} = \frac{{\left( {\sum\limits_{i = 1}^{{K_{\text{U}}}} {{p_{{\text{ul}},i}}T_{k,i}^{}}  + \sigma _{\text{dl}}^2} \right)}}{{MP}}\mathop {\max }\limits_n \left[ {\text{tr}{{\bm{\Psi }}_{{{\text{dl}},n,[\alpha ]}}} - \text{tr}\left( {{\alpha _0}{\bm{\Psi }}_{{{\text{dl}},n,[\alpha ]}}^2} \right) + \sum\limits_{j = 1}^{{K_{\text{D}}}} {{{\underline{\dot c} }_{{\text{dl}},j}}} \text{tr}\left( {{{\bm{\Psi }}_{{{\text{dl}},n,[\alpha ]}}}{{\bm{T}}_{{{\text{dl}},j,n}}}{{\bm{\Psi }}_{{{\text{dl}},n,[\alpha ]}}}} \right)} \right], \\
\label{eq:zf_six_d}
& {\underline{\overline u} _{{\text{dl}},k}} = \overline u_{{\text{dl}},k,[\alpha ]}^{\text{(1)}} - \overline{\dot{u}}_{{\text{dl}},k,[\alpha ]}^{\text{(1)}} - \frac{{2\underline{\overline u} _{{\text{dl}},k}^{\text{(2)}}\left( {\overline u_{{\text{dl}},k,[\alpha ]}^{\text{(2)}} - \overline{\dot{u}}_{{\text{dl}},k,[\alpha ]}^{\text{(2)}}} \right)}}{{\underline{\overline u} _{{\text{dl}},k}^{\text{(1)}}}} + \frac{{{{\left( {\underline{\overline u} _{{\text{dl}},k}^{\text{(2)}}} \right)}^2}\left( {\overline u_{{\text{dl}},k,[\alpha ]}^{\text{(1)}} - \overline{\dot{u}}_{{\text{dl}},k,[\alpha ]}^{\text{(1)}}} \right)}}{{{{\left( {\underline{\overline u} _{{\text{dl}},k}^{\text{(1)}}} \right)}^2}}}, \\
\label{eq:zf_six_e}
& \overline u_{{\text{dl}},k,[\alpha ]}^{\text{(1)}} = \sum\limits_{n = 1}^{{N_{\text{D}}}} {\frac{1}{M}\text{tr}\left( {{{\bm{T}}_{{{\text{dl}},k,n}}}{{\bm{\Psi }}_{{{\text{dl}},n,[\alpha ]}}}} \right)}, \\
\label{eq:zf_six_f}
& \overline u_{{\text{dl}},k,[\alpha ]}^{\text{(2)}} = \sum\limits_{n = 1}^{{N_{\text{D}}}} {\frac{{{\varphi _{{\text{dl}},k,n}}}}{M}\text{tr}\left( {{{\bm{T}}_{{{\text{dl}},k,n}}}{{\bm{\Psi }}_{{{\text{dl}},n,[\alpha ]}}}} \right)}, \\
\label{eq:zf_six_g}
& \overline{\dot{u}}_{{\text{dl}},k,[\alpha ]}^{\text{(1)}} = \sum\limits_{n = 1}^{{N_{\text{D}}}} {\frac{1}{M}\left[ {\text{tr}\left( {{\alpha _0}{{\bm{T}}_{{{\text{dl}},k,n}}}{\bm{\Psi }}_{{{\text{dl}},n,[\alpha ]}}^2} \right) - \sum\limits_{j = 1}^{{K_{\text{D}}}} {{{\underline{\dot c} }_{{\text{dl}},j}}} \text{tr}\left( {{{\bm{T}}_{{{\text{dl}},k,n}}}{{\bm{\Psi }}_{{{\text{dl}},n,[\alpha ]}}}{{\bm{T}}_{{{\text{dl}},j,n}}}{{\bm{\Psi }}_{{{\text{dl}},n,[\alpha ]}}}} \right)} \right]}, \\
\label{eq:zf_six_h}
& \overline{\dot{u}}_{{\text{dl}},k,[\alpha ]}^{\text{(2)}} = \sum\limits_{n = 1}^{{N_{\text{D}}}} {\frac{{{\varphi _{{\text{dl}},k,n}}}}{M}\left[ {\text{tr}\left( {{\alpha _0}{{\bm{T}}_{{{\text{dl}},k,n}}}{\bm{\Psi }}_{{{\text{dl}},n,[\alpha ]}}^2} \right) - \sum\limits_{j = 1}^{{K_{\text{D}}}} {{{\underline{\dot c} }_{{\text{dl}},j}}} \text{tr}\left( {{{\bm{T}}_{{{\text{dl}},k,n}}}{{\bm{\Psi }}_{{{\text{dl}},n,[\alpha ]}}}{{\bm{T}}_{{{\text{dl}},j,n}}}{{\bm{\Psi }}_{{{\text{dl}},n,[\alpha ]}}}} \right)} \right]}.
\end{align}
\end{subequations}

In (\ref{eq:zf_six}), ${\underline{e} _{{\text{dl}},k,n}}$ ($k = 1,\cdots,{K_{\text{D}}},i = 1,\cdots,{N_{\text{D}}}$) form the unique solutions of the following ${K_{\text{D}}} \times {N_{\text{D}}}$ equations
\begin{eqnarray}
\label{eq:pha_dl_ZF}
{{\underline{\bm{\Psi } }}_{{{\text{dl}},n}}}\!\!\! &=&\!\!\! {\left( {\frac{1}{M}\sum\limits_{j = 1}^{{K_{\text{D}}}} {\frac{{{{\bm{T}}_{{{\text{dl}},j,n}}}}}{{\sum\limits_{m = 1}^{{N_{\text{D}}}} {{{\underline{e} }_{{\text{dl}},j,m}}} }} + {{\bm{I}}_{M}}} } \right)^{ - 1}}, \\
\label{eq:e_dl_ZF}
{\underline{e} _{{\text{dl}},k,n}}\!\!\! &=&\!\!\! \frac{1}{M}\text{tr}\left( {{{\bm{T}}_{{{\text{dl}},k,n}}}{{{\underline{\bm{\Psi } }}}_{{{\text{dl}},n}}}} \right),
\end{eqnarray}
${e_{{\text{dl}},k,n,[\alpha ]}}$ ($k = 1,\cdots,{K_{\text{D}}},i = 1,\cdots,{N_{\text{D}}}$) are the unique solutions of
\begin{eqnarray}
\label{eq:pha_dl_ZF_a}
{{\bm{\Psi }}_{{{\text{dl}},n,[\alpha ]}}}\!\!\! &=&\!\!\! {\left( {\frac{1}{M}\sum\limits_{j = 1}^{{K_{\text{D}}}} {\frac{{{{\bm{T}}_{{{\text{dl}},j,n}}}}}{{1 + \sum\limits_{m = 1}^{{N_{\text{D}}}} {{e_{{\text{dl}},j,m,[\alpha ]}}} }} + {\alpha _0}{{\bm{I}}_{M}}} } \right)^{ - 1}}, \\
\label{eq:e_dl_ZF_a}
{e_{{\text{dl}},k,n,[\alpha ]}}\!\!\! &=&\!\!\! \frac{1}{M}\text{tr}\left( {{{\bm{T}}_{{{\text{dl}},k,n}}}{{\bm{\Psi }}_{{{\text{dl}},n,[\alpha ]}}}} \right),
\end{eqnarray}
where ${\alpha _0} = 0$ and
\begin{equation}
\label{eq:c_ZF_under_dot}
{\underline{\dot c} _{{\text{dl}},j}} \triangleq \frac{{ - 1}}{{M\sum\limits_{m = 1}^{{N_{\text{D}}}} {{e_{{\text{dl}},j,m,[\alpha ]}}} }}.
\end{equation}
\end{theorem}

\begin{IEEEproof}
The proof of Theorem \ref{thm:R_dl_zf} is given in Appendix \ref{sec:Appendix_B}.
\end{IEEEproof}

In Theorem \ref{thm:R_dl_zf}, the downlink sum-rate under ZF precoding is derived by setting the regularization parameter $\alpha \to 0$. Under RZF precoding, the downlink sum-rate can be maximized by choosing the optimal regularization parameter. Therefore, RZF system achieves higher sum-rate than ZF system and RZF is more robust than ZF to imperfect CSI \cite{lsanguinetti_interference_2015}.

\section{Deterministic Equivalent Sum-Rate for Uplink}\label{sec:DE_UL}
In this section, to obtain a general result, we will study the uplink sum-rate when the downlink transmission performs RZF precoding rather than ZF precoding. We adopt MMSE receiver at CPU to detect the uplink data streams in (\ref{eq:trans_up_model}). For CF massive MIMO, the centralized processing of both uplink and downlink baseband signals at the CPU allows the CPU to perform mitigation of the DL-to-UL interference. However, due to the imperfect CSI, the interference cancellation is not perfect. In this section, we first study a deterministic equivalent of the covariance matrix for the residual interference plus noise, and then derive a deterministic equivalent of the sum-rate for uplink.

\subsection{Downlink-to-Uplink Interference Cancellation and Joint MMSE detection}
After the cancellation of the DL-to-UL interference at the CPU, the received uplink signal can be rewritten as
\begin{equation}
{{\hat{\bm y}}_{{\text{ul}}}} = \sqrt {{p_{{\text{ul}},i}}} {{\bm{g}}_{{{\text{ul}},i}}}{x_i} + \sum\limits_{j = 1,j \ne i}^{{K_{\text{U}}}} {\sqrt {{p_{{\text{ul}},j}}} {{\bm{g}}_{{{\text{ul}},j}}}{x_j}}  + \sum\limits_{k = 1}^{{K_{\text{D}}}} {{{{\tilde{\bm G}}}_{\text{I}}}{{\bm{w}}_{{{\text{rzf}},k}}}{s_k}}  + {{\bm{z}}_{{\text{ul}}}}.
\end{equation}
where ${\tilde{\bm G}} \triangleq \left( {{{\bm{G}}_{\text{I}}} - {{{\hat{\bm G}}}_{\text{I}}}} \right) $ represents the channel estimation error, ${\tilde{\bm G}}_{\text{I}}^{\text{H}} \triangleq \left[ {{\tilde{\bm g}}_{{{\text{I}},1}}^{},\cdots,{\tilde{\bm g}}_{{{\text{I}},M}{{N}_{\text{U}}}}^{}} \right] \in {\mathbb{C}^{M{N_{\text{D}}} \times M{N_{\text{U}}}}}$, and the third term in the right-hand-side represents the residual interference after DL-to-UL interference cancellation. Due to imperfect cancellation,  the covariance matrix of the residual interference plus noise is given by
\begin{align*}
{\bm{\Sigma }} &= {\text{cov}}\left( {\sum\limits_{k = 1}^{{K_{\text{D}}}} {{{{\tilde{\bm G}}}_{\text{I}}}{{\bm{w}}_{{{\text{rzf}},k}}}{s_k}}  + {{\bm{z}}_{{\text{ul}}}},\sum\limits_{k = 1}^{{K_{\text{D}}}} {{{{\tilde{\bm G}}}_{\text{I}}}{{\bm{w}}_{{{\text{rzf}},k}}}{s_k}}  + {{\bm{z}}_{{\text{ul}}}}} \right) \\
&= \sum\limits_{k = 1}^{{K_{\text{D}}}} {\text{E}\left( {{{{\tilde{\bm G}}}_{\text{I}}}{{\bm{w}}_{{{\text{rzf}},k}}}{\bm{w}}_{{{\text{rzf}},k}}^{\text{H}}{\tilde{\bm G}}_{\text{I}}^{\text{H}}} \right)}  + \sigma _{\text{ul}}^2{{\bm{I}}_{{M}{{N}_{\text{U}}}}}.
\end{align*}
After simple manipulations, it is found that ${\bm{\Sigma }}$ is a main diagonal matrix and ${\bm{\Sigma }} \triangleq \text{diag}\left( {{{\bm{\Sigma }}_{1}},\cdots,{{\bm{\Sigma }}_{{{N}_{\text{U}}}}}} \right) \in {\mathbb{C}^{M{N_{\text{U}}} \times M{N_{\text{U}}}}}$, ${{\bm{\Sigma }}_{i}} \in {\mathbb{C}^{M \times M}}$. The $n$th diagonal element of ${\bm{\Sigma }}$ for ${n = 1,\cdots,M{N_{\text{U}}}}$ is
\begin{equation}
\label{eq:UL_delta_n}
{\delta _n} = \sum\limits_{k = 1}^{{K_{\text{D}}}} {{\bm{w}}_{{{\text{rzf}},k}}^{\text{H}}{{\bm{A}}_{n}}{{\bm{w}}_{{{\text{rzf}},k}}}}  + \sigma _{\text{ul}}^2,
\end{equation}
where ${{\bm{A}}_{n}}$ represents
\begin{equation}
{{\bm{A}}_{n}} \triangleq \text{E}\left( {{{{\tilde{\bm g}}}_{{{\text{I}},n}}}{\tilde{\bm g}}_{{{\text{I}},n}}^{\text{H}}} \right),
\end{equation}
the main diagonal matrix ${{\bm{A}}_{n}} \triangleq \text{diag}\left( {{{\bm{A}}_{{n,1}}},\cdots,{{\bm{A}}_{{n,}{{N}_{\text{D}}}}}} \right)$ and ${{\bm{A}}_{{n,i}}} \in {\mathbb{C}^{M \times M}}$. It should be noted that each element of the residual interference and noise vector are independent from each other and have different distribution properties.

With MMSE receiver\cite{dtse_fudamentals_2005}, we denote the combining matrix as
\begin{equation}
{\bm{R}} = {\left( {\sum\limits_{i = 1}^{{K_{\text{U}}}} {{p_{{\text{ul}},i}}{{{\hat{\bm g}}}_{{{\text{ul}},i}}}{\hat{\bm g}}_{{{\text{ul}},i}}^{\text{H}}}  + {\bm{\Sigma }}} \right)^{ - 1}}{{\hat{\bm G}}_{{\text{ul}}}} = {\bm{C}}_{{\text{ul}}}^{{\text{-1}}}{{\hat{\bm G}}_{{\text{ul}}}},
\end{equation}
the $j$th column for ${j = 1,\cdots,{K_{\text{U}}}}$ of which is
\begin{equation}
{{\bm{r}}_{j}} = {\left( {\sum\limits_{i = 1}^{{K_{\text{U}}}} {{p_{{\text{ul}},i}}{{{\hat{\bm g}}}_{{{\text{ul}},i}}}{\hat{\bm g}}_{{{\text{ul}},i}}^{\text{H}}}  + {\bm{\Sigma }}} \right)^{ - 1}}{{\hat{\bm g}}_{{{\text{ul}},j}}} = {\bm{C}}_{{\text{ul}}}^{{\text{-1}}}{{\hat{\bm g}}_{{{\text{ul}},j}}},
\end{equation}
where we define ${{\hat{\bm G}}_{{\text{ul}}}} \triangleq [{{\hat{\bm g}}_{{{\text{ul}},1}}},\cdots,{{\hat{\bm g}}_{{{\text{ul}},}{{K}_{\text{U}}}}}] \in {\mathbb{C}^{M{N_{\text{U}}} \times {K_{\text{U}}}}}$ and ${{\bm{C}}_{{\text{ul}}}} \triangleq \sum\limits_{i = 1}^{{K_{\text{U}}}} {{p_{{\text{ul}},i}}{{{\hat{\bm g}}}_{{{\text{ul}},i}}}{\hat{\bm g}}_{{{\text{ul}},i}}^{\text{H}}}  + {\bm{\Sigma }}$.

Hence, the SINR of the $k$th uplink channel takes the form
\begin{equation}
\label{eq:r_ul_k}
{\gamma _{{\text{ul}},k}} = \frac{{{p_{{\text{ul}},k}}{{\left| {{\hat{\bm g}}_{{{\text{ul}},k}}^{\text{H}}{\bm{C}}_{{\text{ul}}}^{{\text{-1}}}{{\bm{g}}_{{{\text{ul}},k}}}} \right|}^2}}}{{\sum\limits_{i = 1,i \ne k}^{{K_{\text{U}}}} {{p_{{\text{ul}},i}}{{\left| {{\hat{\bm g}}_{{{\text{ul}},k}}^{\text{H}}{\bm{C}}_{{\text{ul}}}^{{\text{-1}}}{{\bm{g}}_{{{\text{ul}},i}}}} \right|}^2}}  + {\hat{\bm g}}_{{{\text{ul}},k}}^{\text{H}}{\bm{C}}_{{\text{ul}}}^{{\text{-1}}}{\bm{\Sigma C}}_{{\text{ul}}}^{{\text{-1}}}{{{\hat{\bm g}}}_{{{\text{ul}},k}}}}}.
\end{equation}
The uplink ergodic sum-rate with MMSE receiver can be expressed as
\begin{equation}
{R_{{\text{ul}},{\text{sum}}}} \triangleq \sum\limits_{k = 1}^{{K_{\text{U}}}} {{\text{E}_{{{{\hat{\bm G}}}_{{\text{ul}}}}}}\left[ {{{\log }_2}\left( {1 + {\gamma _{{\text{ul}},k}}} \right)} \right]}.
\end{equation}

\subsection{A Deterministic Equivalent for Uplink Sum-rate}
To get a deterministic equivalent for uplink sum-rate, we first calculate a deterministic equivalent of the covariance matrix for the residual interference plus noise according to the following theorem.
\begin{theorem}\label{thm:cov}
Letting Assumption \ref{asm:g_dl} hold true and as $\mathcal{N} \to \infty $, we have
\begin{equation}\label{eq:sigma_as}
{\bm{\Sigma }} - {\overline{\bm \Sigma }}\xrightarrow{{a.s.}}0,
\end{equation}
where
\begin{eqnarray}
\label{eq:sigma_DE}
{\overline{\bm \Sigma }}\!\!\! &=&\!\!\! \text{diag}\left( {{{\overline \delta }_1},\cdots,{{\overline \delta }_{M{N_{\text{U}}}}}} \right), \\
\label{eq:UL_delta_n_DE}
{\overline \delta _n}\!\!\! &=&\!\!\! \sum\limits_{k = 1}^{{K_{\text{D}}}} {\frac{{{{\overline \xi }^2}\overline{\dot{u}}_{{\text{dl}},k,{\text{A}}}^{\text{(1)}}}}{{{{\left( {1 + \overline u_{{\text{dl}},k}^{\text{(1)}}} \right)}^2}}}}  + \sigma _{\text{ul}}^2,
\end{eqnarray}
with $\overline u_{{\text{dl}},k}^{\text{(1)}}$ defined in (\ref{eq:u_dl_k_1}) and
\begin{eqnarray}
\label{eq:kexi_2_DE}
{\overline \xi ^2}\!\!\! &=&\!\!\! \mathop {\min }\limits_i \frac{{MP}}{{\text{tr}{{\bm{\Psi }}_{{{\text{dl}},i}}} - \alpha \text{tr}{\bm{\Psi }}_{{{\text{dl}},i}}^{2} + \alpha \sum\limits_{j = 1}^{{K_{\text{D}}}} {{{\dot c}_{{\text{dl}},j,i}}} \text{tr}\left( {{{\bm{\Psi }}_{{{\text{dl}},i}}}{{\bm{T}}_{{{\text{dl}},j,i}}}{{\bm{\Psi }}_{{{\text{dl}},i}}}} \right)}}, \\
\label{eq:u_dl_k_A_1_DE}
\overline{\dot{u}}_{{\text{dl}},k,{\text{A}}}^{\text{(1)}}\!\!\! &=&\!\!\! \sum\limits_{m = 1}^{{N_{\text{D}}}} {\frac{1}{M}\left[ {\text{tr}\left( {{{\bm{T}}_{{{\text{dl}},k,m}}}{{\bm{\Psi }}_{{{\text{dl}},m}}}{{\bm{A}}_{{n,m}}}{{\bm{\Psi }}_{{{\text{dl}},m}}}} \right) - \sum\limits_{j = 1}^{{K_{\text{D}}}} {\dot c_{{\text{dl}},j,m}^{\text{(A)}}\text{tr}\left( {{{\bm{T}}_{{{\text{dl}},k,m}}}{{\bm{\Psi }}_{{{\text{dl}},m}}}{{\bm{T}}_{{{\text{dl}},j,m}}}{{\bm{\Psi }}_{{{\text{dl}},m}}}} \right)} } \right]}.
\end{eqnarray}

The $j$th diagonal element of ${{\bm{A}}_{n}}$ is given by
\begin{equation}
{a_j} = \frac{2}{{M{N_{\text{U}}}}}\text{tr}\left[ {\left( {{{\bm{I}}_{{M}{{N}_{\text{U}}}}} - {{\bm{\Lambda }}_{{{\text{I}},j}}}} \right){\bm{q}}_{{{\text{I}},j,n}}^{\text{H}}{{\bm{q}}_{{{\text{I}},j,n}}}} \right],
\end{equation}
where ${{\bm{Q}}_{{{\text{I}},j}}} \triangleq {\bm{T}}_{{{\text{I}},j}}^{\frac{{1}}{{2}}}$ and ${{\bm{q}}_{{{\text{I}},j,n}}}$ denotes the $n$th row of matrix ${{\bm{Q}}_{{{\text{I}},j}}}$.

In addition, ${\dot{\bm C}}_{{\text{dl}}}^{{\text{(A)}}} = \left[ {\dot c_{{\text{dl}},k,i}^{\text{(A)}}} \right] \in {\mathbb{C}^{{K_{\text{D}}} \times {N_{\text{D}}}}}$ is a solution to the following linear equation
\begin{equation}
{{\bm{\Theta }}_{{\text{dl}}}}\text{vec}\left( {{\dot{\bm C}}_{{\text{dl}}}^{{\text{(A)}}}} \right) = \text{vec}\left( {{\bm{\Gamma }}_{{\text{dl}}}^{{\text{(A)}}}} \right),
\end{equation}
wherein ${{\bm{\Theta }}_{{\text{dl}}}}$ is defined in (\ref{eq:theta_dl}) and
\begin{equation}
{\left[ {{\bm{\Gamma }}_{{\text{dl}}}^{{\text{(A)}}}} \right]_{k,i}} =  - \frac{1}{M}\sum\limits_{j = 1}^{{N_{\text{D}}}} {\frac{1}{M}\frac{1}{{{{\left( {1 + \sum\limits_{m = 1}^{{N_{\text{D}}}} {{e_{{\text{dl}},k,m}}} } \right)}^2}}}} {\text{tr}}\left( {{{\bm{T}}_{{{\text{dl}},k,j}}}{\bm{\Psi }}_{{{\text{dl}},j}}^{}{{\bm{A}}_{{n,j}}}{{\bm{\Psi }}_{{{\text{dl}},j}}}} \right).
\end{equation}
\end{theorem}

\begin{IEEEproof}
See Appendix \ref{sec:Appendix_C}.
\end{IEEEproof}

Since the DL-to-UL interference is one of the dominant factors degrading the system performance of NAFD, the mitigation of such interference helps to further increase the uplink achievable sum-rate in the CF massive MIMO with NAFD. In Theorem \ref{thm:cov}, ${\overline{\bm \Sigma }}$ could be used to evaluate the covariance matrix of the residual interference plus noise after the cancellation of the DL-to-UL interference at the CPU.  In addition, given statistical information of the channel, the covariance matrix of the residual interference plus noise in the uplink transmission can be approximated by Theorem \ref{thm:cov} without knowing the actual channel realization, which can simplify the computation of ${\bm{\Sigma }}$ in the uplink receiver.

In order to obtain a deterministic equivalent for uplink sum-rate with MMSE receiver, the following assumptions are demanded.

\begin{assumption}\label{asm:g_ul}
For the channel ${{\hat{\bm g}}_{{{\text{ul}},i}}}$ in (\ref{eq:es_up_channel}), where $i = 1,\cdots,{K_{\text{U}}}$, we have the following hypotheses\cite{swagner_large_2012}:

1) ${{\bm{h}}_{{{\text{ul}},i}}}$ and ${{\bm{z}}_{{{\text{ulp}},i}}}$ have finite eighth-order moment.

2) ${\bm{T}}_{{{\text{ul}},i}}^{}$ has uniformly bounded spectral norm.
\end{assumption}

A deterministic equivalent ${\overline R_{{\text{ul}},{\text{sum}}}}$ of ergodic sum-rate with MMSE receiver ${R_{{\text{ul}},{\text{sum}}}}$ is provided in the following theorem.

\begin{theorem}\label{thm:R_ul}
Letting Assumption \ref{asm:g_ul} hold true and as $\mathcal{N} \to \infty $, we have
\begin{equation}\label{eq:R_ul_as}
\frac{1}{{{K_{\text{U}}}}}\left( {{R_{{\text{ul}},{\text{sum}}}} - {{\overline R}_{{\text{ul}},{\text{sum}}}}} \right)\xrightarrow{{a.s.}}0,
\end{equation}
where
\begin{align}
\label{eq:R_ul_DE}
&{\overline R_{{\text{ul}},{\text{sum}}}} = \sum\limits_{k = 1}^{{K_{\text{U}}}} {{{\log }_2}\left( {1 + {{\overline \gamma }_{{\text{ul}},k}}} \right)}, \\
\label{eq:r_ul_DE}
&{\overline \gamma _{{\text{ul}},k}} = \frac{{{p_{{\text{ul}},k}}{{\left( {\overline u_{{\text{ul}},k}^{\text{(2)}}} \right)}^2}}}{{{{\left( {1 + {p_{{\text{ul}},k}}\overline u_{{\text{ul}},k}^{\text{(1)}}} \right)}^2}\sum\limits_{i = 1,i \ne k}^{{K_{\text{U}}}} {{p_{{\text{ul}},i}}{{\overline u}_{{\text{ul}},i}}}  + \overline{\dot{u}}_{{\text{ul}},k,\Sigma }^{\text{(1)}}}},
\end{align}
with
\begin{subequations} \label{eq:ul_six}
\begin{align}
\label{eq:u_ul_k_1}
& \overline u_{{\text{ul}},k}^{\text{(1)}} = \sum\limits_{n = 1}^{{N_{\text{U}}}} {\frac{1}{M}} \text{tr}\left( {{\overline{\bm \Sigma }}_{n}^{{\text{ -1}}}{{\bm{T}}_{{{\text{ul}},k,n}}}{{\bm{\Psi }}_{{{\text{ul}},n}}}} \right), \\
\label{eq:u_ul_k_2}
& \overline u_{{\text{ul}},k}^{\text{(2)}} = \sum\limits_{n = 1}^{{N_{\text{U}}}} {\frac{{{\varphi _{{\text{ul}},k,n}}}}{M}} \text{tr}\left( {{\overline{\bm \Sigma }}_{n}^{{\text{ -1}}}{{\bm{T}}_{{{\text{ul}},k,n}}}{{\bm{\Psi }}_{{{\text{ul}},n}}}} \right), \\
\label{eq:u_ul_i_DE}
& {\overline u_{{\text{ul}},i}} = \frac{1}{{M{N_{\text{U}}}}}\frac{1}{{{{\left( {1 + {p_{{\text{ul}},k}}\overline u_{{\text{ul}},k}^{\text{(1)}}} \right)}^2}}}\left[ {\overline{\dot{u}}_{{\text{ul}},i}^{\text{(1)}} - \frac{{2{p_{{\text{ul}},i}}\overline u_{{\text{ul}},i}^{\text{(2)}}\overline{\dot{u}}_{{\text{ul}},i}^{\text{(2)}}}}{{1 + {p_{{\text{ul}},i}}\overline u_{{\text{ul}},i}^{\text{(1)}}}} + \frac{{p_{{\text{ul}},i}^2{{\left( {\overline u_{{\text{ul}},i}^{\text{(2)}}} \right)}^2}\overline{\dot{u}}_{{\text{ul}},i}^{\text{(1)}}}}{{{{\left( {1 + {p_{{\text{ul}},i}}\overline u_{{\text{ul}},i}^{\text{(1)}}} \right)}^2}}}} \right], \\
\label{eq:u_ul_i_1_DE}
& \overline{\dot{u}}_{{\text{ul}},i}^{\text{(1)}} = \sum\limits_{n = 1}^{{N_{\text{U}}}} {\frac{1}{M}\left[ {\text{tr}\left( {{{\bm{T}}_{{{\text{ul}},i,n}}}{{\bm{\Psi }}_{{{\text{ul}},n}}}{{\bm{T}}_{{{\text{ul}},k,n}}}{{\bm{\Psi }}_{{{\text{ul}},n}}}} \right) - \sum\limits_{j = 1}^{{K_{\text{U}}}} {\dot c_{{\text{ul}},j,n}^{\text{(k)}}\text{tr}\left( {{{\bm{T}}_{{{\text{ul}},i,n}}}{{\bm{\Psi }}_{{{\text{ul}},n}}}{\overline{\bm \Sigma }}_{n}^{{\text{ -1}}}{{\bm{T}}_{{{\text{ul}},j,n}}}{{\bm{\Psi }}_{{{\text{ul}},n}}}} \right)} } \right]}, \\
\label{eq:u_ul_i_2_DE}
& \overline{\dot{u}}_{{\text{ul}},i}^{\text{(2)}} = \sum\limits_{n = 1}^{{N_{\text{U}}}} {\frac{{{\varphi _{{\text{ul}},i,n}}}}{M}\left[ {\text{tr}\left( {{{\bm{T}}_{{{\text{ul}},i,n}}}{{\bm{\Psi }}_{{{\text{ul}},n}}}{{\bm{T}}_{{{\text{ul}},k,n}}}{{\bm{\Psi }}_{{{\text{ul}},n}}}} \right) - \sum\limits_{j = 1}^{{K_{\text{U}}}} {\dot c_{{\text{ul}},j,n}^{\text{(k)}}\text{tr}\left( {{{\bm{T}}_{{{\text{ul}},i,n}}}{{\bm{\Psi }}_{{{\text{ul}},n}}}{\overline{\bm \Sigma }}_{n}^{{\text{ -1}}}{{\bm{T}}_{{{\text{ul}},j,n}}}{{\bm{\Psi }}_{{{\text{ul}},n}}}} \right)} } \right]}, \\
\label{eq:u_ul_k_sigma_1_DE}
& \overline{\dot{u}}_{{\text{ul}},k,\Sigma }^{\text{(1)}} = \sum\limits_{n = 1}^{{N_{\text{U}}}} {\frac{1}{M}\left[ {\text{tr}\left( {{{\bm{T}}_{{{\text{ul}},k,n}}}{{\bm{\Psi }}_{{{\text{ul}},n}}}{{{\overline{\bm \Sigma }}}_{n}}{{\bm{\Psi }}_{{{\text{ul}},n}}}} \right) - \sum\limits_{j = 1}^{{K_{\text{U}}}} {\dot c_{{\text{ul}},j,n}^{(\Sigma )}\text{tr}\left( {{{\bm{T}}_{{{\text{ul}},k,n}}}{{\bm{\Psi }}_{{{\text{ul}},n}}}{\overline{\bm \Sigma }}_{n}^{{\text{ -1}}}{{\bm{T}}_{{{\text{ul}},j,n}}}{{\bm{\Psi }}_{{{\text{ul}},n}}}} \right)} } \right]}.
\end{align}
\end{subequations}
In (\ref{eq:u_ul_k_1})-(\ref{eq:u_ul_k_sigma_1_DE}), ${e_{{\text{ul}},k,i}}$ ($k = 1,\cdots,{K_{\text{U}}},i = 1,\cdots,{N_{\text{U}}}$) form the unique solutions of the following ${K_{\text{U}}} \times {N_{\text{U}}}$ equations
\begin{eqnarray}
\label{eq:pha_ul}
{{\bm{\Psi }}_{{{\text{ul}},j}}}\!\!\! &=&\!\!\! {\left( {\frac{1}{M}\sum\limits_{i = 1}^{{K_{\text{U}}}} {{p_{{\text{ul}},i}}\frac{{{\overline{\bm \Sigma }}_{j}^{{\text{ -1}}}{{\bm{T}}_{{{\text{ul}},i,j}}}}}{{1 + \sum\limits_{m = 1}^{{N_{\text{U}}}} {{e_{{\text{ul}},i,m}}} }} + {{\bm{I}}_{M}}} } \right)^{ - 1}}, \\
\label{eq:e_ul}
{e_{{\text{ul}},k,i}}\!\!\! &=&\!\!\! \frac{1}{M}\text{tr}\left( {{\overline{\bm \Sigma }}_{i}^{{\text{ -1}}}{{\bm{T}}_{{{\text{ul}},k,i}}}{{\bm{\Psi }}_{{{\text{ul}},i}}}} \right).
\end{eqnarray}
In addition, ${\dot{\bm C}}_{{\text{ul}}}^{{\text{(k)}}} = \left[ {\dot c_{{\text{ul}},n,i}^{\text{(k)}}} \right] \in {\mathbb{C}^{{K_{\text{U}}} \times {N_{\text{U}}}}}$ is a solution to the linear equation
\begin{equation}\label{eq:li_eq_Cul}
{{\bm{\Theta }}_{{\text{ul}}}}\text{vec}\left( {{\dot{\bm C}}_{{\text{ul}}}^{{\text{(k)}}}} \right) = \text{vec}\left( {{\bm{\Gamma }}_{{\text{ul}}}^{{\text{(k)}}}} \right),
\end{equation}
wherein ${{\bm{\Theta }}_{{\text{ul}}}} = \left[ {{{\bm{\Theta }}_{{{\text{ul}},in}}}} \right] \in {\mathbb{C}^{{N_{\text{U}}}{K_{\text{U}}} \times {N_{\text{U}}}{K_{\text{U}}}}}$ and ${\bm{\Gamma }}_{{\text{ul}}}^{{\text{(k)}}} \in {\mathbb{C}^{{K_{\text{U}}} \times {N_{\text{U}}}}}$ take the form
\begin{subequations}
\begin{align}
\label{eq:theta_ul}
{\left[ {{{\bm{\Theta }}_{{{\text{ul}},in}}}} \right]_{jl}} &= \left\{ {\begin{array}{*{20}{l}}
  { - \frac{{{p_{{\text{ul}},n}}}}{{{M^2}}}\frac{1}{{{{\left( {1 + \sum\limits_{m = 1}^{{N_{\text{U}}}} {{e_{{\text{ul}},n,m}}} } \right)}^2}}}\text{tr}\left( {{{\bm{T}}_{{{\text{ul}},n,j}}}{{\bm{\Psi }}_{{{\text{ul}},j}}}{\overline{\bm \Sigma }}_{j}^{{\text{ -1}}}{{\bm{T}}_{{{\text{ul}},l,j}}}{{\bm{\Psi }}_{{{\text{ul}},j}}}} \right),}&{\left( {i,n} \right) \ne \left( {j,l} \right)} \\
  {1 - \frac{{{p_{{\text{ul}},n}}}}{{{M^2}}}\frac{1}{{{{\left( {1 + \sum\limits_{m = 1}^{{N_{\text{U}}}} {{e_{{\text{ul}},n,m}}} } \right)}^2}}}\text{tr}\left[ {{{\left( {{{\bm{T}}_{{{\text{ul}},n,i}}}{{\bm{\Psi }}_{{{\text{ul}},i}}}} \right)}^2}{\overline{\bm \Sigma }}_{i}^{{\text{ -1}}}} \right],}&{\left( {i,n} \right) = \left( {j,l} \right)}
\end{array}} \right., \\
{\left[ {{\bm{\Gamma }}_{{\text{ul}}}^{{\text{(k)}}}} \right]_{n,i}} &=  - \frac{{{p_{{\text{ul}},n}}}}{M}\sum\limits_{j = 1}^{{N_{\text{U}}}} {\frac{1}{M}\frac{1}{{{{\left( {1 + \sum\limits_{m = 1}^{{N_{\text{U}}}} {{e_{{\text{ul}},n,m}}} } \right)}^2}}}} {\text{tr}}\left( {{{\bm{T}}_{{{\text{ul}},n,j}}}{\bm{\Psi }}_{{{\text{ul}},j}}^{}{{\bm{T}}_{{{\text{ul}},k,j}}}{{\bm{\Psi }}_{{{\text{ul}},j}}}} \right).
\end{align}
\end{subequations}

${\dot{\bm C}}_{{\text{ul}}}^{{(\Sigma )}} = \left[ {\dot c_{{\text{ul}},k,i}^{(\Sigma )}} \right] \in {\mathbb{C}^{{K_{\text{U}}} \times {N_{\text{U}}}}}$ is a solution to the following equation
\begin{equation}
{{\bm{\Theta }}_{{\text{ul}}}}\text{vec}\left( {{\dot{\bm C}}_{{\text{ul}}}^{{(\Sigma )}}} \right) = \text{vec}\left( {{\bm{\Gamma }}_{{\text{ul}}}^{{(\Sigma )}}} \right),
\end{equation}
where
\begin{equation}
{\left[ {{\bm{\Gamma }}_{{\text{ul}}}^{{(\Sigma )}}} \right]_{k,i}} =  - \frac{{{p_{{\text{ul}},k}}}}{M}\sum\limits_{j = 1}^{{N_{\text{U}}}} {\frac{1}{M}\frac{1}{{{{\left( {1 + \sum\limits_{m = 1}^{{N_{\text{U}}}} {{e_{{\text{ul}},k,m}}} } \right)}^2}}}} {\text{tr}}\left( {{{\bm{T}}_{{{\text{ul}},k,j}}}{\bm{\Psi }}_{{{\text{ul}},j}}^{}{{{\overline{\bm \Sigma }}}_{j}}{{\bm{\Psi }}_{{{\text{ul}},j}}}} \right).
\end{equation}
\end{theorem}

\begin{IEEEproof}
The proof of Theorem \ref{thm:R_ul} is given in Appendix \ref{sec:Appendix_D}.
\end{IEEEproof}

In Theorem \ref{thm:R_ul}, $\overline{\dot{u}}_{{\text{ul}},k,\Sigma }^{\text{(1)}}$ denotes the impact of residual interference and noise on the uplink achievable rate after imperfect cancellation of the DL-to-UL interference. Improving the accuracy of the channel estimation of the channels between transmitting RAUs and receiving RAUs will improve the spectral efficiency of the uplink.

\begin{remark}
In conclusion, Theorems \ref{thm:R_dl_rzf}-\ref{thm:R_ul} indicate that the spectral efficiency in the CF massive MIMO with NAFD can be approximated by its deterministic equivalent without knowing the actual channel realizations. The deterministic equivalents are analytical and much easier to compute than the corresponding ergodic result, which requires time-consuming Monte-Carlo simulations.
Besides, the effects of spatial correlations, large-scale fading, imperfect CSI, as well as imperfect DL-to-UL interference cancellation are taking into account in the CF massive MIMO with NAFD. Since our system model is quite general and the derived results in Theorem \ref{thm:R_dl_rzf}-\ref{thm:R_ul} can be interpreted as a unified formula that encompasses the results of numerous system models, which can be treated as the special cases of our system model. For instance, massive MIMO full-duplex and C-RAN based full-duplex are the special case of the considered system configuration.
\end{remark}

\section{User Scheduling}
In this section, we propose a novel user scheduling strategy to alleviate the UL-to-DL interference and further improve the spectral efficiency of NAFD. Consider a CF massive MIMO where the total number of users that wait on queue to be served is much larger than the number of users that can be served simultaneously, which conforms to the practical situation. The total numbers of UEs waiting for uplink and downlink transmissions are represented as ${K_{{\text{U}},{\text{ALL}}}}$ and ${K_{{\text{D}},{\text{ALL}}}}$ respectively with ${K_{{\text{U}},{\text{ALL}}}} > {K_{\text{U}}}$ and  ${K_{{\text{D}},{\text{ALL}}}} > {K_{\text{D}}}$. For simplicity, we assume that $\frac{{{K_{{\text{U}},{\text{ALL}}}}}}{{{K_{\text{U}}}}} = \frac{{{K_{{\text{D}},{\text{ALL}}}}}}{{{K_{\text{D}}}}} = L$, where $L=2,3,\cdots.$ Let ${\mathcal{K}_{{\text{ALL}}}}$ denote the set of all users and $\left| {{\mathcal{K}_{{\text{ALL}}}}} \right| = {K_{{\text{U}},{\text{ALL}}}} + {K_{{\text{D}},{\text{ALL}}}}$. In addition, we define ${\bm{\Delta }}$ as a family of sets over ${{\mathcal{K}_{{\text{ALL}}}}}$ and ${\bm{\Delta }} = \left\{ {{\mathcal{A}_1},\cdots,{\mathcal{A}_L}} \right\}$, where ${{\mathcal{A}_l}}$ denotes the $l$th group of UEs in uplink and downlink who wait to be served.

The results in \cite{osimeone_fullduplex_2014} suggested that to improve downlink performance of full-duplex systems, the UL-to-DL interference should be small or large enough so that the users in downlink can perform interference cancellation. Thus, an appropriate user scheduling algorithm should ensure one of these two conditions is satisfied. For the second condition, when the channel capacity between the $i$th user in uplink and the $k$th user in downlink is no smaller than the uplink data rate of the $i$th user, the $k$th user could remove the interference from the $i$th user. Accordingly, we define the SINR of the $i$th user active in uplink at the $k$th user active in downlink in the $l$th group under RZF precoding as
\begin{equation}
{\gamma _{{\text{U}},{\text{rzf}},l,k,i}} = \frac{{{p_{{\text{ul}},l,i}}{{\left| {{u_{l,k,i}}} \right|}^2}}}{{{\xi ^2}\sum\limits_{j = 1}^{{K_{\text{D}}}} {{{\left| {{\bm{g}}_{{\text{dl,}}l,k}^{\text{H}}{\bm{C}}_{{\text{dl}},l}^{{\text{-1}}}{{{\hat{\bm g}}}_{{\text{dl}},l,j}}} \right|}^2} + \sum\limits_{j = 1,j \ne i}^{{K_{\text{U}}}} {{p_{{\text{ul}},l,j}}{{\left| {{u_{l,k,j}}} \right|}^2}}  + \sigma _{{\text{dl}},l}^2} }}.
\end{equation}
The second condition is satisfied with
\begin{equation}
\label{eq:cancel_condition}
{\log _2}\left( {1 + {\gamma _{{\text{U}},{\text{rzf}},l,k,i}}} \right) \geqslant {\log _2}\left( {1 + {\gamma _{{\text{ul}},l,i}}} \right),
\end{equation}
where ${\gamma _{{\text{ul}},l,i}}$ is expressed as
\begin{equation}
{\gamma _{{\text{ul}},l,i}} = \frac{{{p_{{\text{ul}},l,i}}{{\left| {{\hat{\bm g}}_{{\text{ul}},l,i}^{\text{H}}{\bm{C}}_{{\text{ul}},l}^{{\text{-1}}}{{\bm{g}}_{{\text{ul}},l,i}}} \right|}^2}}}{{\sum\limits_{j = 1,j \ne i}^{{K_{\text{U}}}} {{p_{{\text{ul}},l,j}}{{\left| {{\hat{\bm g}}_{{\text{ul}},l,i}^{\text{H}}{\bm{C}}_{{\text{ul}},l}^{{\text{-1}}}{{\bm{g}}_{{\text{ul}},l,j}}} \right|}^2}}  + {\hat{\bm g}}_{{\text{ul}},l,i}^{\text{H}}{\bm{C}}_{{\text{ul}},l}^{{\text{-1}}}{{\bm{\Sigma }}_l}{\bm{C}}_{{\text{ul}},l}^{{\text{-1}}}{{{\hat{\bm g}}}_{{\text{ul}},l,i}}}}.
\end{equation}
Furthermore, a sign function is defined as
\begin{equation}
{\delta _{l,k,i}} = \left\{ {\begin{array}{*{20}{l}}
  {1,}&{{\text{the}} \ {\text{received}} \ {\text{interference}} \ {\text{in}} \ {\text{downlink}} \ {\text{user}} \ k \ {\text{from}} \ {\text{uplink}} \ {\text{user}} \ i \ {\text{in}} \ {\text{the}} \ l{\text{th}} \ {\text{group}} \ {\text{exists}}} \\
  {0,}&{{\text{the}} \ {\text{received}} \ {\text{interference}} \ {\text{in}} \ {\text{downlink}} \ {\text{user}} \ k \ {\text{from}} \ {\text{uplink}} \ {\text{user}} \ i \ {\text{in}} \ {\text{the}} \ l{\text{th}} \ {\text{group}} \ {\text{is}} \ {\text{cancelled}}}.
\end{array}} \right.
\end{equation}

A proper user scheduling scheme should be designed to minimize the interference between uplink and downlink, who communicate with RAUs simultaneously. Therefore, we formulate the UL-to-DL interference minimization problem in NAFD as follows
\begin{subequations} \label{eq:min_up_to_down_interference}
\begin{align}
\label{eq:min_interference_a}
& \mathop {\min }\limits_{\bm{\Delta }} \quad {\text{IUD}}\left( {\bm{\Delta }} \right) = \sum\limits_{l = 1}^L {\sum\limits_{k = 1}^{{K_{\text{D}}}} {\sum\limits_{i = 1}^{{K_{\text{U}}}} {{\delta _{l,k,i}}{{\log }_2}\left( {1 + {\gamma _{{\text{U}},{\text{rzf}},l,k,i}}} \right)} } } \\
\label{eq:min_interference_b}
& \ s.t. \quad {\delta _{l,k,i}} \in \left\{ {0,1} \right\}, \quad \forall l \in \left[ {1,L} \right],\forall k \in \left[ {1,{K_{\text{D}}}} \right],\forall i \in \left[ {1,{K_{\text{U}}}} \right], \\
\label{eq:min_interference_c}
& \qquad \ \ {\mathcal{A}_l} \subseteq {\mathcal{K}_{{\text{ALL}}}},\bigcup\limits_{l = 1}^L {{\mathcal{A}_l}}  = {\mathcal{K}_{{\text{ALL}}}}, \quad \forall l \in \left[ {1,L} \right], \\
\label{eq:min_interference_d}
& \qquad \ \ {\mathcal{A}_i} \cap {\mathcal{A}_j} = \emptyset , \quad \forall i \ne j.
\end{align}
\end{subequations}
(\ref{eq:min_interference_b}) follows the fact that the received interference at the $k$th user active in downlink from the $i$th user in the $l$th group of uplink can be cancelled when it satisfies (\ref{eq:cancel_condition}). The combination of all $L$ groups of users is the universal set ${\mathcal{K}_{{\text{ALL}}}}$, as described in (\ref{eq:min_interference_c}). (\ref{eq:min_interference_d}) shows that since each user needs to be served for only one time, different groups of users are independent from each other. The objective function in (\ref{eq:min_interference_a}) couples all users in ${\mathcal{K}_{{\text{ALL}}}}$ together, which makes the UL-to-DL interference minimization problem extremely cumbersome. One achievable method to find the optimal $\bm{\Delta }$ is the exhaustive search in the entire search space. Obviously, in a CF massive MIMO, the entire search space of all the users will be prohibitively large and pursuing the optimal system performance would lead to high computational complexity. Therefore, the optimal exhaustive search can not be implemented in practice.

In the field of artificial intelligence, a genetic algorithm is a search heuristic inspired by the process of natural selection. Genetic algorithms are commonly used to generate high-quality solutions to optimization and search problems by relying on bio-inspired operators such as mutation, crossover and selection \cite{msrinivas_adaptive_1994,jzhang_clustering_2007}. Therefore, aiming at minimizing the UL-to-DL interference in a CF massive MIMO with NAFD, a user scheduling method based genetic algorithm is proposed, named after genetic algorithm based user scheduling strategy (GAS). Similar to the process of a genetic algorithm \cite{mmelanie_introduction_1996}, the proposed GAS includes the following three steps:

\emph{Step 1: Initialization.} ${K_{{\text{U}},{\text{ALL}}}}$ users and ${K_{{\text{D}},{\text{ALL}}}}$ users are randomly distributed in a circular area with radius $R$. ${N_{{\text{U}}}}$ receiving RAUs and ${N_{{\text{D}}}}$ transmitting RAUs, each equipped with $M$ antennas, are alternately placed on a circle of radius $r$. The population size of each iteration ${S_{\text{P}}}$ and the maximum number of iterations ${S_{\text{I}}}$ are chosen properly to ensure fast convergence to an optimal solution and prevent premature convergence to a local optimum.

\emph{Step 2: Selection.} Create ${S_{\text{P}}}$ candidate solutions randomly. During each successive iteration, every candidate solution ${\bm{\Delta }}$ to the objective function (\ref{eq:min_interference_a}) is selected from the current generation with probability inversely proportional to its corresponding calculated IUD(${\bm{\Delta }}$).

\emph{Step 3: Genetic Operators.} Each selected ${\bm{\Delta }}$ exchange part of its elements with other selected solution randomly. Afterwards, we select certain existing ${\bm{\Delta }}$s and exchange the groups of some users within a user grouping solution randomly. Note that a validation adjustment must be adopted to ensure each derived ${\bm{\Delta }}$ satisfies (\ref{eq:min_interference_c}) and (\ref{eq:min_interference_d}).

The algorithm terminates when IUD(${\bm{\Delta }}$) reaches to a satisfactory value and fluctuates within an acceptable range, or ${S_{\text{I}}}$ goes to the maximum number of iterations. Thus, we obtain the optimal user grouping scheme that can minimize the UL-to-DL interference in the whole system.

\section{Simulations and Discussions}
As described above, the deterministic equivalents in Theorems \ref{thm:R_dl_rzf}-\ref{thm:R_ul} provide an effective and efficient technique to approximate the ergodic sum-rate for the CF massive MIMO with NAFD. In Section \ref{sec:accuracy}, Monte-Carlo simulations are used to validate the accuracy of the above asymptotic characterization. The simulation results are obtained by averaging over 10,000 i.i.d. Rayleigh block-fading channels. It will be confirmed that although the analytical results are derived in the large-system regime, the deterministic equivalents is useful even for small-dimensional systems. In Section \ref{sec:CCFD}, we compare the spectral efficiency of NAFD and CCFD systems to demonstrate that NAFD offers a more spectral-efficient solution in comparison with massive MIMO CCFD as well as C-RAN CCFD. Finally, in Section \ref{sec:user_scheduling}, a random user scheduling scheme is simulated as a comparison to illustrate the performance improvement of GAS in a CF massive MIMO with NAFD.

\subsection{Accuracy of the Deterministic Equivalent Results} \label{sec:accuracy}
In this subsection, numerical results are presented to confirm the accuracy of the derived deterministic equivalents of downlink ergodic sum-rate under RZF and ZF precoding as well as uplink ergodic sum-rate under MMSE estimation. It is assumed that each RAU is equipped with $M = 8$ antennas and the number of receiving RAUs and transmitting RAUs are equal to ${N_{\text{U}}} = {N_{\text{D}}} = 4$. There are 64 active users and ${K_{\text{U}}} = {K_{\text{D}}} = 32$. The regularization parameter $\alpha $ is chosen optimally \cite{jzhang_large_2013}. Since the channel between transmitting RAUs and receiving RAUs is static and imperfect CSI is considered, we set ${\left\{ {\tau _{{\text{I}},k,n}^2 = 0.1} \right\}_{\forall k,n}}$. In addition, assuming transmit powers of all users are the same, i.e., ${\left\{ {{p_{{\text{ul}},i}}} \right\}_{\forall i}} = {p_{{\text{ul}}}}$, we define downlink SNR and uplink SNR as ${\rho _{\text{dl}}} \triangleq \frac{P}{{\sigma _{\text{dl}}^2}}$ and ${\rho _{\text{ul}}} \triangleq \frac{{{p_{\text{ul}}}}}{{\sigma _{\text{ul}}^2}}$, respectively.

\begin{figure}[htbp]
\centering
\begin{minipage}[t]{0.48\linewidth}
\centering
\includegraphics[scale=.45]{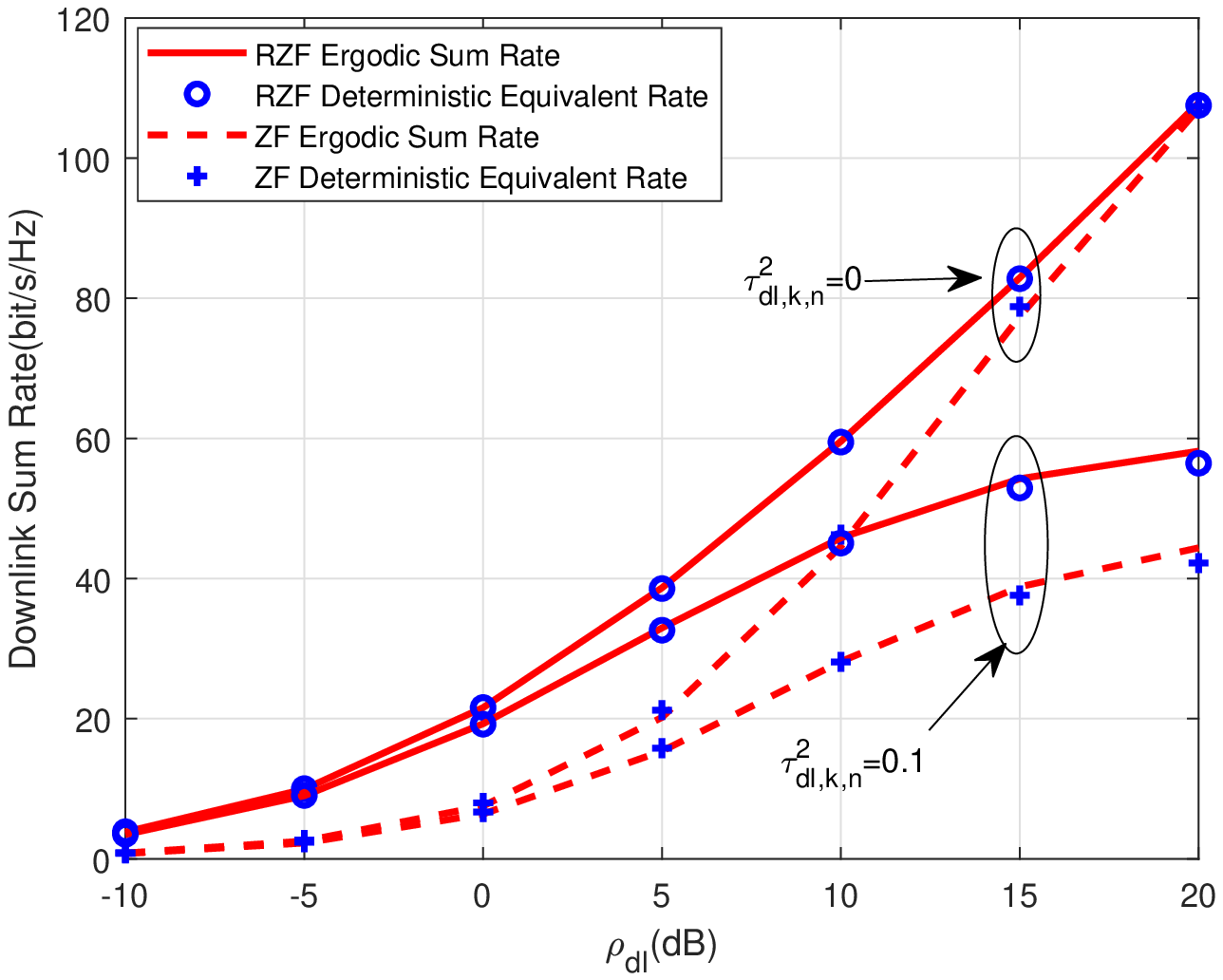}
\caption{Downlink ergodic sum-rate and the deterministic equivalents under RZF and ZF precoding for a fixed uplink SNR ${\rho _{{\text{ul}}}} =  {-\text{10dB}}$ with ${{\bm{T}}_{{k,n}}} = {{\bm{I}}_{M}}$.}
\label{Figure2}
\end{minipage}
\begin{minipage}[t]{0.48\linewidth}
\centering
\includegraphics[scale=.45]{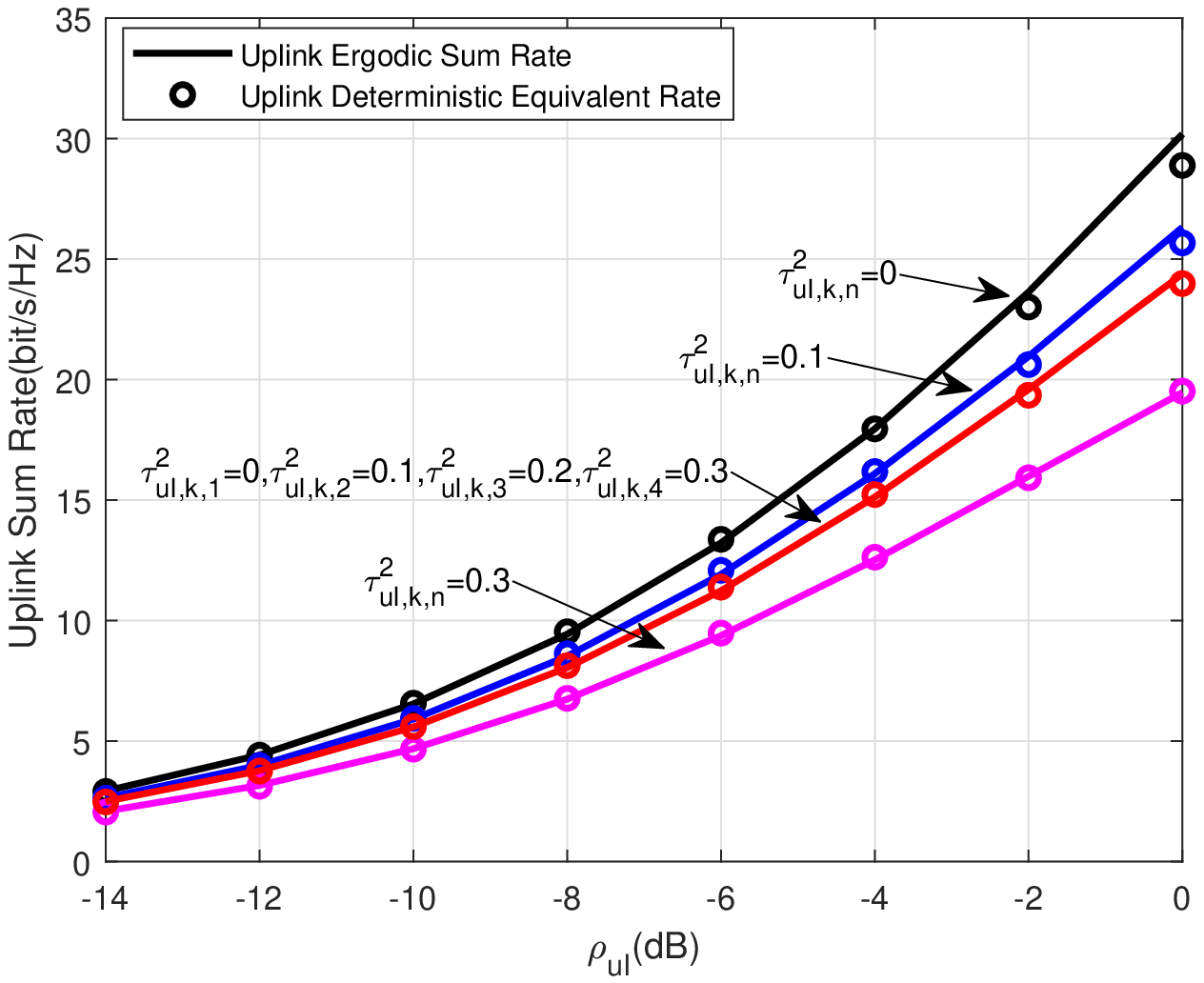}
\caption{Uplink ergodic sum-rate and the deterministic equivalents for a fixed downlink SNR ${\rho _{\text{dl}}} = {\text{5dB}}$ with ${\left\{ {\tau _{{\text{dl}},k,n}^2 = 0.1} \right\}_{\forall k,n}}$, ${{\bm{T}}_{{k,n}}} = {{\bm{I}}_{M}}$.}
\label{Figure3}
\end{minipage}
\end{figure}

Fig.{\ref{Figure2}} compares the downlink ergodic sum-rate and its corresponding deterministic equivalents under RZF and ZF precoding for two cases: 1) ${\left\{ {\tau _{{\text{dl}},k,n}^2 = 0} \right\}_{\forall k,n}}$; 2) ${\left\{ {\tau _{{\text{dl}},k,n}^2 = 0.1} \right\}_{\forall k,n}}$, where ${{\bm{T}}_{{k,n}}} = {{\bm{I}}_{M}}$ and ${\rho _{\text{ul}}} = {-\text{10dB}}$. It is shown in Fig.{\ref{Figure2}} that the derived approximation of downlink sum-rate under RZF and ZF precoding is even accurate for systems with finite number of antennas.

In Fig.{\ref{Figure2}}, it can be seen that the deterministic approximation becomes less accurate for high
SNR. The reason is that in the derivation of the approximated sum-rate, the utilized upper-bounds in \cite[Proposition 12]{swagner_large_2012} are proportional to the SNR. Hence, larger system dimensions are required in the high SNR regime to improve the accuracy of approximated value. In addition, it can be seen that when the channel is ill-conditioned, such as when $\tau _{{\text{dl}},k,n}^2 = 0.1$ or ${\rho _{\text{dl}}}$ is small, the achievable rate under RZF precoding is much larger than that of ZF system. This is because RZF introduces a regularization parameter in the channel inversion to mitigate the ill-condition problem. The regularization parameter can control the amount of the introduced interference by choosing the parameter properly. Meanwhile, when the channel condition is better, e.g., $\tau _{{\text{dl}},k,n}^2 = 0$ and ${\rho _{\text{dl}}}$ is large, the achievable sum-rate under RZF and ZF precoding are literally the same.

%

\begin{figure}
\centering
\includegraphics[scale=.55]{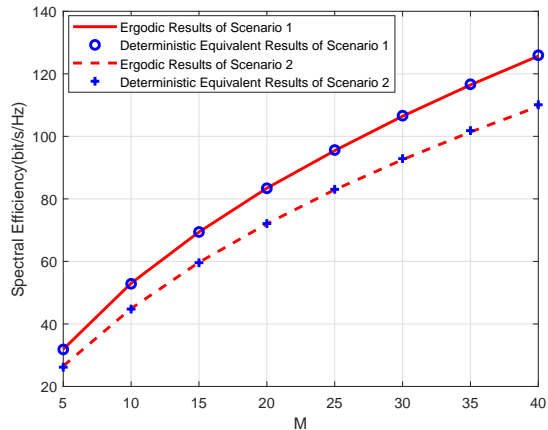}
\caption{The simulation results and the deterministic equivalents of spectral efficiency in two scenarios: 1) ${{\bm{T}}_{{k,n}}} = {{\bm{I}}_{M}}$, ${\left\{ {\tau _{{\text{ul}},k,n}^2 = \tau _{{\text{dl}},k,n}^2 = 0} \right\}_{\forall k,n}}$; 2) ${{\bm{T}}_{{k,n}}} \ne {{\bm{I}}_{M}}$, ${\left\{ {\tau _{{\text{ul}},k,n}^2 = \tau _{{\text{dl}},k,n}^2 = 0.1} \right\}_{\forall k,n}}$.}
\label{Figure4}
\end{figure}

Fig.{\ref{Figure3}} compares the uplink ergodic sum-rate and its corresponding deterministic equivalents in (\ref{eq:R_ul_DE}) for the following four different cases: 1) ${\left\{ {\tau _{{\text{ul}},k,n}^2 = 0} \right\}_{\forall k,n}}$; 2) ${\left\{ {\tau _{{\text{ul}},k,n}^2 = 0.1} \right\}_{\forall k,n}}$; 3) ${\left\{ {\tau _{{\text{ul}},k,n}^2 = 0.3} \right\}_{\forall k,n}}$; 4) $\left\{ {\tau _{{\text{ul}},k,1}^2 = 0,} \right.$
${\left. {\tau _{{\text{ul}},k,2}^2 = 0.1, \tau _{{\text{ul}},k,3}^2 = 0.2, \tau _{{\text{ul}},k,4}^2 = 0.3} \right\}_{\forall k}}$, where ${\left\{ {\tau _{{\text{dl}},k,n}^2 = 0.1} \right\}_{\forall k,n}}$, ${{\bm{T}}_{{k,n}}} = {{\bm{I}}_{M}}$ and ${\rho _{\text{dl}}} = {\text{5dB}}$. Fig.{\ref{Figure3}} demonstrates the accuracy of the derived deterministic equivalents. Furthermore, the achievable rate in the uplink transmission becomes lower as the channel estimation error gets larger, which is consistent with the practical situations.

Next, we examine the accuracy of the deterministic equivalents for general system settings. The co-existence of imperfect CSI and independent spatial correlations are considered. The spectral efficiency under RZF precoding and MMSE uplink receiver is the sum of uplink rate and downlink rate, i.e., ${R_{{\text{dl}},{\text{rzf}},{\text{sum}}}} + {R_{{\text{ul}},{\text{sum}}}}$, the analytical approximation of which is ${\overline R_{{\text{dl}},{\text{rzf}},{\text{sum}}}} + {\overline R_{{\text{ul}},{\text{sum}}}}$. Fig.{\ref{Figure4}} demonstrates the simulation results and the analytical results of spectral efficiency under two different system settings: 1) The channels are independent and uncorrelated, i.e., ${{\bm{T}}_{{k,n}}} = {{\bm{I}}_{M}}$, and perfect CSI is available, i.e., ${\left\{ {\tau _{{\text{ul}},k,n}^2 = \tau _{{\text{dl}},k,n}^2 = 0} \right\}_{\forall k,n}}$; 2) The channels are independent but correlated, i.e., ${{\bm{T}}_{{k,n}}} \ne {{\bm{I}}_{M}}$, and there exists channel estimation errors, i.e., ${\left\{ {\tau _{{\text{ul}},k,n}^2 = \tau _{{\text{dl}},k,n}^2 = 0.1} \right\}_{\forall k,n}}$.

As illustrated in Fig.{\ref{Figure4}}, the asymptotic results are accurate in both ideal systems and non-ideal systems. Besides, the spectral efficiency in ideal systems is distinctly better than that in non-ideal systems, and it improves if the number of antennas increases.

\subsection{Comparison between NAFD and CCFD} \label{sec:CCFD}
In this subsection, we compare the spectral efficiency of NAFD and CCFD systems, where CCFD includes CCFD massive MIMO  as well as CCFD C-RAN. In order to demonstrate the comparison of the two different systems properly and clearly, we model the large-scale fading matrix of uplink, ${{\bm{T}}_{{\text{ul}},k,n}} \in {\mathbb{C}^{M \times M}}$, as
$
{{\bm{T}}_{{\text{ul}},k,n}} = {c_{\text{r}}}d_{{\text{ul,}}k,n}^{ - \eta }{{\bm{I}}_M},
$
where $d_{{\text{ul,}}k,n}$ denotes the distance between the $k$th UE active in the uplink and the $n$th receiving RAU, $\eta$ is the path loss exponent, and $c_{\text{r}}$ is the median of the mean path gain at a reference distance $d_{{\text{ul,}}k,n} = {\text{1km}}$. Similarly, the large-scale fading matrix between the $k$th UE and the $n$th transmitting RAU, the large-scale fading matrix between the $k$th antenna of transmitting RAUs and the $n$th receiving RAU and the large-scale fading between the $i$th UE active in the uplink and the $k$th UE active in the downlink are respectively modelled as
$
{{\bm{T}}_{{\text{dl}},k,n}} = {c_{\text{r}}}d_{{\text{dl,}}k,n}^{ - \eta }{{\bm{I}}_M}, \;\;{{\bm{T}}_{{\text{I}},k,n}} = {c_{\text{r}}}d_{{\text{I,}}k,n}^{ - \eta }{{\bm{I}}_M}, \;\; {T_{k,i}} = {c_{\text{r}}}d_{k,i}^{ - \eta }.
$

\begin{figure}[htbp]
\centering
\begin{minipage}[t]{0.45\linewidth}
\centering
\includegraphics[scale=.45]{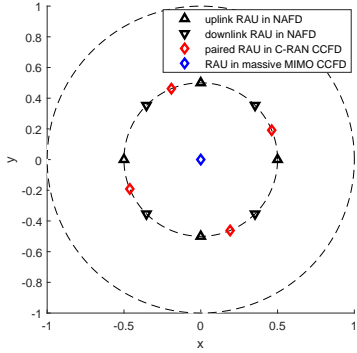}
\caption{RAUs layout in NAFD, massive MIMO CCFD, C-RAN CCFD with ${N_{\text{U}}} = {N_{\text{D}}} = 4$.}
\label{Figure5}
\end{minipage}
\begin{minipage}[t]{0.45\linewidth}
\centering
\includegraphics[scale=.45]{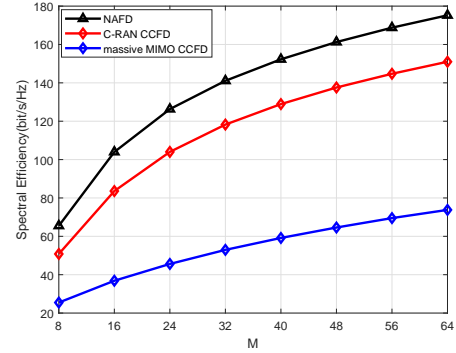}
\caption{Spectral efficiency against $M$ with RZF precoder for downlink and MMSE receiver for uplink in NAFD, massive MIMO CCFD and C-RAN CCFD systems.}
\label{Figure6}
\end{minipage}
\centering
\end{figure}

Consider a circular area with radius $R = {\text{1km}}$ and all users are randomly distributed in this area. The minimum distance from users to RAUs is set as $r_{\text{0}} = {\text{30m}}$. The receiving RAUs and transmitting RAUs are alternately placed on a circle of radius $r = {\text{500m}}$ in NAFD systems. Meanwhile, in CCFD massive MIMO, all RAUs are co-located in the center of this area. In CCFD C-RAN systems, one receiving RAU and one transmitting RAU are paired and located in the same place. Fig. \ref{Figure5} shows one special deployment of RAUs for the three different systems. In CCFD systems, the self-interference from the $k$th antenna of transmitting RAUs to the $i$th antenna of receiving RAUs in the same location is modelled as i.i.d. ${\cal C}{\cal N}\left( {0,{\sigma _{\text{SI}}^2}} \right)$ random variable \cite{hqngo_multipair_2014}. Other system settings are assumed in Section \ref{sec:accuracy}. The path loss exponent $\eta = \text{3.7}$. For the fairness of the comparison, the cancellation of self-interference in CCFD systems is considered and $\sigma _{\text{SI}}^2$ is set to be $\frac{1}{M}$.

%

Fig.\ref{Figure6} depicts the spectral efficiency with RZF precoder for downlink and MMSE receiver for uplink in NAFD, CCFD massive MIMO  and CCFD C-RAN systems against the number of antannas each RAU equipped with.
It is seen from Fig.\ref{Figure6} that the spectral efficiency in NAFD, CCFD massive MIMO  and CCFD C-RAN systems increases as the number of antennas $M$ becomes larger. Furthermore, NAFD systems are more spectral-efficient than both CCFD massive MIMO and CCFD C-RAN systems. This is in line with the theory that the performance comparison beteween NAFD and CCFD is similar to that between co-located MIMO and distributed MIMO. Since the RAUs are placed in a more advantageous position, distributed antennas can achieve additional power gains and macro diversity\cite{dwang_spectral_2013,ldai_uplink_2014}. In addition, the spectral efficiency in massive MIMO CCFD is fairly small. This is because all RAUs are located in the center of the area and the signal quality of cell edge users is rather poor.

\begin{figure}[htbp]
\centering
\subfigure[]{
\begin{minipage}[t]{0.4\linewidth}
\centering
\includegraphics[scale=.4]{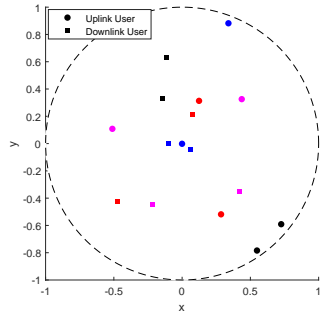}
\label{fig:area_opt}
\end{minipage}
}
\subfigure[]{
\begin{minipage}[t]{0.4\linewidth}
\centering
\includegraphics[scale=.4]{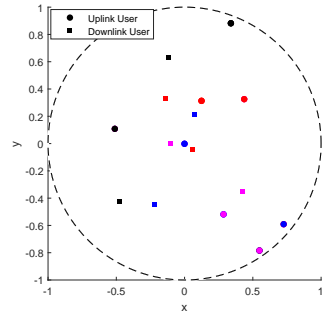}
\label{fig:area_ran}
\end{minipage}
}
\centering
\caption{Illustration of various user scheduling schemes with ${K_{{\text{U}},{\text{ALL}}}} = {K_{{\text{D}},{\text{ALL}}}} = 8$, ${K_{\text{U}}} = {K_{\text{D}}} = 2$, and $L = 4$. Four different colours denote four groups of users and each group of users will be served simultaneously. (a) Optimal user scheduling scheme. (b) Random user scheduling Scheme.}\label{fig:area}
\end{figure}


\subsection{Comparison between GAS and Random User Scheduling} \label{sec:user_scheduling}
In this subsection, we compare the achievable downlink sum-rate applying GAS and random user scheduling in a CF massive MIMO with NAFD. For simple demonstration, we consider a system with ${N_{\text{U}}} = {N_{\text{D}}} = 2$ and each RAU is equipped with $M = 1$ antenna. ${K_{\text{U}}} = {K_{\text{D}}} = 2$ and there are ${K_{{\text{U}},{\text{ALL}}}} = {K_{{\text{D}},{\text{ALL}}}} = 8$ users who wait to be served. Therefore, all users can be divided into $L = \frac{{{K_{{\text{U}},{\text{ALL}}}}}}{{{K_{\text{U}}}}} = \frac{{{K_{{\text{D}},{\text{ALL}}}}}}{{{K_{\text{D}}}}} = 4$ groups.
Other system settings are assumed in Section \ref{sec:CCFD}. Applying GAS, the derived optimal user grouping scheme and one random user grouping scheme are depicted in Fig. {\ref{fig:area}}.
It can be seen from Fig.{\ref{fig:area_opt}} that UEs active in both uplink and downlink in the same group are geographically distant from each other to mitigate the UL-to-DL interference. Meanwhile, in some special cases, such as in the blue group, one UE active in uplink is quite close to the two UEs active in downlink. This is because the interference of the UE in uplink to the two UEs in downlink is large enough, so that the two UEs can perform successive interference cancellation. Fig. {\ref{fig:area_ran}} illustrates a random user scheduling scheme where the UEs in uplink and UEs in downlink may be close to each other but not close enough to do interference cancellation successfully.

\begin{figure}
\centering
\includegraphics[scale=.55]{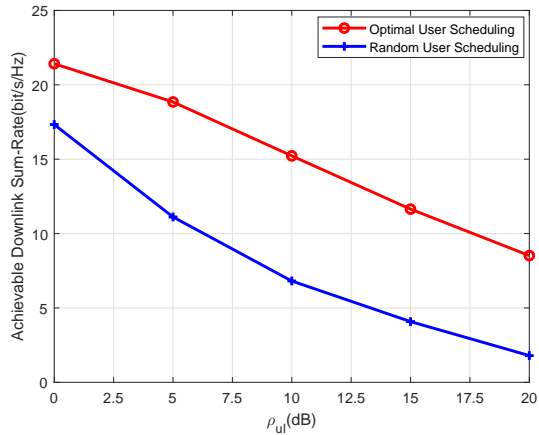}
\caption{The achievable downlink sum-rate of the optimal user scheduling scheme and the random user scheduling scheme against the uplink SNR ${\rho _{\text{ul}}}$ in CF massive MIMO with NAFD for a fixed downlink SNR ${\rho _{\text{dl}}} = {\text{5dB}}$.}
\label{fig:scheduling_rate}
\end{figure}

Fig. {\ref{fig:scheduling_rate}} demonstrates the achievable downlink sum-rate of the optimal user scheduling scheme obtained by GAS and the random user scheduling scheme in NAFD. As seen from Fig. {\ref{fig:scheduling_rate}}, the achievable downlink sum-rate decreases as the uplink SNR ${\rho _{\text{ul}}}$ increases in both systems. Furthermore, the achievable sum-rate with optimal user scheduling scheme is greatly improved in comparison with the achievable sum-rate under random user scheduling. This validates the effectiveness of the proposed GAS, which can further improve the spectral efficiency in NAFD.

\section{Conclusions}
In this paper, we have presented a novel spatial domain duplexing method named NAFD for CF massive MIMO network. The spectral efficiency of NAFD was also investigated with imperfect CSI and spatial correlations. Utilizing large dimensional random matrix theory, the deterministic equivalents of the sum-rate for uplink with MMSE receiver as well as downlink with RZF and ZF precoders were derived. A user scheduling scheme named as GAS was proposed to help alleviating the UL-to-DL interference and the simulation results have validated that in CF massive MIMO with NAFD, higher sum-rate can be achieved applying GAS rather than random user scheduling. Numerical simulations have revealed that under various environment settings, the analytical results have provided reliable performance predictions in both large-scale system and system with a finite number of antennas. Furthermore, it was verified that NAFD with half-duplexing RAUs offers a more spectral-efficient solution than CCFD massive MIMO as well as CCFD C-RAN.

\appendices

\setcounter{equation}{0}
\renewcommand\theequation{\Alph{section}-\arabic{equation}}

\section{Important lemmas}\label{sec:Appendix_A}
We first introduce five important lemmas, which serve as the mathematical basis for the following derivations of deterministic equivalent.
\begin{lemma} \label{matrix_inverse}
(Matrix Inversion Lemma): \cite[Lemma 2.2]{jwsilverstein_empirical_1995}: For any ${\bm{U}} \in {\mathbb{C}^{N \times N}}$, ${\bm{x}} \in {\mathbb{C}^N}$ and $c \in \mathbb{C}$, when ${\bm{U}}$ and ${\bm{U}} + c{\bm{x}}{{\bm{x}}^{\text{H}}}$ are invertible, we have
\[
{{\bm{x}}^{\text{H}}}{\left( {{\bm{U}} + c{\bm{x}}{{\bm{x}}^{\text{H}}}} \right)^{{\text{-1}}}} = \frac{{{{\bm{x}}^{\text{H}}}{{\bm{U}}^{{\text{-1}}}}}}{{1 + c{{\bm{x}}^{\text{H}}}{{\bm{U}}^{{\text{-1}}}}{\bm{x}}}}.
\]
\end{lemma}

\begin{lemma} \label{same_matrix}
\cite[Lemma 14.2]{rcouillet_random_2011}: Let ${{\bm{A}}_{1}},{{\bm{A}}_{2}},\cdots,{{\bm{A}}_{N}} \in {\mathbb{C}^{N \times N}}$ be a series of random matrices and ${{\bm{x}}_{1}},{{\bm{x}}_{2}},\cdots,{{\bm{x}}_{N}} \in {\mathbb{C}^N}$ be random vectors of i.i.d. entries with zero mean, variance $\frac{1}{N}$, and finite eighth-order moment, independent of ${{\bm{A}}_{N}}$. When $N \to \infty $, we have
\[
{\bm{x}}_{N}^{\text{H}}{{\bm{A}}_{N}}{{\bm{x}}_{N}} - \frac{1}{N}\text{tr}{{\bm{A}}_{N}}\xrightarrow{{a.s.}}0.
\]
\end{lemma}

\begin{lemma} \label{diff_matrix}
\cite[Lemma 5]{swagner_large_2012}: Let ${{\bm{A}}_{N}}$ be as in Lemma {\ref{same_matrix}} and ${{\bm{x}}_{N}},{{\bm{y}}_{N}} \in {\mathbb{C}^N}$ be random, mutually independent vectors with standard i.i.d. entries of zero mean, variance $\frac{1}{N}$, and finite eighth-order moment, independent of ${{\bm{A}}_{N}}$. When $N \to \infty $, we have
\[
{\bm{y}}_{N}^{\text{H}}{{\bm{A}}_{N}}{{\bm{x}}_{N}}\xrightarrow{{a.s.}}0.
\]
\end{lemma}

\begin{lemma} \label{plus}
\cite[Lemma 14.3]{rcouillet_random_2011}: Let ${{\bm{A}}_{1}},{{\bm{A}}_{2}},\cdots,{{\bm{A}}_{N}} \in {\mathbb{C}^{N \times N}}$ be deterministic matrices with uniformly bounded spectral norm and ${{\bm{B}}_{1}},{{\bm{B}}_{2}},\cdots,{{\bm{B}}_{N}} \in {\mathbb{C}^{N \times N}}$ be random Hermitian matrices with eigenvalues $\lambda _1^{{{\bm{B}}_{N}}} \leqslant \cdots \leqslant \lambda _N^{{{\bm{B}}_{N}}}$. When $N$ is large, there exists $\varepsilon  > 0$ for which $\lambda _1^{{{\bm{B}}_{N}}} > \varepsilon $ with probability 1. Then, for any ${\bm{v}} \in {\mathbb{C}^N}$, ${\bm{B}}_{N}^{{\text{-1}}}$ and ${\left( {{{\bm{B}}_{N}} + {\bm{v}}{{\bm{v}}^{\text{H}}}} \right)^{{\text{-1}}}}$ exist with probability 1 and when $N \to \infty $, we have
\[
\frac{1}{N}\text{tr}\left( {{{\bm{A}}_{N}}{\bm{B}}_{N}^{{\text{-1}}}} \right) - \frac{1}{N}\text{tr}\left[ {{{\bm{A}}_{N}}{{\left( {{{\bm{B}}_{N}} + {\bm{v}}{{\bm{v}}^{\text{H}}}} \right)}^{{\text{-1}}}}} \right]\xrightarrow{{a.s.}}0.
\]
\end{lemma}

\begin{lemma} \label{resolvent_identity}
(Resolvent Identity): Letting ${\bm{U}},{\bm{V}} \in {\mathbb{C}^{N \times N}}$ be two invertible matrices, we have
\[
{{\bm{U}}^{{\text{-1}}}} - {{\bm{V}}^{{\text{-1}}}} =  - {{\bm{U}}^{{\text{-1}}}}\left( {{\bm{U}} - {\bm{V}}} \right){{\bm{V}}^{{\text{-1}}}}.
\]
\end{lemma}

\section{Proof of Theorem 2}\label{sec:Appendix_B}
We proceed by deriving the approximation ${\overline \gamma _{\text{dl},\text{zf},k}}$ of downlink SINR under ZF precoding ${\gamma _{\text{dl},\text{zf},k}}$ first. Rewrite ${\overline \gamma _{\text{dl},\text{zf},k}}$ in (\ref{eq:r_dl_rzf_DE}) as
\begin{equation}
\label{a2_b1}
{\overline \gamma _{\text{dl},\text{rzf},k}} = \frac{{{{\left( {\alpha \overline u_{\text{dl},k}^{(2)}} \right)}^2}}}{{{{\left( {\alpha  + \alpha \overline u_{\text{dl},k}^{(1)}} \right)}^2}\left( {{{\overline u}_{\text{dl},k}} + {{\overline v}_{\text{dl}}}} \right)}}.
\end{equation}
According to \cite[Appendix III]{swagner_large_2012}, under Assumptions {\ref{asm:g_dl}}, {\ref{asm:eigen}} and {\ref{asm:e}}, we have ${\overline \gamma _{\text{dl},\text{zf},k}} = {\lim _{\alpha  \to 0}}{\overline \gamma _{\text{dl},\text{rzf},k}}$. Therefore, ${\overline \gamma _{\text{dl},\text{zf},k}}$ in (\ref{a2_b1}) consists of three terms: 1) $\underline{\overline u} _{\text{dl},k}^{(1)} \triangleq {\lim _{\alpha  \to 0}}\alpha \overline u_{\text{dl},k}^{(1)}$, $\underline{\overline u} _{\text{dl},k}^{(2)} \triangleq {\lim _{\alpha  \to 0}}\alpha \overline u_{\text{dl},k}^{(2)}$; 2) ${\underline{\overline u} _{\text{dl},k}} \triangleq {\lim _{\alpha  \to 0}}{\overline u_{\text{dl},k}}$; 3) ${\underline{\overline v} _{\text{dl}}} \triangleq {\lim _{\alpha  \to 0}}{\overline v_{\text{dl}}}$. We now subsequently derive the deterministic equivalents for each of these three terms, which together constitute the final expression for ${\overline \gamma _{\text{dl},\text{zf},k}}$.

\begin{lemma}\label{lemma:signal_dl_ZF}
Letting Assumptions {\ref{asm:g_dl}}-{\ref{asm:e}} hold true and as $\mathcal{N} \to \infty $, we have
\begin{align}
\label{a2_b2}
\underline{\overline u} _{\text{dl},k}^{(1)} &= \sum\limits_{n = 1}^{{N_\text{D}}} {\frac{1}{M}\text{tr}\left( {{{\bm{T}}_{{{\text{dl}},k,n}}}{{{\underline{\bm{\Psi } }}}_{{{\text{dl}},n}}}} \right)}, \\
\label{a2_b3}
\underline{\overline u} _{\text{dl},k}^{(2)} &= \sum\limits_{n = 1}^{{N_\text{D}}} {\frac{{{\varphi _{\text{dl},k,n}}}}{M}\text{tr}\left( {{{\bm{T}}_{{{\text{dl}},k,n}}}{{{\underline{\bm{\Psi } }}}_{{{\text{dl}},n}}}} \right)}.
\end{align}
where ${{\underline{\bm{\Psi } }}_{{{\text{dl}},n}}}$ is given by (\ref{eq:pha_dl_ZF}).
\end{lemma}

\begin{IEEEproof}
Defining ${\underline{e} _{\text{dl},k,n}} \triangleq {\lim _{\alpha  \to 0}}\alpha {e_{\text{dl},k,n}}$, we have
\begin{equation}
\label{a2_b4}
{\underline{e} _{\text{dl},k,n}} = \frac{1}{M}\text{tr}\left( {{{\bm{T}}_{{{\text{dl}},k,n}}}{{{\underline{\bm{\Psi } }}}_{{{\text{dl}},n}}}} \right),
\end{equation}
where ${{\underline{\bm{\Psi } }}_{{{\text{dl}},n}}}$ is given by (\ref{eq:pha_dl_ZF}). Thus, we can easily obtain
\begin{align*}
& \underline{\overline u} _{\text{dl},k}^{(1)} = {\lim _{\alpha  \to 0}}\alpha \overline u_{\text{dl},k}^{(1)} = {\lim _{\alpha  \to 0}}\sum\limits_{n = 1}^{{N_\text{D}}} {\alpha {e_{\text{dl},k,n}}}  = \sum\limits_{n = 1}^{{N_\text{D}}} {\frac{1}{M}\text{tr}\left( {{{\bm{T}}_{{{\text{dl}},k,n}}}{{{\underline{\bm{\Psi } }}}_{{{\text{dl}},n}}}} \right)}, \\
& \underline{\overline u} _{\text{dl},k}^{(2)} = {\lim _{\alpha  \to 0}}\alpha \overline u_{\text{dl},k}^{(2)} = {\lim _{\alpha  \to 0}}\sum\limits_{n = 1}^{{N_\text{D}}} {\alpha {\varphi _{\text{dl},k,n}}{e_{\text{dl},k,n}}}  = \sum\limits_{n = 1}^{{N_\text{D}}} {\frac{{{\varphi _{\text{dl},k,n}}}}{M}\text{tr}\left( {{{\bm{T}}_{{{\text{dl}},k,n}}}{{{\underline{\bm{\Psi } }}}_{{{\text{dl}},n}}}} \right)}.
\end{align*}
\end{IEEEproof}

\begin{lemma}\label{lemma:noise_dl_ZF}
Letting Assumptions {\ref{asm:g_dl}}-{\ref{asm:e}} hold true and as $\mathcal{N} \to \infty $, we have
\begin{equation}
\label{a2_b5}
{\underline{\overline u} _{\text{dl},k}} = \overline u_{\text{dl},k,[\alpha ]}^{(1)} - \overline{\dot{u}}_{\text{dl},k,[\alpha ]}^{(1)} - \frac{{2\underline{\overline u} _{\text{dl},k}^{(2)}\left( {\overline u_{\text{dl},k,[\alpha ]}^{(2)} - \overline{\dot{u}}_{\text{dl},k,[\alpha ]}^{(2)}} \right)}}{{\underline{\overline u} _{\text{dl},k}^{(1)}}} + \frac{{{{\left( {\underline{\overline u} _{\text{dl},k}^{(2)}} \right)}^2}\left( {\overline u_{\text{dl},k,[\alpha ]}^{(1)} - \overline{\dot{u}}_{\text{dl},k,[\alpha ]}^{(1)}} \right)}}{{{{\left( {\underline{\overline u} _{\text{dl},k}^{(1)}} \right)}^2}}},
\end{equation}
where $\underline{\overline u} _{\text{dl},k}^{(1)}$, $\underline{\overline u} _{\text{dl},k}^{(2)}$, $\overline u_{\text{dl},k,[\alpha ]}^{(1)}$, $\overline u_{\text{dl},k,[\alpha ]}^{(2)}$, $\overline{\dot{u}}_{\text{dl},k,[\alpha ]}^{(1)}$ and $\overline{\dot{u}}_{\text{dl},k,[\alpha ]}^{(2)}$ are given by (\ref{eq:zf_six_a}), (\ref{eq:zf_six_b}), (\ref{eq:zf_six_e}), (\ref{eq:zf_six_f}), (\ref{eq:zf_six_g}) and (\ref{eq:zf_six_h}), respectively.
\end{lemma}

\begin{IEEEproof}
Defining $\overline u_{\text{dl},k,[\alpha ]}^{(1)} \triangleq {\lim _{\alpha  \to 0}}\overline u_{\text{dl},k}^{(1)}$, we have
\begin{align*}
\overline u_{\text{dl},k,[\alpha ]}^{(1)} &= {\lim _{\alpha  \to 0}}\sum\limits_{n = 1}^{{N_\text{D}}} {\frac{1}{M}\text{tr}\left[ {{{\bm{T}}_{{{\text{dl}},k,n}}}{{\left( {\frac{1}{M}\sum\limits_{j = 1}^{{K_\text{D}}} {\frac{{{{\bm{T}}_{{{\text{dl}},j,n}}}}}{{1 + \sum\limits_{m = 1}^{{N_\text{D}}} {{e_{\text{dl},j,m}}} }} + \alpha {{\bm{I}}_{M}}} } \right)}^{ - 1}}} \right]} \\
&= \sum\limits_{n = 1}^{{N_\text{D}}} {\frac{1}{M}\text{tr}\left( {{{\bm{T}}_{{{\text{dl}},k,n}}}{{\bm{\Psi }}_{{{\text{dl}},n,[\alpha ]}}}} \right)},
\end{align*}
where ${{\bm{\Psi }}_{{{\text{dl}},n,[\alpha ]}}}$ is defined in (\ref{eq:pha_dl_ZF_a}). Similarly, we can obtain
\[
\overline u_{\text{dl},k,[\alpha ]}^{(2)} \triangleq {\lim _{\alpha  \to 0}}\overline u_{\text{dl},k}^{(2)} = \sum\limits_{n = 1}^{{N_\text{D}}} {\frac{{{\varphi _{\text{dl},k,n}}}}{M}\text{tr}\left( {{{\bm{T}}_{{{\text{dl}},k,n}}}{{\bm{\Psi }}_{{{\text{dl}},n,[\alpha ]}}}} \right)}.
\]

In addition, we define $\overline{\dot{u}}_{\text{dl},k,[\alpha ]}^{(1)} \triangleq {\lim _{\alpha  \to 0}}\alpha \overline{\dot{u}}_{\text{dl},k}^{(1)}$, and
\begin{align*}
\overline{\dot{u}}_{\text{dl},k,[\alpha ]}^{(1)} &= {\lim _{\alpha  \to 0}}\alpha \sum\limits_{n = 1}^{{N_\text{D}}} {\frac{1}{M}{\text{tr}}\left[ {{{\bm{T}}_{{{\text{dl}},k,n}}}{\bm{\Psi }}_{{{\text{dl}},n}}^{}\left( {\frac{1}{M}\sum\limits_{j = 1}^{{K_\text{D}}} {\frac{{{{\bm{T}}_{{{\text{dl}},j,n}}}{\alpha ^2}\sum\limits_{m = 1}^{{N_\text{D}}} {\dot e_{\text{dl},j,m}^{}} }}{{{{\left( {\alpha  + \alpha \sum\limits_{m = 1}^{{N_\text{D}}} {{e_{\text{dl},j,m}}} } \right)}^2}}} + {{\bm{I}}_{M}}} } \right){\bm{\Psi }}_{{{\text{dl}},n}}^{}} \right]} \\
&= \sum\limits_{n = 1}^{{N_\text{D}}} {\frac{1}{M}{\text{tr}}\left[ {{{\bm{T}}_{{{\text{dl}},k,n}}}{{\bm{\Psi }}_{{{\text{dl}},n,[\alpha ]}}}\left( {\frac{1}{M}\sum\limits_{j = 1}^{{K_\text{D}}} {\frac{{{{\bm{T}}_{{{\text{dl}},j,n}}}}}{{\sum\limits_{m = 1}^{{N_\text{D}}} {{e_{\text{dl},j,m,[\alpha ]}}} }} + {\alpha _0}{{\bm{I}}_{M}}} } \right){{\bm{\Psi }}_{{{\text{dl}},n,[\alpha ]}}}} \right]} \\
&= \sum\limits_{n = 1}^{{N_\text{D}}} {\frac{1}{M}\left[ {\text{tr}\left( {{\alpha _0}{{\bm{T}}_{{{\text{dl}},k,n}}}{\bm{\Psi }}_{{{\text{dl}},n,[\alpha ]}}^2} \right) - \sum\limits_{j = 1}^{{K_\text{D}}} {{{\underline{\dot c} }_{\text{dl},j}}} \text{tr}\left( {{{\bm{T}}_{{{\text{dl}},k,n}}}{{\bm{\Psi }}_{{{\text{dl}},n,[\alpha ]}}}{{\bm{T}}_{{{\text{dl}},j,n}}}{{\bm{\Psi }}_{{{\text{dl}},n,[\alpha ]}}}} \right)} \right]},
\end{align*}
where ${\alpha _0} = 0$, ${\underline{\dot c} _{\text{dl},j}}$ is given by (\ref{eq:c_ZF_under_dot}) and the second equality follows by
\begin{equation}
\label{a2_b6}
{\lim _{\alpha  \to 0}}{\alpha ^2}\dot e_{\text{dl},j,m}^{} = {\underline{e} _{\text{dl},j,m}}.
\end{equation}
In a similar way, we have
\[
\overline{\dot{u}}_{\text{dl},k,[\alpha ]}^{(2)} \triangleq {\lim _{\alpha  \to 0}}\alpha \overline{\dot{u}}_{\text{dl},k}^{(2)} = \sum\limits_{n = 1}^{{N_\text{D}}} {\frac{{{\varphi _{\text{dl},k,n}}}}{M}\left[ {\text{tr}\left( {{\alpha _0}{{\bm{T}}_{{{\text{dl}},k,n}}}{\bm{\Psi }}_{{{\text{dl}},n,[\alpha ]}}^2} \right) - \sum\limits_{j = 1}^{{K_\text{D}}} {{{\underline{\dot c} }_{\text{dl},j}}} \text{tr}\left( {{{\bm{T}}_{{{\text{dl}},k,n}}}{{\bm{\Psi }}_{{{\text{dl}},n,[\alpha ]}}}{{\bm{T}}_{{{\text{dl}},j,n}}}{{\bm{\Psi }}_{{{\text{dl}},n,[\alpha ]}}}} \right)} \right]}.
\]

Now we derive ${\underline{\overline u} _{\text{dl},k}}$ as
\begin{align*}
{{\underline{\overline u} }_{\text{dl},k}} &= {\lim _{\alpha  \to 0}}{{\overline u}_{\text{dl},k}} \\
&= {\lim _{\alpha  \to 0}}\overline u_{\text{dl},k}^{(1)} - \alpha \overline{\dot{u}}_{\text{dl},k}^{(1)} - \frac{{2\alpha \overline u_{\text{dl},k}^{(2)}\left( {\overline u_{\text{dl},k}^{(2)} - \alpha \overline{\dot{u}}_{\text{dl},k}^{(2)}} \right)}}{{\alpha  + \alpha \overline u_{\text{dl},k}^{(1)}}} + \frac{{{{\left( {\alpha \overline u_{\text{dl},k}^{(2)}} \right)}^2}\left( {\overline u_{\text{dl},k}^{(1)} - \alpha \overline{\dot{u}}_{\text{dl},k}^{(1)}} \right)}}{{{{\left( {\alpha  + \alpha \overline u_{\text{dl},k}^{(1)}} \right)}^2}}} \\
&= \overline u_{\text{dl},k,[\alpha ]}^{(1)} - \overline{\dot{u}}_{\text{dl},k,[\alpha ]}^{(1)} - \frac{{2\underline{\overline u} _{\text{dl},k}^{(2)}\left( {\overline u_{\text{dl},k,[\alpha ]}^{(2)} - \overline{\dot{u}}_{\text{dl},k,[\alpha ]}^{(2)}} \right)}}{{\underline{\overline u} _{\text{dl},k}^{(1)}}} + \frac{{{{\left( {\underline{\overline u} _{\text{dl},k}^{(2)}} \right)}^2}\left( {\overline u_{\text{dl},k,[\alpha ]}^{(1)} - \overline{\dot{u}}_{\text{dl},k,[\alpha ]}^{(1)}} \right)}}{{{{\left( {\underline{\overline u} _{\text{dl},k}^{(1)}} \right)}^2}}}.
\end{align*}
This completes the proof.
\end{IEEEproof}

\begin{lemma}\label{lemma:v_dl_ZF}
Letting Assumptions {\ref{asm:g_dl}}-{\ref{asm:e}} hold true and as $\mathcal{N} \to \infty $, we have
\begin{equation}
\label{a2_b7}
{\underline{\overline v} _{\text{dl}}} = \frac{{\left( {\sum\limits_{i = 1}^{{K_\text{U}}} {{p_{{\text{ul}},i}}T_{k,i}^{}}  + \sigma _{\text{dl}}^2} \right)}}{{MP}}\mathop {\max }\limits_n \left[ {\text{tr}{{\bm{\Psi }}_{{{\text{dl}},n,[\alpha ]}}} - \text{tr}\left( {{\alpha _0}{\bm{\Psi }}_{{{\text{dl}},n,[\alpha ]}}^2} \right) + \sum\limits_{j = 1}^{{K_\text{D}}} {{{\underline{\dot c} }_{\text{dl},j}}} \text{tr}\left( {{{\bm{\Psi }}_{{{\text{dl}},n,[\alpha ]}}}{{\bm{T}}_{{{\text{dl}},j,n}}}{{\bm{\Psi }}_{{\text{dl},n,[\alpha ]}}}} \right)} \right].
\end{equation}
\end{lemma}

\begin{IEEEproof}
From (\ref{eq:zf_six_c}), we know
\begin{equation}
\label{a2_b8}
{\overline v_{\text{dl}}} = \frac{{\left( {\sum\limits_{i = 1}^{{K_\text{U}}} {{p_{{\text{ul}},i}}T_{k,i}^{}}  + \sigma _{\text{dl}}^2} \right)}}{{MP}}\mathop {\max }\limits_n \left( {\text{tr}{{\bm{\Psi }}_{{{\text{dl}},n}}} - \alpha \text{tr}{\bm{\dot \Psi }}_{{{\text{dl}},n}}^{}} \right).
\end{equation}
As $\alpha  \to 0$, we have
\begin{eqnarray}
\label{a2_b9}
{\lim _{\alpha  \to 0}}\text{tr}{{\bm{\Psi }}_{{{\text{dl}},n}}}\!\!\! &=&\!\!\! \text{tr}{{\bm{\Psi }}_{{{\text{dl}},n,[\alpha ]}}}, \\
\label{a2_b10}
{\lim _{\alpha  \to 0}}\alpha \text{tr}{{\bm{\dot \Psi }}_{{{\text{dl}},n}}}\!\!\! &=&\!\!\! \text{tr}\left[ {{{\bm{\Psi }}_{{{\text{dl}},n,[\alpha ]}}}\left( {\frac{1}{M}\sum\limits_{j = 1}^{{K_\text{D}}} {\frac{{{{\bm{T}}_{{{\text{dl}},j,n}}}}}{{\sum\limits_{m = 1}^{{N_\text{D}}} {{e_{\text{dl},j,m,[\alpha ]}}} }} + {\alpha _0}{{\bm{I}}_{M}}} } \right){{\bm{\Psi }}_{{{\text{dl}},n,[\alpha ]}}}} \right].
\end{eqnarray}
Substituting (\ref{a2_b9}) and (\ref{a2_b10}) into (\ref{a2_b8}), the result in (\ref{a2_b7}) can be derived.
\end{IEEEproof}

From Lemmas \ref{lemma:signal_dl_ZF}, \ref{lemma:noise_dl_ZF} and \ref{lemma:v_dl_ZF}, we derive the deterministic equivalent ${\overline \gamma _{\text{dl},\text{zf},k}}$, as shown in (\ref{eq:r_dl_zf_DE}), of downlink SINR under ZF precoding ${\gamma _{\text{dl},\text{zf},k}}$. Similar to the proof in Theorem \ref{thm:R_dl_rzf}, we know that when $\mathcal{N} \to \infty $, it satisfies that $\frac{1}{{{K_\text{D}}}}\left( {{R_{\text{dl},\text{zf},\text{sum}}} - {{\overline R}_{\text{dl},\text{zf},\text{sum}}}} \right)\xrightarrow{{a.s.}}0$, where ${\overline R_{\text{dl},\text{zf},\text{sum}}}$ is given by (\ref{eq:R_dl_zf_DE}). This completes the proof.

\setcounter{equation}{0}

\section{Proof of Theorem 3}\label{sec:Appendix_C}
We derive the deterministic equivalent ${{\bm{A}}_{n}}$ first. Since ${{\bm{A}}_{n}}$ is a main diagonal matrix, the $j$th diagonal element is
\begin{align}
\begin{split}
\label{a3_c1}
{a_j} &= \text{E}\left[ {{{\left( {{{\bm{q}}_{{{\text{I}},j,n}}}{{{\hat{\bm h}}}_{{{\text{I}},j}}}{ - }{{\bm{q}}_{{{\text{I}},j,n}}}{{\bm{h}}_{{{\text{I}},j}}}} \right)}^{\text{H}}}\left( {{{\bm{q}}_{{{\text{I}},j,n}}}{{{\hat{\bm h}}}_{{{\text{I}},j}}}{ - }{{\bm{q}}_{{{\text{I}},j,n}}}{{\bm{h}}_{{{\text{I}},j}}}} \right)} \right] \\
&= \text{E}\left\{ {{{\left[ {{{\bm{q}}_{{{\text{I}},j,n}}}\left( {{{\bm{\Lambda }}_{{{\text{I}},j}}}{ - }{{\bm{I}}_{{M}{{N}_{\text{U}}}}}} \right){{\bm{h}}_{{{\text{I}},j}}} + {{\bm{q}}_{{{\text{I}},j,n}}}{{\bm{\Omega }}_{{{\text{I}},j}}}{{\bm{z}}_{{{\text{I}},j}}}} \right]}^{\text{H}}}\left[ {{{\bm{q}}_{{{\text{I}},j,n}}}\left( {{{\bm{\Lambda }}_{{{\text{I}},j}}}{ - }{{\bm{I}}_{{M}{{N}_{\text{U}}}}}} \right){{\bm{h}}_{{{\text{I}},j}}} + {{\bm{q}}_{{{\text{I}},j,n}}}{{\bm{\Omega }}_{{{\text{I}},j}}}{{\bm{z}}_{{{\text{I}},j}}}} \right]} \right\} \\
&= \text{E}\left[ {{\bm{h}}_{{{\text{I}},j}}^{\text{H}}\left( {{{\bm{\Lambda }}_{{{\text{I}},j}}}{ - }{{\bm{I}}_{{M}{{N}_{\text{U}}}}}} \right){\bm{q}}_{{{\text{I}},j,n}}^{\text{H}}{{\bm{q}}_{{{\text{I}},j,n}}}\left( {{{\bm{\Lambda }}_{{{\text{I}},j}}}{ - }{{\bm{I}}_{{M}{{N}_{\text{U}}}}}} \right){{\bm{h}}_{{{\text{I}},j}}} + {\bm{z}}_{{{\text{I}},j}}^{\text{H}}{{\bm{\Omega }}_{{{\text{I}},j}}}{\bm{q}}_{{{\text{I}},j,n}}^{\text{H}}{{\bm{q}}_{{{\text{I}},j,n}}}{{\bm{\Omega }}_{{{\text{I}},j}}}{{\bm{z}}_{{{\text{I}},j}}}} \right] \\
&= \frac{2}{{M{N_\text{U}}}}\text{tr}\left[ {\left( {{{\bm{I}}_{{M}{{N}_{\text{U}}}}}{ - }{{\bm{\Lambda }}_{{{\text{I}},j}}}} \right){\bm{q}}_{{{\text{I}},j,n}}^{\text{H}}{{\bm{q}}_{{{\text{I}},j,n}}}} \right],
\end{split}
\end{align}
where ${{\bm{Q}}_{{{\text{I}},j}}} \triangleq {\bm{T}}_{{{\text{I}},j}}^{\frac{{1}}{{2}}}$, ${{\bm{q}}_{{{\text{I}},j,n}}}$ is the $n$th row of ${{\bm{Q}}_{{{\text{I}},j}}}$. The second equality follows from (\ref{eq:es_RAUI_channel}) and the fourth equality follows from Lemma \ref{same_matrix}.

We proceed to derive the deterministic equivalent ${\bm{w}}_{{{\text{rzf}},k}}^{\text{H}}{{\bm{A}}_{n}}{{\bm{w}}_{{{\text{rzf}},k}}}$. According to the definition of RZF precoding vector in (\ref{eq:RZF_vector}), we know that
\[
{\bm{w}}_{{{\text{rzf}},k}}^{\text{H}}{{\bm{A}}_{n}}{{\bm{w}}_{{{\text{rzf}},k}}} = {\xi ^2}{\hat{\bm g}}_{{{\text{dl}},k}}^{\text{H}}{\bm{C}}_{{\text{dl}}}^{{\text{-1}}}{{\bm{A}}_{n}}{\bm{C}}_{{\text{dl}}}^{{\text{-1}}}{{\hat{\bm g}}_{{{\text{dl}},k}}}.
\]
Applying Lemma \ref{matrix_inverse}, we have
\begin{equation}
\label{a3_c2}
{\bm{w}}_{{{\text{rzf}},k}}^{\text{H}}{{\bm{A}}_{n}}{{\bm{w}}_{{{\text{rzf}},k}}} = \frac{{{\xi ^2}}}{{{{\left( {1 + {\hat{\bm h}}_{{{\text{dl}},k}}^{\text{H}}{\bm{T}}_{{{\text{dl}},k}}^{\frac{{1}}{{2}}}{\bm{C}}_{{{\text{dl}},[k]}}^{{\text{-1}}}{\bm{T}}_{{{\text{dl}},k}}^{\frac{{1}}{{2}}}{{{\hat{\bm h}}}_{{{\text{dl}},k}}}} \right)}^{\text{2}}}}}{\hat{\bm h}}_{{{\text{dl}},k}}^{\text{H}}{\bm{T}}_{{{\text{dl}},k}}^{\frac{{1}}{{2}}}{\bm{C}}_{{{\text{dl}},[k]}}^{{\text{-1}}}{{\bm{A}}_{n}}{\bm{C}}_{{{\text{dl}},[k]}}^{{\text{-1}}}{\bm{T}}_{{{\text{dl}},k}}^{\frac{{1}}{{2}}}{{\hat{\bm h}}_{{{\text{dl}},k}}}.
\end{equation}
From (\ref{eq:xi_i_2}), (\ref{eq:v_dl}) and (\ref{eq:v_dl_DE}), we can obtain the deterministic equivalent ${\xi ^2}$, which is
\begin{equation}
\label{a3_c3}
{\overline \xi ^2} = \mathop {\min }\limits_i \frac{{MP}}{{\text{tr}{{\bm{\Psi }}_{{{\text{dl}},i}}} - \alpha \text{tr}{\bm{\Psi }}_{{{\text{dl}},i}}^{\text{2}} + \alpha \sum\limits_{j = 1}^{{K_{\text{D}}}} {{{\dot c}_{{\text{dl}},j,i}}} \text{tr}\left( {{{\bm{\Psi }}_{{{\text{dl}},i}}}{{\bm{T}}_{{{\text{dl}},j,i}}}{{\bm{\Psi }}_{{{\text{dl}},i}}}} \right)}},
\end{equation}
where ${{\bm{\Psi }}_{{{\text{dl}},i}}}$ and ${\dot c_{{\text{dl}},j,i}}$ are given by (\ref{eq:pha_dl}) and (\ref{eq:li_eq_Crzf}), respectively.

In the proof of Theorem \ref{thm:R_dl_rzf}, the deterministic equivalent of ${\hat{\bm h}}_{{{\text{dl}},k}}^{\text{H}}{\bm{T}}_{{{\text{dl}},k}}^{\frac{{1}}{{2}}}{\bm{C}}_{{{\text{dl}},[k]}}^{{\text{-1}}}{\bm{T}}_{{{\text{dl}},k}}^{\frac{{1}}{{2}}}{{\hat{\bm h}}_{{{\text{dl}},k}}}$ is $\overline u_{{\text{dl}},k}^{(1)}$ in (\ref{eq:u_dl_k_1}). Furthermore, similar to the derivations of $\overline{\dot{u}}_{{\text{dl}},k}^{(1)}$, we derive
\begin{equation}
\label{a3_c4}
{\hat{\bm h}}_{{{\text{dl}},k}}^{\text{H}}{\bm{T}}_{{{\text{dl}},k}}^{\frac{{1}}{{2}}}{\bm{C}}_{{{\text{dl}},[k]}}^{{\text{-1}}}{{\bm{A}}_{n}}{\bm{C}}_{{{\text{dl}},[k]}}^{{\text{-1}}}{\bm{T}}_{{{\text{dl}},k}}^{\frac{{1}}{{2}}}{{\hat{\bm h}}_{{{\text{dl}},k}}} - \overline{\dot{u}}_{{\text{dl}},k,{\text{A}}}^{(1)}\xrightarrow{{a.s.}}0,
\end{equation}
where
\begin{equation}
\label{a3_c5}
\overline{\dot{u}}_{{\text{dl}},k,{\text{A}}}^{(1)} = \sum\limits_{m = 1}^{{N_{\text{D}}}} {\frac{1}{M}\left[ {\text{tr}\left( {{{\bm{T}}_{{{\text{dl}},k,m}}}{{\bm{\Psi }}_{{{\text{dl}},m}}}{{\bm{A}}_{{n,m}}}{{\bm{\Psi }}_{{{\text{dl}},m}}}} \right) - \sum\limits_{j = 1}^{{K_{\text{D}}}} {\dot c_{{\text{dl}},j,m}^{\text{(A)}}\text{tr}\left( {{{\bm{T}}_{{{\text{dl}},k,m}}}{{\bm{\Psi }}_{{{\text{dl}},m}}}{{\bm{T}}_{{{\text{dl}},j,m}}}{{\bm{\Psi }}_{{{\text{dl}},m}}}} \right)} } \right]}.
\end{equation}
${\dot{\bm C}}_{{\text{dl}}}^{{\text{(A)}}} = \left[ {\dot c_{{\text{dl}},k,i}^{\text{(A)}}} \right] \in {\mathbb{C}^{{K_{\text{D}}} \times {N_{\text{D}}}}}$ is a solution to the following linear equation
\[
{{\bm{\Theta }}_{{\text{dl}}}}\text{vec}\left( {{\dot{\bm C}}_{{\text{dl}}}^{{\text{(A)}}}} \right) = \text{vec}\left( {{\bm{\Gamma }}_{{\text{dl}}}^{{\text{(A)}}}} \right),
\]
where ${{\bm{\Theta }}_{{\text{dl}}}}$ is given by (\ref{eq:theta_dl}), ${\bm{\Gamma }}_{{\text{dl}}}^{{\text{(A)}}} \in {\mathbb{C}^{{K_{\text{D}}} \times {N_{\text{D}}}}}$ takes the form
\[
{\left[ {{\bm{\Gamma }}_{{\text{dl}}}^{{\text{(A)}}}} \right]_{k,i}} =  - \frac{1}{M}\sum\limits_{j = 1}^{{N_{\text{D}}}} {\frac{1}{M}\frac{1}{{{{\left( {1 + \sum\limits_{m = 1}^{{N_{\text{D}}}} {{e_{{\text{dl}},k,m}}} } \right)}^2}}}} {\text{tr}}\left( {{{\bm{T}}_{{{\text{dl}},k,j}}}{\bm{\Psi }}_{{{\text{dl}},j}}^{}{{\bm{A}}_{{n,j}}}{{\bm{\Psi }}_{{{\text{dl}},j}}}} \right).
\]
Plugging (\ref{a3_c3}), (\ref{eq:u_dl_k_1}) and (\ref{a3_c4}) into (\ref{a3_c2}), we have the deterministic equivalent of ${\bm{w}}_{{{\text{rzf}},k}}^{\text{H}}{{\bm{A}}_{n}}{{\bm{w}}_{{{\text{rzf}},k}}}$ as follows
\begin{equation}
\label{a3_c6}
{\bm{w}}_{{{\text{rzf}},k}}^{\text{H}}{{\bm{A}}_{n}}{{\bm{w}}_{{{\text{rzf}},k}}} - \frac{{{{\overline \xi }^2}\overline{\dot{u}}_{{\text{dl}},k,{\text{A}}}^{(1)}}}{{{{\left( {1 + \overline u_{{\text{dl}},k}^{(1)}} \right)}^2}}}\xrightarrow{{a.s.}}0.
\end{equation}

Substituting (\ref{a3_c6}) into (\ref{eq:UL_delta_n}), we obtain the deterministic equivalent for the $n$th diagonal element of ${\bm{\Sigma }}$ given by (\ref{eq:UL_delta_n_DE}). Therefore, the deterministic equivalents for the  covariance matrix of the residual interference and noise is derived in (\ref{eq:sigma_DE}). This completes the proof.

\setcounter{equation}{0}

\section{Proof of Theorem 4}\label{sec:Appendix_D}
This appendix aims to derive a deterministic equivalent of the sum-rate with MMSE receiver. To begin with, the uplink SINR ${\gamma _{{\text{ul}},k}}$ in (\ref{eq:r_ul_k}) consists of three terms: 1) ${{p_{{\text{ul}},k}}{{\left| {{\hat{\bm g}}_{{{\text{ul}},k}}^{\text{H}}{\bm{C}}_{{\text{ul}}}^{{\text{-1}}}{{\bm{g}}_{{{\text{ul}},k}}}} \right|}^2}}$; 2)$\sum\limits_{i = 1,i \ne k}^{{K_{\text{U}}}} {{p_{{\text{ul}},i}}\left| {{\hat{\bm g}}_{{{\text{ul}},k}}^{\text{H}}{\bm{C}}_{{\text{ul}}}^{{\text{-1}}}} \right.} $ ${\left. {{{\bm{g}}_{{{\text{ul}},i}}}} \right|^2}$; 3) ${\hat{\bm g}}_{{{\text{ul}},k}}^{\text{H}}{\bm{C}}_{{\text{ul}}}^{{\text{-1}}}{\bm{\Sigma C}}_{{\text{ul}}}^{{\text{-1}}}{{\hat{\bm g}}_{{{\text{ul}},k}}}$. We will subsequently derive the deterministic equivalents for each of these three terms, which together constitute the final expression for ${\overline \gamma _{{\text{ul}},k}}$.

\begin{lemma}\label{lemma:UL_signal}
Letting Assumption \ref{asm:g_ul} hold true and as $\mathcal{N} \to \infty $, we have
\begin{equation}
\label{a4_d1}
{\hat{\bm g}}_{{{\text{ul}},k}}^{\text{H}}{\bm{C}}_{{\text{ul}}}^{{\text{-1}}}{{\bm{g}}_{{{\text{ul}},k}}} - \frac{{\overline u_{{\text{ul}},k}^{(2)}}}{{1 + {p_{{\text{ul}},k}}\overline u_{{\text{ul}},k}^{(1)}}}\xrightarrow{{a.s.}}0,
\end{equation}
where $\overline u_{{\text{ul}},k}^{(1)}$ and $\overline u_{{\text{ul}},k}^{(2)}$ are expressed as (\ref{eq:u_ul_k_1}) and (\ref{eq:u_ul_k_2}), respectively.
\end{lemma}

\begin{IEEEproof}
For signal power ${\left| {{\hat{\bm g}}_{{{\text{ul}},k}}^{\text{H}}{\bm{C}}_{{\text{ul}}}^{{\text{-1}}}{{\bm{g}}_{{{\text{ul}},k}}}} \right|^2}$, we have
\begin{align}
\begin{split}
\label{a4_d2}
{\bm{g}}_{{{\text{ul}},k}}^{\text{H}}{\bm{C}}_{{\text{ul}}}^{{\text{-1}}}{{{\hat{\bm g}}}_{{{\text{ul}},k}}} &= \frac{1}{{1 + {p_{{\text{ul}},k}}{\hat{\bm g}}_{{{\text{ul}},k}}^{\text{H}}{\bm{C}}_{{{\text{ul}},[k]}}^{{\text{-1}}}{{{\hat{\bm g}}}_{{{\text{ul}},k}}}}}{\bm{g}}_{{{\text{ul}},k}}^{\text{H}}{\bm{C}}_{{{\text{ul}},[k]}}^{{\text{-1}}}{{{\hat{\bm g}}}_{{{\text{ul}},k}}} \\
&= \frac{1}{{1 + {p_{{\text{ul}},k}}{\hat{\bm h}}_{{{\text{ul}},k}}^{\text{H}}{\bm{T}}_{{{\text{ul}},k}}^{\frac{{1}}{{2}}}{\bm{C}}_{{{\text{ul}},[k]}}^{{\text{-1}}}{\bm{T}}_{{{\text{ul}},k}}^{\frac{{1}}{{2}}}{{{\hat{\bm h}}}_{{{\text{ul}},k}}}}}{\bm{h}}_{{{\text{ul}},k}}^{\text{H}}{\bm{T}}_{{{\text{ul}},k}}^{\frac{{1}}{{2}}}{\bm{C}}_{{{\text{ul}},[k]}}^{{\text{-1}}}{\bm{T}}_{{{\text{ul}},k}}^{\frac{{1}}{{2}}}{{\bm{\Lambda }}_{{{\text{ul}},k}}}{{\bm{h}}_{{{\text{ul}},k}}} \\
& \quad + \frac{1}{{1 + {p_{{\text{ul}},k}}{\hat{\bm h}}_{{{\text{ul}},k}}^{\text{H}}{\bm{T}}_{{{\text{ul}},k}}^{\frac{{1}}{{2}}}{\bm{C}}_{{{\text{ul}},[k]}}^{{\text{-1}}}{\bm{T}}_{{{\text{ul}},k}}^{\frac{{1}}{{2}}}{{{\hat{\bm h}}}_{{{\text{ul}},k}}}}}{\bm{h}}_{{{\text{ul}},k}}^{\text{H}}{\bm{T}}_{{{\text{ul}},k}}^{\frac{{1}}{{
2}}}{\bm{C}}_{{{\text{ul}},[k]}}^{{\text{-1}}}{\bm{T}}_{{{\text{ul}},k}}^{\frac{{1}}{{2}}}{{\bm{\Omega }}_{{{\text{ul}},k}}}{{\bm{z}}_{{{\text{ulp}},k}}},
\end{split}
\end{align}
where we define ${{\bm{C}}_{{{\text{ul}},[k]}}} \triangleq \sum\limits_{i = 1,i \ne k}^{{K_{\text{U}}}} {{p_{{\text{ul}},i}}{{{\hat{\bm g}}}_{{{\text{ul}},i}}}{\hat{\bm g}}_{{{\text{ul}},i}}^{\text{H}}}  + {\bm{\Sigma }}$. The first equality follows from Lemma \ref{matrix_inverse}.  Since ${{\bm{h}}_{{{\text{ul}},k}}}$ and ${{\bm{z}}_{{{\text{ulp}},k}}}$ are independent, from Lemma \ref{diff_matrix} we know that ${\bm{h}}_{{{\text{ul}},k}}^{\text{H}}{\bm{T}}_{{{\text{ul}},k}}^{\frac{{1}}{{2}}}{\bm{C}}_{{{\text{ul}},[k]}}^{{\text{-1}}}{\bm{T}}_{{{\text{ul}},k}}^{\frac{{1}}{{2}}}{{\bm{\Omega }}_{{{\text{ul}},k}}}{{\bm{z}}_{{{\text{ulp}},k}}}$ almost surely converges to zero. Besides, applying Lemma \ref{same_matrix}, ${\hat{\bm h}}_{{{\text{ul}},k}}^{\text{H}}{\bm{T}}_{{{\text{ul}},k}}^{\frac{{1}}{{2}}}{\bm{C}}_{{{\text{ul}},[k]}}^{{\text{-1}}}{\bm{T}}_{{{\text{ul}},k}}^{\frac{{1}}{{2}}}{{\hat{\bm h}}_{{{\text{ul}},k}}}$ almost surely converges to $u_{{\text{ul}},k}^{(1)} = \frac{1}{{M{N_{\text{U}}}}}\text{tr}\left( {{{\bm{T}}_{{{\text{ul}},k}}}{\bm{C}}_{{{\text{ul}},[k]}}^{{\text{-1}}}} \right)$ and ${\bm{h}}_{{{\text{ul}},k}}^{\text{H}}{\bm{T}}_{{{\text{ul}},k}}^{\frac{{1}}{{2}}}{\bm{C}}_{{{\text{ul}},[k]}}^{{\text{-1}}}{\bm{T}}_{{{\text{ul}},k}}^{\frac{{1}}{{2}}}{{\bm{\Lambda }}_{{{\text{ul}},k}}}{{\bm{h}}_{{{\text{ul}},k}}}$ almost surely converges to $u_{{\text{ul}},k}^{(2)} = \frac{1}{{M{N_{\text{U}}}}}\text{tr}\left( {{\bm{T}}_{{{\text{ul}},k}}^{\frac{{1}}{{2}}}} {{{\bm{\Lambda }}_{{{\text{ul}},k}}}{\bm{T}}_{{{\text{ul}},k}}^{\frac{{1}}{{2}}}{\bm{C}}_{{{\text{ul}},[k]}}^{{\text{-1}}}} \right)$.

Rewrite ${\bm{C}}_{{\text{ul}}}^{{\text{-1}}}$ as
\begin{align*}
{\bm{C}}_{{\text{ul}}}^{{\text{-1}}} &= {\left( {\sum\limits_{i = 1}^{{K_{\text{U}}}} {{p_{{\text{ul}},i}}{{{\hat{\bm g}}}_{{{\text{ul}},i}}}{\hat{\bm g}}_{{{\text{ul}},i}}^{\text{H}}}  + {\bm{\Sigma }}} \right)^{ - 1}}{\bm{\Sigma }}{{\bm{\Sigma }}^{{\text{-1}}}} \\
&= {\left( {{{\bm{\Sigma }}^{{\text{-1}}}}\sum\limits_{i = 1}^{{K_{\text{U}}}} {{p_{{\text{ul}},i}}{{{\hat{\bm g}}}_{{{\text{ul}},i}}}{\hat{\bm g}}_{{{\text{ul}},i}}^{\text{H}}}  + {{\bm{I}}_{{M}{{N}_{\text{U}}}}}} \right)^{ - 1}}{{\bm{\Sigma }}^{{\text{-1}}}} \\
&= {\bm{C}}_{{{\text{ul}},\Sigma }}^{{\text{-1}}}{{\bm{\Sigma }}^{{\text{-1}}}},
\end{align*}
where ${\bm{C}}_{{{\text{ul}},\Sigma }}^{} \triangleq {{\bm{\Sigma }}^{{\text{-1}}}}\sum\limits_{i = 1}^{{K_{\text{U}}}} {{p_{{\text{ul}},i}}{{{\hat{\bm g}}}_{{{\text{ul}},i}}}{\hat{\bm g}}_{{{\text{ul}},i}}^{\text{H}}} + {{\bm{I}}_{{M}{{N}_{\text{U}}}}}$. Then, applying Lemma \ref{plus} and \cite[Theorem 2]{jzhang_large_2013}, we have
\begin{equation}
\label{a4_d3}
u_{{\text{ul}},k}^{(1)} - \overline u_{{\text{ul}},k}^{(1)}\xrightarrow{{a.s.}}0, u_{{\text{ul}},k}^{(2)} - \overline u_{{\text{ul}},k}^{(2)}\xrightarrow{{a.s.}}0,
\end{equation}
where $\overline u_{{\text{ul}},k}^{(1)}$ and $\overline u_{{\text{ul}},k}^{(2)}$ are expressed as (\ref{eq:u_ul_k_1}) and (\ref{eq:u_ul_k_2}), respectively. From (\ref{a4_d2}) and (\ref{a4_d3}), we derive (\ref{a4_d1}).
\end{IEEEproof}

\begin{lemma}\label{lemma:UL_interference}
Letting Assumption \ref{asm:g_ul} hold true and as $\mathcal{N} \to \infty $, we have
\begin{equation}
\label{a4_d4}
{\left| {{\hat{\bm g}}_{{{\text{ul}},k}}^{\text{H}}{\bm{C}}_{{\text{ul}}}^{{\text{-1}}}{{\bm{g}}_{{{\text{ul}},i}}}} \right|^2} - {\overline u_{{\text{ul}},i}}\xrightarrow{{a.s.}}0,
\end{equation}
where ${\overline u_{{\text{ul}},i}}$ is given by (\ref{eq:u_ul_i_DE}).
\end{lemma}

\begin{IEEEproof}
We rewrite ${\left| {{\hat{\bm g}}_{{{\text{ul}},k}}^{\text{H}}{\bm{C}}_{{\text{ul}}}^{{\text{-1}}}{{\bm{g}}_{{{\text{ul}},i}}}} \right|^2}$ as
\begin{align}
\begin{split}
\label{a4_d5}
{\left| {{\hat{\bm g}}_{{{\text{ul}},k}}^{\text{H}}{\bm{C}}_{{\text{ul}}}^{{\text{-1}}}{{\bm{g}}_{{{\text{ul}},i}}}} \right|^2} &= {\hat{\bm g}}_{{{\text{ul}},k}}^{\text{H}}{\bm{C}}_{{\text{ul}}}^{{\text{-1}}}{{\bm{g}}_{{{\text{ul}},i}}}{\bm{g}}_{{{\text{ul}},i}}^{\text{H}}{\bm{C}}_{{\text{ul}}}^{{\text{-1}}}{{{\hat{\bm g}}}_{{{\text{ul}},k}}} \\
&= {\hat{\bm g}}_{{{\text{ul}},k}}^{\text{H}}{\bm{C}}_{{{\text{ul}},[k]}}^{{\text{-1}}}{{\bm{g}}_{{{\text{ul}},i}}}{\bm{g}}_{{{\text{ul}},i}}^{\text{H}}{\bm{C}}_{{\text{ul}}}^{{\text{-1}}}{{{\hat{\bm g}}}_{{{\text{ul}},k}}} + {\hat{\bm g}}_{{{\text{ul}},k}}^{\text{H}}\left( {{\bm{C}}_{{\text{ul}}}^{{\text{-1}}} - {\bm{C}}_{{{\text{ul}},[k]}}^{{\text{-1}}}} \right){{\bm{g}}_{{{\text{ul}},i}}}{\bm{g}}_{{{\text{ul}},i}}^{\text{H}}{\bm{C}}_{{\text{ul}}}^{{\text{-1}}}{{{\hat{\bm g}}}_{{{\text{ul}},k}}}. \\
\end{split}
\end{align}
Utilizing \cite[Lemma 7]{swagner_large_2012} and Lemma \ref{plus}, ${\hat{\bm g}}_{{{\text{ul}},k}}^{\text{H}}{\bm{C}}_{{{\text{ul}},[k]}}^{{\text{-1}}}{{\bm{g}}_{{{\text{ul}},i}}}{\bm{g}}_{{{\text{ul}},i}}^{\text{H}}{\bm{C}}_{{\text{ul}}}^{{\text{-1}}}{{\hat{\bm g}}_{{{\text{ul}},k}}}$ almost surely converges to
\begin{equation}
\label{a4_d6}
{\hat{\bm g}}_{{{\text{ul}},k}}^{\text{H}}{\bm{C}}_{{{\text{ul}},[k]}}^{{\text{-1}}}{{\bm{g}}_{{{\text{ul}},i}}}{\bm{g}}_{{{\text{ul}},i}}^{\text{H}}{\bm{C}}_{{\text{ul}}}^{{\text{-1}}}{{\hat{\bm g}}_{{{\text{ul}},k}}} - \frac{1}{{M{N_{\text{U}}}}}\frac{1}{{1 + {p_{{\text{ul}},k}}\overline u_{{\text{ul}},k}^{(1)}}}{\bm{g}}_{{{\text{ul}},i}}^{\text{H}}{\bm{C}}_{{\text{ul}}}^{{\text{-1}}}{{\bm{T}}_{{{\text{ul}},k}}}{\bm{C}}_{{\text{ul}}}^{{\text{-1}}}{{\bm{g}}_{{{\text{ul}},i}}}\xrightarrow{{a.s.}}0.
\end{equation}
Besides, applying Lemma \ref{resolvent_identity}, we have
\begin{equation}
\label{a4_d7}
{\hat{\bm g}}_{{{\text{ul}},k}}^{\text{H}}\left( {{\bm{C}}_{{\text{ul}}}^{{\text{-1}}} - {\bm{C}}_{{{\text{ul}},[k]}}^{{\text{-1}}}} \right){{\bm{g}}_{{{\text{ul}},i}}}{\bm{g}}_{{{\text{ul}},i}}^{\text{H}}{\bm{C}}_{{\text{ul}}}^{{\text{-1}}}{{\hat{\bm g}}_{{{\text{ul}},k}}} =  - {p_{{\text{ul}},k}}{\hat{\bm g}}_{{{\text{ul}},k}}^{\text{H}}{\bm{C}}_{{\text{ul}}}^{{\text{-1}}}{{\hat{\bm g}}_{{{\text{ul}},k}}}{\hat{\bm g}}_{{{\text{ul}},k}}^{\text{H}}{\bm{C}}_{{{\text{ul}},[k]}}^{{\text{-1}}}{{\bm{g}}_{{{\text{ul}},i}}}{\bm{g}}_{{{\text{ul}},i}}^{\text{H}}{\bm{C}}_{{\text{ul}}}^{{\text{-1}}}{{\hat{\bm g}}_{{{\text{ul}},k}}}.
\end{equation}
According to Lemma \ref{matrix_inverse}, we can obtain
\[
{\hat{\bm g}}_{{{\text{ul}},k}}^{\text{H}}{\bm{C}}_{{\text{ul}}}^{{\text{-1}}}{{\hat{\bm g}}_{{{\text{ul}},k}}}{ = }\frac{1}{{1 + {p_{{\text{ul}},k}}{\hat{\bm g}}_{{{\text{ul}},k}}^{\text{H}}{\bm{C}}_{{{\text{ul}},[k]}}^{{\text{-1}}}{{{\hat{\bm g}}}_{{{\text{ul}},k}}}}}{\hat{\bm g}}_{{{\text{ul}},k}}^{\text{H}}{\bm{C}}_{{{\text{ul}},[k]}}^{{\text{-1}}}{{\hat{\bm g}}_{{{\text{ul}},k}}},
\]
which can be used together with (\ref{a4_d3}) to determine the approximation of ${\hat{\bm g}}_{{{\text{ul}},k}}^{\text{H}}{\bm{C}}_{{\text{ul}}}^{{\text{-1}}}{{\hat{\bm g}}_{{{\text{ul}},k}}}$
\begin{equation}
\label{a4_d8}
{\hat{\bm g}}_{{{\text{ul}},k}}^{\text{H}}{\bm{C}}_{{\text{ul}}}^{{\text{-1}}}{{\hat{\bm g}}_{{{\text{ul}},k}}} - \frac{{\overline u_{{\text{ul}},k}^{(1)}}}{{1 + {p_{{\text{ul}},k}}\overline u_{{\text{ul}},k}^{(1)}}}\xrightarrow{{a.s.}}0.
\end{equation}
Substituting (\ref{a4_d6}) and (\ref{a4_d8}) into (\ref{a4_d7}), and then plugging (\ref{a4_d6}) and (\ref{a4_d7}) into (\ref{a4_d5}), ${\left| {{\hat{\bm g}}_{{{\text{ul}},k}}^{\text{H}}{\bm{C}}_{{\text{ul}}}^{{\text{-1}}}{{\bm{g}}_{{{\text{ul}},i}}}} \right|^2}$ almost surely converges to
\begin{equation}
\label{a4_d9}
{\left| {{\hat{\bm g}}_{{{\text{ul}},k}}^{\text{H}}{\bm{C}}_{{\text{ul}}}^{{\text{-1}}}{{\bm{g}}_{{{\text{ul}},i}}}} \right|^2} - \frac{1}{{M{N_{\text{U}}}}}\frac{1}{{{{\left( {1 + {p_{{\text{ul}},k}}\overline u_{{\text{ul}},k}^{(1)}} \right)}^2}}}{\bm{g}}_{{{\text{ul}},i}}^{\text{H}}{\bm{C}}_{{\text{ul}}}^{{\text{-1}}}{{\bm{T}}_{{{\text{ul}},k}}}{\bm{C}}_{{\text{ul}}}^{{\text{-1}}}{{\bm{g}}_{{{\text{ul}},i}}}\xrightarrow{{a.s.}}0.
\end{equation}

Next, we will derive the deterministic equivalent ${\bm{g}}_{{{\text{ul}},i}}^{\text{H}}{\bm{C}}_{{\text{ul}}}^{{\text{-1}}}{{\bm{T}}_{{{\text{ul}},k}}}{\bm{C}}_{{\text{ul}}}^{{\text{-1}}}{{\bm{g}}_{{{\text{ul}},i}}}$. It can be decomposed as
\begin{align}
\begin{split}
\label{a4_d10}
& {\bm{g}}_{{{\text{ul}},i}}^{\text{H}}{\bm{C}}_{{\text{ul}}}^{{\text{-1}}}{{\bm{T}}_{{{\text{ul}},k}}}{\bm{C}}_{{\text{ul}}}^{{\text{-1}}}{{\bm{g}}_{{{\text{ul}},i}}} \\
&= {\bm{g}}_{{{\text{ul}},i}}^{\text{H}}{\bm{C}}_{{{\text{ul}},[i]}}^{{\text{-1}}}{{\bm{T}}_{{{\text{ul}},k}}}{\bm{C}}_{{\text{ul}}}^{{\text{-1}}}{{\bm{g}}_{{{\text{ul}},i}}} + {\bm{g}}_{{{\text{ul}},i}}^{\text{H}}\left( {{\bm{C}}_{{\text{ul}}}^{{\text{-1}}} - {\bm{C}}_{{{\text{ul}},[i]}}^{{\text{-1}}}} \right){{\bm{T}}_{{{\text{ul}},k}}}{\bm{C}}_{{\text{ul}}}^{{\text{-1}}}{{\bm{g}}_{{{\text{ul}},i}}} \\
&= {\bm{h}}_{{{\text{ul}},i}}^{\text{H}}{\bm{T}}_{{{\text{ul}},i}}^{\frac{{1}}{{2}}}{\bm{C}}_{{{\text{ul}},[i]}}^{{\text{-1}}}{{\bm{T}}_{{{\text{ul}},k}}}{\bm{C}}_{{\text{ul}}}^{{\text{-1}}}{\bm{T}}_{{{\text{ul}},i}}^{\frac{{1}}{{2}}}{{\bm{h}}_{{{\text{ul}},i}}} - {\bm{h}}_{{{\text{ul}},i}}^{\text{H}}{\bm{T}}_{{{\text{ul}},i}}^{\frac{{1}}{{2}}}{\bm{C}}_{{\text{ul}}}^{{\text{-1}}}\left( {{{\bm{C}}_{{\text{ul}}}} - {{\bm{C}}_{{{\text{ul}},[i]}}}} \right){\bm{C}}_{{{\text{ul}},[i]}}^{{\text{-1}}}{{\bm{T}}_{{{\text{ul}},k}}}{\bm{C}}_{{\text{ul}}}^{{\text{-1}}}{\bm{T}}_{{{\text{ul}},i}}^{\frac{{1}}{{2}}}{{\bm{h}}_{{{\text{ul}},i}}} \\
&= {\bm{h}}_{{{\text{ul}},i}}^{\text{H}}{\bm{T}}_{{{\text{ul}},i}}^{\frac{{1}}{{2}}}{\bm{C}}_{{{\text{ul}},[i]}}^{{\text{-1}}}{{\bm{T}}_{{{\text{ul}},k}}}{\bm{C}}_{{\text{ul}}}^{{\text{-1}}}{\bm{T}}_{{{\text{ul}},i}}^{\frac{{1}}{{2}}}{{\bm{h}}_{{{\text{ul}},i}}} - {p_{{\text{ul}},i}}{\bm{h}}_{{{\text{ul}},i}}^{\text{H}}{\bm{T}}_{{{\text{ul}},i}}^{\frac{{1}}{{2}}}{\bm{C}}_{{\text{ul}}}^{{\text{-1}}}{{{\hat{\bm h}}}_{{{\text{ul}},i}}}{\hat{\bm h}}_{{{\text{ul}},i}}^{\text{H}}{\bm{C}}_{{{\text{ul}},[i]}}^{{\text{-1}}}{{\bm{T}}_{{{\text{ul}},k}}}{\bm{C}}_{{\text{ul}}}^{{\text{-1}}}{\bm{T}}_{{{\text{ul}},i}}^{\frac{{1}}{{2}}}{{\bm{h}}_{{{\text{ul}},i}}},
\end{split}
\end{align}
where
\begin{equation}
\label{a4_d11}
{{\hat{\bm h}}_{{{\text{ul}},i}}}{\hat{\bm h}}_{{{\text{ul}},i}}^{\text{H}} = {\bm{T}}_{{{\text{ul}},i}}^{\frac{{1}}{{2}}}\left( {{{\bm{\Lambda }}_{{{\text{ul}},i}}}{{\bm{h}}_{{{\text{ul}},i}}}{\bm{h}}_{{{\text{ul}},i}}^{\text{H}}{{\bm{\Lambda }}_{{{\text{ul}},i}}} + {{\bm{\Omega }}_{{{\text{ul}},i}}}{{\bm{z}}_{{{\text{ulp}},i}}}{\bm{z}}_{{{\text{ulp}},i}}^{\text{H}}{{\bm{\Omega }}_{{{\text{ul}},i}}} + {{\bm{\Lambda }}_{{{\text{ul}},i}}}{{\bm{h}}_{{{\text{ul}},i}}}{\bm{z}}_{{{\text{ulp}},i}}^{\text{H}}{{\bm{\Omega }}_{{{\text{ul}},i}}} + {{\bm{\Omega }}_{{{\text{ul}},i}}}{{\bm{z}}_{{{\text{ulp}},i}}}{\bm{h}}_{{{\text{ul}},i}}^{\text{H}}{{\bm{\Lambda }}_{{{\text{ul}},i}}}} \right){\bm{T}}_{{{\text{ul}},i}}^{\frac{{1}}{{2}}}.
\end{equation}
Substituting (\ref{a4_d11}) into (\ref{a4_d10}), we obtain
\begin{align}
\begin{split}
\label{a4_d12}
& {\bm{g}}_{{{\text{ul}},i}}^{\text{H}}{\bm{C}}_{{\text{ul}}}^{{\text{-1}}}{{\bm{T}}_{{{\text{ul}},k}}}{\bm{C}}_{{\text{ul}}}^{{\text{-1}}}{{\bm{g}}_{{{\text{ul}},i}}} \\
& = {\bm{h}}_{{{\text{ul}},i}}^{\text{H}}{\bm{T}}_{{{\text{ul}},i}}^{\frac{{1}}{{2}}}{\bm{C}}_{{{\text{ul}},[i]}}^{{\text{-1}}}{{\bm{T}}_{{{\text{ul}},k}}}{\bm{C}}_{{\text{ul}}}^{{\text{-1}}}{\bm{T}}_{{{\text{ul}},i}}^{\frac{{1}}{{2}}}{{\bm{h}}_{{{\text{ul}},i}}} \\
& \quad - \left[ {{p_{{\text{ul}},i}}{\bm{h}}_{{{\text{ul}},i}}^{\text{H}}{\bm{T}}_{{{\text{ul}},i}}^{\frac{{1}}{{2}}}{\bm{C}}_{{\text{ul}}}^{{\text{-1}}}{\bm{T}}_{{{\text{ul}},i}}^{\frac{{1}}{{2}}}{{\bm{\Lambda }}_{{{\text{ul}},i}}}{{\bm{h}}_{{{\text{ul}},i}}}\left( {{\bm{h}}_{{{\text{ul}},i}}^{\text{H}}{{\bm{\Lambda }}_{{{\text{ul}},i}}}{\bm{T}}_{{{\text{ul}},i}}^{\frac{{1}}{{2}}}{\bm{C}}_{{{\text{ul}},[i]}}^{{\text{-1}}}{{\bm{T}}_{{{\text{ul}},k}}}{\bm{C}}_{{\text{ul}}}^{{\text{-1}}}{\bm{T}}_{{{\text{ul}},i}}^{\frac{{1}}{{2}}}{{\bm{h}}_{{{\text{ul}},i}}}} \right)} \right. \\
& \quad \qquad + {p_{{\text{ul}},i}}{\bm{h}}_{{{\text{ul}},i}}^{\text{H}}{\bm{T}}_{{{\text{ul}},i}}^{\frac{{1}}{{2}}}{\bm{C}}_{{\text{ul}}}^{{\text{-1}}}{\bm{T}}_{{{\text{ul}},i}}^{\frac{{1}}{{2}}}{{\bm{\Omega }}_{{{\text{ul}},i}}}{{\bm{z}}_{{{\text{ulp}},i}}}\left( {{\bm{z}}_{{{\text{ulp}},i}}^{\text{H}}{{\bm{\Omega }}_{{{\text{ul}},i}}}{\bm{T}}_{{{\text{ul}},i}}^{\frac{{1}}{{2}}}{\bm{C}}_{{{\text{ul}},[i]}}^{{\text{-1}}}{{\bm{T}}_{{{\text{ul}},k}}}{\bm{C}}_{{\text{ul}}}^{{\text{-1}}}{\bm{T}}_{{{\text{ul}},i}}^{\frac{{1}}{{2}}}{{\bm{h}}_{{{\text{ul}},i}}}} \right) \\
& \quad \qquad + {p_{{\text{ul}},i}}{\bm{h}}_{{{\text{ul}},i}}^{\text{H}}{\bm{T}}_{{{\text{ul}},i}}^{\frac{{1}}{{2}}}{\bm{C}}_{{\text{ul}}}^{{\text{-1}}}{\bm{T}}_{{{\text{ul}},i}}^{\frac{{1}}{{2}}}{{\bm{\Lambda }}_{{{\text{ul}},i}}}{{\bm{h}}_{{{\text{ul}},i}}}\left( {{\bm{z}}_{{{\text{ulp}},i}}^{\text{H}}{{\bm{\Omega }}_{{{\text{ul}},i}}}{\bm{T}}_{{{\text{ul}},i}}^{\frac{{1}}{{2}}}{\bm{C}}_{{{\text{ul}},[i]}}^{{\text{-1}}}{{\bm{T}}_{{{\text{ul}},k}}}{\bm{C}}_{{\text{ul}}}^{{\text{-1}}}{\bm{T}}_{{{\text{ul}},i}}^{\frac{{1}}{{2}}}{{\bm{h}}_{{{\text{ul}},i}}}} \right) \\
& \quad \qquad + \left. {{p_{{\text{ul}},i}}{\bm{h}}_{{{\text{ul}},i}}^{\text{H}}{\bm{T}}_{{{\text{ul}},i}}^{\frac{{1}}{{2}}}{\bm{C}}_{{\text{ul}}}^{{\text{-1}}}{\bm{T}}_{{{\text{ul}},i}}^{\frac{{1}}{{2}}}{{\bm{\Omega }}_{{{\text{ul}},i}}}{{\bm{z}}_{{{\text{ulp}},i}}}\left( {{\bm{h}}_{{{\text{ul}},i}}^{\text{H}}{{\bm{\Lambda }}_{{{\text{ul}},i}}}{\bm{T}}_{{{\text{ul}},i}}^{\frac{{1}}{{2}}}{\bm{C}}_{{{\text{ul}},[i]}}^{{\text{-1}}}{{\bm{T}}_{{{\text{ul}},k}}}{\bm{C}}_{{\text{ul}}}^{{\text{-1}}}{\bm{T}}_{{{\text{ul}},i}}^{\frac{{1}}{{2}}}{{\bm{h}}_{{{\text{ul}},i}}}} \right)} \right].
\end{split}
\end{align}

Applying \cite[Lemma 4]{jzhang_large_2013}, the deterministic equivalents of each term in (\ref{a4_d12}) can be obtained as
\begin{flalign}
\label{a4_d13}
{\bm{h}}_{{{\text{ul}},i}}^{\text{H}}{\bm{T}}_{{{\text{ul}},i}}^{\frac{{1}}{{2}}}{\bm{C}}_{{{\text{ul}},[i]}}^{{\text{-1}}}{{\bm{T}}_{{{\text{ul}},k}}}{\bm{C}}_{{\text{ul}}}^{{\text{-1}}}{\bm{T}}_{{{\text{ul}},i}}^{\frac{{1}}{{2}}}{{\bm{h}}_{{{\text{ul}},i}}} - \left( {\dot u_{{\text{ul}},i}^{(1)} - \frac{{{p_{{\text{ul}},i}}u_{{\text{ul}},i}^{(2)}\dot u_{{\text{ul}},i}^{(2)}}}{{1 + {p_{{\text{ul}},i}}u_{{\text{ul}},i}^{(3)} + {p_{{\text{ul}},i}}u_{{\text{ul}},i}^{(4)}}}} \right)\xrightarrow{{a.s.}}0, \\
\label{a4_d14}
{\bm{h}}_{{{\text{ul}},i}}^{\text{H}}{\bm{T}}_{{{\text{ul}},i}}^{\frac{{1}}{{2}}}{\bm{C}}_{{\text{ul}}}^{{\text{-1}}}{\bm{T}}_{{{\text{ul}},i}}^{\frac{{1}}{{2}}}{{\bm{\Lambda }}_{{{\text{ul}},i}}}{{\bm{h}}_{{{\text{ul}},i}}} - \left( {u_{{\text{ul}},i}^{(2)} - \frac{{{p_{{\text{ul}},i}}u_{{\text{ul}},i}^{(2)}u_{{\text{ul}},i}^{(3)}}}{{1 + {p_{{\text{ul}},i}}u_{{\text{ul}},i}^{(3)} + {p_{{\text{ul}},i}}u_{{\text{ul}},i}^{(4)}}}} \right)\xrightarrow{{a.s.}}0, \\
\label{a4_d15}
{\bm{h}}_{{{\text{ul}},i}}^{\text{H}}{\bm{T}}_{{{\text{ul}},i}}^{\frac{{1}}{{2}}}{\bm{C}}_{{\text{ul}}}^{{\text{-1}}}{\bm{T}}_{{{\text{ul}},i}}^{\frac{{1}}{{2}}}{{\bm{\Omega }}_{{{\text{ul}},i}}}{{\bm{z}}_{{{\text{ulp}},i}}} - \frac{{ - {p_{{\text{ul}},i}}u_{{\text{ul}},i}^{(2)}u_{{\text{ul}},i}^{(4)}}}{{1 + {p_{{\text{ul}},i}}u_{{\text{ul}},i}^{(3)} + {p_{{\text{ul}},i}}u_{{\text{ul}},i}^{(4)}}}\xrightarrow{{a.s.}}0, \\
\label{a4_d16}
{\bm{h}}_{{{\text{ul}},i}}^{\text{H}}{{\bm{\Lambda }}_{{{\text{ul}},i}}}{\bm{T}}_{{{\text{ul}},i}}^{\frac{{1}}{{2}}}{\bm{C}}_{{{\text{ul}},[i]}}^{{\text{-1}}}{{\bm{T}}_{{{\text{ul}},k}}}{\bm{C}}_{{\text{ul}}}^{{\text{-1}}}{\bm{T}}_{{{\text{ul}},i}}^{\frac{{1}}{{2}}}{{\bm{h}}_{{{\text{ul}},i}}} - \left( {\dot u_{{\text{ul}},i}^{(2)} - \frac{{{p_{{\text{ul}},i}}u_{{\text{ul}},i}^{(2)}\dot u_{{\text{ul}},i}^{(3)}}}{{1 + {p_{{\text{ul}},i}}u_{{\text{ul}},i}^{(3)} + {p_{{\text{ul}},i}}u_{{\text{ul}},i}^{(4)}}}} \right)\xrightarrow{{a.s.}}0, \\
\label{a4_d17}
{\bm{z}}_{{{\text{ulp}},i}}^{\text{H}}{{\bm{\Omega }}_{{{\text{ul}},i}}}{\bm{T}}_{{{\text{ul}},i}}^{\frac{{1}}{{2}}}{\bm{C}}_{{{\text{ul}},[i]}}^{{\text{-1}}}{{\bm{T}}_{{{\text{ul}},k}}}{\bm{C}}_{{\text{ul}}}^{{\text{-1}}}{\bm{T}}_{{{\text{ul}},i}}^{\frac{{1}}{{2}}}{{\bm{h}}_{{{\text{ul}},i}}} - \frac{{ - {p_{{\text{ul}},i}}u_{{\text{ul}},i}^{(2)}\dot u_{{\text{ul}},i}^{(4)}}}{{1 + {p_{{\text{ul}},i}}u_{{\text{ul}},i}^{(3)} + {p_{{\text{ul}},i}}u_{{\text{ul}},i}^{(4)}}}\xrightarrow{{a.s.}}0,
\end{flalign}
where
\begin{align*}
&u_{{\text{ul}},i}^{(3)} = \frac{1}{{M{N_{\text{U}}}}}\text{tr}\left( {{\bm{T}}_{{{\text{ul}},i}}^{\frac{{1}}{{2}}}{\bm{\Lambda }}_{{{\text{ul}},i}}^{2}{\bm{T}}_{{{\text{ul}},i}}^{\frac{{1}}{{2}}}{\bm{C}}_{{\text{ul}}}^{{\text{-1}}}} \right), u_{{\text{ul}},i}^{(4)} = \frac{1}{{M{N_{\text{U}}}}}\text{tr}\left( {{\bm{T}}_{{{\text{ul}},i}}^{\frac{{1}}{{2}}}{\bm{\Omega }}_{{{\text{ul}},i}}^{2}{\bm{T}}_{{{\text{ul}},i}}^{\frac{{1}}{{2}}}{\bm{C}}_{{\text{ul}}}^{{\text{-1}}}} \right), \\
&\dot u_{{\text{ul}},i}^{(1)} = \frac{1}{{M{N_{\text{U}}}}}\text{tr}\left( {{{\bm{T}}_{{{\text{ul}},i}}}{\bm{C}}_{{\text{ul}}}^{{\text{-1}}}{{\bm{T}}_{{{\text{ul}},k}}}{\bm{C}}_{{\text{ul}}}^{{\text{-1}}}} \right), \dot u_{{\text{ul}},i}^{(2)} = \frac{1}{{M{N_{\text{U}}}}}\text{tr}\left( {{\bm{T}}_{{{\text{ul}},i}}^{\frac{{1}}{{2}}}{{\bm{\Lambda }}_{{{\text{ul}},i}}}{\bm{T}}_{{{\text{ul}},i}}^{\frac{{1}}{{2}}}{\bm{C}}_{{\text{ul}}}^{{\text{-1}}}{{\bm{T}}_{{{\text{ul}},k}}}{\bm{C}}_{{\text{ul}}}^{{\text{-1}}}} \right), \\
&\dot u_{{\text{ul}},i}^{(3)} = \frac{1}{{M{N_{\text{U}}}}}\text{tr}\left( {{\bm{T}}_{{{\text{ul}},i}}^{\frac{{1}}{{2}}}{\bm{\Lambda }}_{{{\text{ul}},i}}^{2}{\bm{T}}_{{{\text{ul}},i}}^{\frac{{1}}{{2}}}{\bm{C}}_{{\text{ul}}}^{{\text{-1}}}{{\bm{T}}_{{{\text{ul}},k}}}{\bm{C}}_{{\text{ul}}}^{{\text{-1}}}} \right), \dot u_{{\text{ul}},i}^{(4)} = \frac{1}{{M{N_{\text{U}}}}}\text{tr}\left( {{\bm{T}}_{{{\text{ul}},i}}^{\frac{{1}}{{2}}}{\bm{\Omega }}_{{{\text{ul}},i}}^{2}{\bm{T}}_{{{\text{ul}},i}}^{\frac{{1}}{{2}}}{\bm{C}}_{{\text{ul}}}^{{\text{-1}}}{{\bm{T}}_{{{\text{ul}},k}}}{\bm{C}}_{{\text{ul}}}^{{\text{-1}}}} \right).
\end{align*}
Applying Lemma \ref{plus} and \cite[Theorem 2]{jzhang_large_2013}, we have
\begin{align}
\label{a4_d18}
\begin{split}
u_{{\text{ul}},k}^{(3)} - \overline u_{{\text{ul}},k}^{(3)}\xrightarrow{{a.s.}}0, u_{{\text{ul}},k}^{(4)} - \overline u_{{\text{ul}},k}^{(4)}\xrightarrow{{a.s.}}0, \\
\dot u_{{\text{ul}},k}^{(1)} - \overline{\dot{u}}_{{\text{ul}},k}^{(1)}\xrightarrow{{a.s.}}0, \dot u_{{\text{ul}},k}^{(2)} - \overline{\dot{u}}_{{\text{ul}},k}^{(2)}\xrightarrow{{a.s.}}0, \\
\dot u_{{\text{ul}},k}^{(3)} - \overline{\dot{u}}_{{\text{ul}},k}^{(3)}\xrightarrow{{a.s.}}0, \dot u_{{\text{ul}},k}^{(4)} - \overline{\dot{u}}_{{\text{ul}},k}^{(4)}\xrightarrow{{a.s.}}0,
\end{split}
\end{align}
where $\overline{\dot{u}}_{{\text{ul}},i}^{(1)}$ and $\overline{\dot{u}}_{{\text{ul}},i}^{(2)}$ are given by (\ref{eq:u_ul_i_1_DE}) and (\ref{eq:u_ul_i_2_DE}), respectively, and
\begin{align*}
&\overline u_{{\text{ul}},i}^{(3)} = \sum\limits_{n = 1}^{{N_{\text{U}}}} {\frac{{\varphi _{{\text{ul}},i,n}^2}}{M}\text{tr}\left( {{\overline{\bm \Sigma }}_{n}^{{\text{ -1}}}{{\bm{T}}_{{{\text{ul}},i,n}}}{{\bm{\Psi }}_{{{\text{ul}},n}}}} \right)}, \overline u_{{\text{ul}},i}^{(4)} = \sum\limits_{n = 1}^{{N_{\text{U}}}} {\frac{{\tau _{{\text{ul}},i,n}^2}}{M}\text{tr}\left( {{\overline{\bm \Sigma }}_{n}^{{\text{ -1}}}{{\bm{T}}_{{{\text{ul}},i,n}}}{{\bm{\Psi }}_{{{\text{ul}},n}}}} \right)}, \\
&\overline{\dot{u}}_{{\text{ul}},i}^{(3)} = \sum\limits_{n = 1}^{{N_{\text{U}}}} {\frac{{\varphi _{{\text{ul}},i,n}^2}}{M}\left[ {\text{tr}\left( {{{\bm{T}}_{{{\text{ul}},i,n}}}{{\bm{\Psi }}_{{{\text{ul}},n}}}{{\bm{T}}_{{{\text{ul}},k,n}}}{{\bm{\Psi }}_{{{\text{ul}},n}}}} \right) - \sum\limits_{j = 1}^{{K_{\text{U}}}} {\dot c_{{\text{ul}},j,n}^{\text{(k)}}\text{tr}\left( {{{\bm{T}}_{{{\text{ul}},i,n}}}{{\bm{\Psi }}_{{{\text{ul}},n}}}{\overline{\bm \Sigma }}_{n}^{{\text{ -1}}}{{\bm{T}}_{{{\text{ul}},j,n}}}{{\bm{\Psi }}_{{{\text{ul}},n}}}} \right)} } \right]}, \\
&\overline{\dot{u}}_{{\text{ul}},i}^{(4)} = \sum\limits_{n = 1}^{{N_{\text{U}}}} {\frac{{\tau _{{\text{ul}},i,n}^2}}{M}\left[ {\text{tr}\left( {{{\bm{T}}_{{{\text{ul}},i,n}}}{{\bm{\Psi }}_{{{\text{ul}},n}}}{{\bm{T}}_{{{\text{ul}},k,n}}}{{\bm{\Psi }}_{{{\text{ul}},n}}}} \right) - \sum\limits_{j = 1}^{{K_{\text{U}}}} {\dot c_{{\text{ul}},j,n}^{\text{(k)}}\text{tr}\left( {{{\bm{T}}_{{{\text{ul}},i,n}}}{{\bm{\Psi }}_{{{\text{ul}},n}}}{\overline{\bm \Sigma }}_{n}^{{\text{ -1}}}{{\bm{T}}_{{{\text{ul}},j,n}}}{{\bm{\Psi }}_{{{\text{ul}},n}}}} \right)} } \right]}.
\end{align*}
In the above equations, ${{\bm{\Psi }}_{{{\text{ul}},n}}}$ is given by (\ref{eq:pha_ul}) and $\dot c_{{\text{ul}},j,n}^{\text{(k)}}$ is
\begin{align}
\label{a4_d19}
&\dot c_{{\text{ul}},j,n}^{\text{(k)}} = \frac{1}{M}\frac{{ - {p_{{\text{ul}},j}}\sum\limits_{m = 1}^{{N_{\text{U}}}} {\dot e_{{\text{ul}},j,m}^{\text{(k)}}} }}{{{{\left( {1 + \sum\limits_{m = 1}^{{N_{\text{U}}}} {{e_{{\text{ul}},j,m}}} } \right)}^2}}}, \\
\label{a4_d20}
&\dot e_{{\text{ul}},i,n}^{\text{(k)}} = \frac{1}{M}\left[ {\text{tr}\left( {{{\bm{T}}_{{{\text{ul}},i,n}}}{{\bm{\Psi }}_{{{\text{ul}},n}}}{{\bm{T}}_{{{\text{ul}},k,n}}}{{\bm{\Psi }}_{{{\text{ul}},n}}}} \right) - \sum\limits_{j = 1}^{{K_{\text{U}}}} {\dot c_{{\text{ul}},j,n}^{\text{(k)}}\text{tr}\left( {{{\bm{T}}_{{{\text{ul}},i,n}}}{{\bm{\Psi }}_{{{\text{ul}},n}}}{\overline{\bm \Sigma }}_{n}^{{\text{ -1}}}{{\bm{T}}_{{{\text{ul}},j,n}}}{{\bm{\Psi }}_{{{\text{ul}},n}}}} \right)} } \right].
\end{align}
Substituting (\ref{a4_d20}) into (\ref{a4_d19}), we find that ${\dot{\bm C}}_{{\text{ul}}}^{{\text{(k)}}} = \left[ {\dot c_{{\text{ul}},n,i}^{\text{(k)}}} \right] \in {\mathbb{C}^{{K_{\text{U}}} \times {N_{\text{U}}}}}$ is a solution to the linear equation (\ref{eq:li_eq_Cul}). Plugging (\ref{a4_d13}) - (\ref{a4_d17}) into (\ref{a4_d12}), and then substituting (\ref{a4_d12}) into (\ref{a4_d9}), ${\overline u_{{\text{ul}},i}}$ is explicitly expressed in (\ref{eq:u_ul_i_DE}).
\end{IEEEproof}

\begin{lemma}\label{lemma:UL_noise}
Letting Assumption \ref{asm:g_ul} hold true and as $\mathcal{N} \to \infty $, we have
\begin{equation}
\label{a4_d21}
{\hat{\bm g}}_{{{\text{ul}},k}}^{\text{H}}{\bm{C}}_{{\text{ul}}}^{{\text{-1}}}{\bm{\Sigma C}}_{{\text{ul}}}^{{\text{-1}}}{{\hat{\bm g}}_{{{\text{ul}},k}}} - \frac{{\overline{\dot{u}}_{{\text{ul}},k,\Sigma }^{\text{(1)}}}}{{{{\left( {1 + {p_{{\text{ul}},k}}\overline u_{{\text{ul}},k}^{\text{(1)}}} \right)}^2}}}\xrightarrow{{a.s.}}0,
\end{equation}
where $\overline u_{{\text{ul}},k}^{\text{(1)}}$ and $\overline{\dot{u}}_{{\text{ul}},k,\Sigma }^{\text{(1)}}$ are given by (\ref{eq:u_ul_k_1}) and (\ref{eq:u_ul_k_sigma_1_DE}), respectively.
\end{lemma}

\begin{IEEEproof}
Applying Lemma \ref{matrix_inverse}, we obtain
\begin{equation}
\label{a4_d22}
{\hat{\bm g}}_{{{\text{ul}},k}}^{\text{H}}{\bm{C}}_{{\text{ul}}}^{{\text{-1}}}{\bm{\Sigma C}}_{{\text{ul}}}^{{\text{-1}}}{{\hat{\bm g}}_{{{\text{ul}},k}}} = \frac{1}{{{{\left( {1 + {p_{{\text{ul}},k}}{\hat{\bm g}}_{{{\text{ul}},k}}^{\text{H}}{\bm{C}}_{{{\text{ul}},[k]}}^{{\text{-1}}}{{{\hat{\bm g}}}_{{{\text{ul}},k}}}} \right)}^2}}}{\hat{\bm g}}_{{{\text{ul}},k}}^{\text{H}}{\bm{C}}_{{{\text{ul}},[k]}}^{{\text{-1}}}{\bm{\Sigma C}}_{{{\text{ul}},[k]}}^{{\text{-1}}}{{\hat{\bm g}}_{{{\text{ul}},k}}}.
\end{equation}
From (\ref{a4_d3}), we know that the deterministic equivalent ${\hat{\bm g}}_{{{\text{ul}},k}}^{\text{H}}{\bm{C}}_{{{\text{ul}},[k]}}^{{\text{-1}}}{{\hat{\bm g}}_{{{\text{ul}},k}}}$ is $\overline u_{{\text{ul}},k}^{\text{(1)}}$. Besides, based on Lemma \ref{same_matrix}, ${\hat{\bm g}}_{{{\text{ul}},k}}^{\text{H}}{\bm{C}}_{{{\text{ul}},[k]}}^{{\text{-1}}}{\bm{\Sigma C}}_{{{\text{ul}},[k]}}^{{\text{-1}}}{{\hat{\bm g}}_{{{\text{ul}},k}}}$ almost surely converges to $\dot u_{{\text{ul}},k,\Sigma }^{\text{(1)}} = \frac{1}{{M{N_{\text{U}}}}}\text{tr}\left( {{\bm{T}}_{{{\text{ul}},k}}^{}{\bm{C}}_{{{\text{ul}},[k]}}^{{\text{-1}}}{\bm{\Sigma C}}_{{{\text{ul}},[k]}}^{{\text{-1}}}} \right)$. Similar to the derivations of $\overline{\dot{u}}_{{\text{ul}},k}^{\text{(1)}}$, we have
\begin{equation}
\label{a4_d23}
\dot u_{{\text{ul}},k,\Sigma }^{\text{(1)}} - \overline{\dot{u}}_{{\text{ul}},k,\Sigma }^{\text{(1)}}\xrightarrow{{a.s.}}0,
\end{equation}
where $\overline{\dot{u}}_{{\text{ul}},k,\Sigma }^{\text{(1)}}$ is given by (\ref{eq:u_ul_k_sigma_1_DE}). Thus, the deterministic equivalent ${\hat{\bm g}}_{{{\text{ul}},k}}^{\text{H}}{\bm{C}}_{{\text{ul}}}^{{\text{-1}}}{\bm{\Sigma C}}_{{\text{ul}}}^{{\text{-1}}}{{\hat{\bm g}}_{{{\text{ul}},k}}}$ can be derived as in (\ref{a4_d21}).
\end{IEEEproof}

From Lemmas \ref{lemma:UL_signal}, \ref{lemma:UL_interference} and \ref{lemma:UL_noise}, we derive the deterministic equivalent ${\overline \gamma _{{\text{ul}},k}}$, as shown in (\ref{eq:r_ul_DE}), of uplink SINR with MMSE receiver ${\gamma _{{\text{ul}},k}}$. Similar to the proof of Theorem \ref{thm:R_dl_rzf}, we know that when $\mathcal{N} \to \infty $, it satisfies that $\frac{1}{{{K_{\text{U}}}}}\left( {{R_{{\text{ul}},{\text{sum}}}} - {{\overline R}_{{\text{ul}},{\text{sum}}}}} \right)\xrightarrow{{a.s.}}0$, where ${\overline R_{{\text{ul}},{\text{sum}}}}$ is given by (\ref{eq:R_ul_DE}). This completes the proof.

%

\bibliographystyle{IEEEtran}
\bibliography{NAFD_EV}

\end{document}